\documentclass[twocolumn]{aastex631}
\usepackage{CJK}

\newcommand{\dotarcsec}{\rlap{.}''}
\newcommand{\lya}{Ly$\alpha$}
\newcommand{\m}{\ensuremath{m_\mathrm{F277W}}}

\begin{document}
\begin{CJK*}{UTF8}{gbsn}

\title{NGDEEP Epoch 1: The Faint-End of the Luminosity Function at $z \sim$ 9--12 from Ultra-Deep JWST Imaging}

\author[0000-0002-9393-6507]{Gene C. K. Leung}
\affiliation{Department of Astronomy, The University of Texas at Austin, Austin, TX, USA}

\author[0000-0002-9921-9218]{Micaela B. Bagley}
\affiliation{Department of Astronomy, The University of Texas at Austin, Austin, TX, USA}

\author[0000-0001-8519-1130]{Steven L. Finkelstein}
\affiliation{Department of Astronomy, The University of Texas at Austin, Austin, TX, USA}

\author[0000-0001-7113-2738]{Henry C. Ferguson}
\affiliation{Space Telescope Science Institute, 3700 San Martin Drive, Baltimore, MD 21218, USA}

\author[0000-0002-6610-2048]{Anton M. Koekemoer}
\affiliation{Space Telescope Science Institute, 3700 San Martin Drive, Baltimore, MD 21218, USA} 

\author[0000-0003-4528-5639]{Pablo G. P\'erez-Gonz\'alez}
\affiliation{Centro de Astrobiolog\'{\i}a (CAB), CSIC-INTA, Ctra. de Ajalvir km 4, Torrej\'on de Ardoz, E-28850, Madrid, Spain}

\author[0000-0003-4965-0402]{Alexa Morales}
\affiliation{Department of Astronomy, The University of Texas at Austin, Austin, TX, USA}

\author[0000-0002-8360-3880]{Dale D. Kocevski}
\affiliation{Department of Physics and Astronomy, Colby College, Waterville, ME 04901, USA}

\author[0000-0001-8835-7722]{Yang, G. (杨光)}
\affiliation{Kapteyn Astronomical Institute, University of Groningen, P.O. Box 800, 9700 AV Groningen, The Netherlands}
\affiliation{SRON Netherlands Institute for Space Research, Postbus 800, 9700 AV Groningen, The Netherlands}

\author[0000-0002-6748-6821]{Rachel S. Somerville}
\affiliation{Center for Computational Astrophysics, Flatiron Institute, 162 5th Avenue, New York, NY, 10010, USA}

\author[0000-0003-3903-6935]{Stephen M. Wilkins}
\affiliation{Astronomy Centre, Department of Physics and Astronomy, University of Sussex, Brighton, BN1 9QH, UK}

\author[0000-0003-3466-035X]{{L. Y. Aaron} {Yung}}
\altaffiliation{NASA Postdoctoral Fellow}
\affiliation{Astrophysics Science Division, NASA Goddard Space Flight Center, 8800 Greenbelt Rd, Greenbelt, MD 20771, USA}

\author[0000-0001-7201-5066]{Seiji Fujimoto}
\altaffiliation{Hubble Fellow}
\affiliation{Department of Astronomy, The University of Texas at Austin, Austin, TX, USA}

\author[0000-0003-2366-8858]{Rebecca L. Larson}
\altaffiliation{NSF Graduate Fellow}
\affiliation{Department of Astronomy, The University of Texas at Austin, Austin, TX, USA}

\author[0000-0001-7503-8482]{Casey Papovich}
\affiliation{Department of Physics and Astronomy, Texas A\&M University, College Station, TX, 77843-4242 USA}
\affiliation{George P.\ and Cynthia Woods Mitchell Institute for Fundamental Physics and Astronomy, Texas A\&M University, College Station, TX, 77843-4242 USA}

\author[0000-0003-3382-5941]{Nor Pirzkal}
\affiliation{ESA/AURA Space Telescope Science Institute, 3700 San Martin Drive, Baltimore, MD, 21218}

\author[0000-0002-4153-053X]{Danielle A. Berg}
\affiliation{Department of Astronomy, The University of Texas at Austin, Austin, TX, USA}

\author[0000-0003-3130-5643]{Jennifer M. Lotz}\affiliation{International Gemini Observatory/NSF's National Optical-Infrared Astronomy Research Laboratory, 950 N. Cherry Ave., Tucson, AZ 85719, USA}

\author[0000-0001-9875-8263]{Marco Castellano}
\affiliation{INAF - Osservatorio Astronomico di Roma, via di Frascati 33, 00078 Monte Porzio Catone, Italy}

\author[0000-0002-0786-7307]{\'{O}scar A. Ch\'{a}vez Ortiz}
\affiliation{Department of Astronomy, The University of Texas at Austin, Austin, TX, USA}

\author[0000-0001-8551-071X]{Yingjie Cheng}
\affiliation{Department of Astronomy, University of Massachusetts Amherst, 710 North Pleasant Street, Amherst, MA 01003-9305, USA}

\author[0000-0001-5414-5131]{Mark Dickinson}\affiliation{NSF's National Optical-Infrared Astronomy Research Laboratory, 950 N. Cherry Ave., Tucson, AZ 85719, USA}

\author[0000-0002-7831-8751]{Mauro Giavalisco}
\affiliation{Department of Astronomy, University of Massachusetts Amherst, 710 North Pleasant Street, Amherst, MA 01003-9305, USA}

\author[0000-0001-6145-5090]{Nimish P. Hathi}
\affiliation{Space Telescope Science Institute, 3700 San Martin Drive, Baltimore, MD 21218, USA}

\author[0000-0001-6251-4988]{Taylor A. Hutchison}
\altaffiliation{NASA Postdoctoral Fellow}
\affiliation{Astrophysics Science Division, NASA Goddard Space Flight Center, 8800 Greenbelt Rd, Greenbelt, MD 20771, USA}

\author[0000-0003-1187-4240]{Intae Jung}
\affiliation{Space Telescope Science Institute, 3700 San Martin Drive, Baltimore, MD 21218, USA}

\author[0000-0001-9187-3605]{Jeyhan S. Kartaltepe}
\affiliation{Laboratory for Multiwavelength Astrophysics, School of Physics and Astronomy, Rochester Institute of Technology, 84 Lomb Memorial Drive, Rochester, NY 14623, USA}\

\author[0000-0002-5554-8896]{Priyamvada Natarajan}
\affiliation{Department of Astronomy, Yale University, 52 Hillhouse Avenue, New Haven, CT 06511, USA}
\affiliation{Department of Physics, Yale University, P.O. Box 208121, New Haven, CT 06520, USA}
\affiliation{Black Hole Initiative at Harvard University, 20 Garden Street, Cambridge, MA 02138, USA}

\author[0000-0003-2283-2185]{Barry Rothberg}
\affiliation{Department of Physics and Astronomy, George Mason University, 4400 University Drive, MSN 3F3, Fairfax, VA 22030, USA}
\affiliation{U.S. Naval Observatory, 3450 Massachusetts Avenue NW, Washington, DC 20392, USA}

\begin{abstract}
We present a robust sample of very high-redshift galaxy candidates from the first epoch of {\it JWST}/NIRCam imaging from the Next Generation Extragalactic Exploratory Deep (NGDEEP) Survey.  The NGDEEP NIRCam imaging in the Hubble Ultra Deep Field Parallel Field 2 (HUDF-Par2) reaches $m=30.4$ (5$\sigma$, point-source) in F277W, making it the deepest public {\it JWST} GO imaging dataset to date.  We describe our detailed data reduction process of the six-filter broad-band {\it JWST}/NIRCam imaging, incorporating custom corrections for systematic effects to produce high-quality calibrated images. 
Using robust photometric redshift selection criteria, we identify a sample of 38 $z \gtrsim 9$ galaxy candidates. These objects span a redshift range of $z=8.5-15.8$, and apparent magnitudes of $\m = 27-30.5$ AB mag, reaching $\sim 1.5$ mag deeper than previous public {\it JWST} imaging surveys. We calculate the rest-frame ultraviolet (UV) luminosity function at $z \sim$ 9 and 11, and present a new measurement of the luminosity function faint-end slope at $z \sim 11$. There is no significant evolution in the faint-end slope and number density from $z=9$ to 11. Comparing our results with theoretical predictions, we find that some models produce better agreement at the faint end than the bright end. These results will help to constrain how stellar feedback impacts star formation at these early epochs.

\end{abstract}


\section{Introduction} \label{sec:intro}

The study of early galaxies is key to our understanding of the universe. Crucial questions remain unanswered, including how galaxies initially formed and evolved at early times when physical conditions were vastly different from today, how the first early supermassive black holes (SMBHs) formed and grew, as well as what kinds of sources dominated the cosmic reionization of the intergalactic medium. In the past decade, observations with the {\it Hubble Space Telescope} ({\it HST}) have advanced our understanding of the physical properties and demographics of galaxies to up to $z \simeq 10$ \citep[e.g.,][]{ish15, mcl16, roj20, bou21, bag22a, fin22a}. {\it HST} has only scratched the surface of the $z\sim 11$ universe, and its discoveries were limited to bright galaxies \citep{oes16}. Therefore, the $z>10$ universe, notably the faint galaxy population, remains largely unexplored due to {\it HST}'s moderate light collecting area and lack of sufficient wavelength coverage in the infrared into which the bulk of the rest-UV-optical emission from galaxies is redshifted.

The commissioning of {\it JWST} \citep{gar23} in 2022 has quickly transformed the frontier in the study of early galaxies thanks to its $7\times$ light-collecting area, superior sensitivity and large imaging field of view of NIRCam \citep{rie03, rie05}. Shortly after its commissioning, numerous studies have used early imaging to identify a large number of high-redshift galaxy candidates from $z\simeq 9-18$ \citep[e.g.][]{bou22a, cas22a, fin22b, fin23, nai22, adams23, atek23, don23b, har23, pg23, rob23, yan23}.

Historically, public deep field observations have been a driving force in advancing the redshift frontier of astronomical observations, with deep field imaging datasets proving to be a treasure trove for the study of galaxies at early times. The iconic Hubble Deep Field \citep[HDF,][]{wil96, thom99, dic00} has resulted in the detection of galaxies out to $z\sim 3$ using its deep near-infrared (NIR) imaging, while the succeeding Hubble Ultra Deep Field \citep[HUDF,][]{bec06} and its WFC3 NIR addition (HUDF09; \citealt{oes10}) has led to the discovery of hundreds of galaxies at $z>6$ \citep{bou06, bou10, fin10, oes10}. The combination of various legacy {\it HST} deep field data has enabled constraints of galaxy evolution up to $z\simeq 10$ \citep[e.g.,][]{ell13, bou15, fin15, bou21}. While early public {\it JWST} imaging datasets from Early Release Science programs \citep{tre22, fin23} have probed higher redshifts than {\it HST} due to their redder wavelength coverage, they have yet to exceed the depths reached by the {\it HST} HUDF.

A crucial advantage of deep field observations is their ability to detect faint objects, such as low-mass galaxies. Galaxies of different masses and luminosities constrain different physical processes. At low redshift, it is well established that the number density of massive galaxies is mainly driven by feedback from accreting black holes \citep{SD15}, while at high redshift, theory predicts that the abundance of massive, high-luminosity galaxies is mainly sensitive to the efficiency of converting gas into stars (gas depletion time; \citealt{yun19a,yun19b}). 
At both high and low redshift, the number of low luminosity galaxies is shaped mainly by how efficiently stellar driven winds can heat and eject gas from galaxies and their halos, and how the mass and energy loading of these winds scale with global galaxy properties like circular velocity or halo mass \citep{yun19a}. Many cosmological simulations of galaxy formation assume phenomenological functions for these scaling relations, which are typically tuned to match galaxy number densities in the local universe. Testing whether these same scaling relations can also reproduce the number density of faint galaxies in the early universe is a critical stress test for the stellar feedback recipes in these models. Furthermore, models predict that faint galaxies played a major role in reionizing the Universe \citep{yun20, yun20b}.

In this paper, we present a study of galaxies at $z>9$ using new ultra-deep NIRCam observations from the first half of the Next Generation Deep Extragalactic Exploratory Public (NGDEEP) Survey. We have obtained and reduced $\simeq 50$ hours of NIRCam imaging data in the HUDF-Par2 field in six broad-band filters. Our data reaches $5\sigma$ detection limits of $29.9-30.4$ mag, making it the deepest public {\it JWST} GO imaging dataset to date. This allows us to probe faint ($M_\mathrm{UV} \gtrsim -18$) galaxies in the early universe unexplored by previous early public {\it JWST} programs, placing important constraints on this population of faint galaxies at $z \gtrsim 9$.

This paper is organized as follows. In Section \ref{sec:data}, we describe our observations and data reduction process. We explain our methodology for the selection of $z \gtrsim 9$ galaxies in Section \ref{sec:method}. Section \ref{sec:results} presents the main results from our galaxy sample, including the luminosity function at $z\sim$ 9 and 11. We compare our results with recent observational studies and theoretical predictions in Section \ref{sec:discuss}. In Section \ref{sec:conclusions}, we summarize our findings.

Throughout this paper, we assume a \citet{pla20} cosmology of $H_0=67.4 ~ \mathrm{km~s}^{-1} ~\mathrm{Mpc}^{-1}$, $\Omega_\mathrm{m}=0.315$ and $\Omega_\mathrm{\Lambda} = 0.685$. All magnitudes are in the AB system.

\section{Data}
\label{sec:data}

NGDEEP is a deep slitless spectroscopic and imaging {\it JWST} Cycle 1 treasury program (\citealt{bag23}, PID 2079, PIs: S. Finkelstein, C. Papovich, N. Pirzkal) designed to study feedback mechanisms in galaxies through cosmic time. NGDEEP utilizes parallel {\it JWST} observations to simultaneously target the HUDF with NIRISS and the HUDF-Par2 field with NIRCam. NGDEEP consists of two observations with identical configurations, except for a position angle rotated by $3^\circ$ to allow for improved contamination subtraction in the NIRISS data. While the full program was planned for January to February 2023, only half of the program (one observation) was performed due to a temporary suspension of operations for NIRISS causing the NGDEEP observations to be pushed to the edge of the visibility window. The next visibility window satisfying the PA requirement of the parallel observations will occur in early 2024, when the remaining observations are expected to be taken. In this study, we report results using NIRCam data from the first half of the NGDEEP program. We supplement our NIRCam data with legacy {\it HST}/ACS F814W imaging in the HUDF-Par2 field.

\begin{figure*}[thbp]
	\centering
		\includegraphics[width=0.99\textwidth]{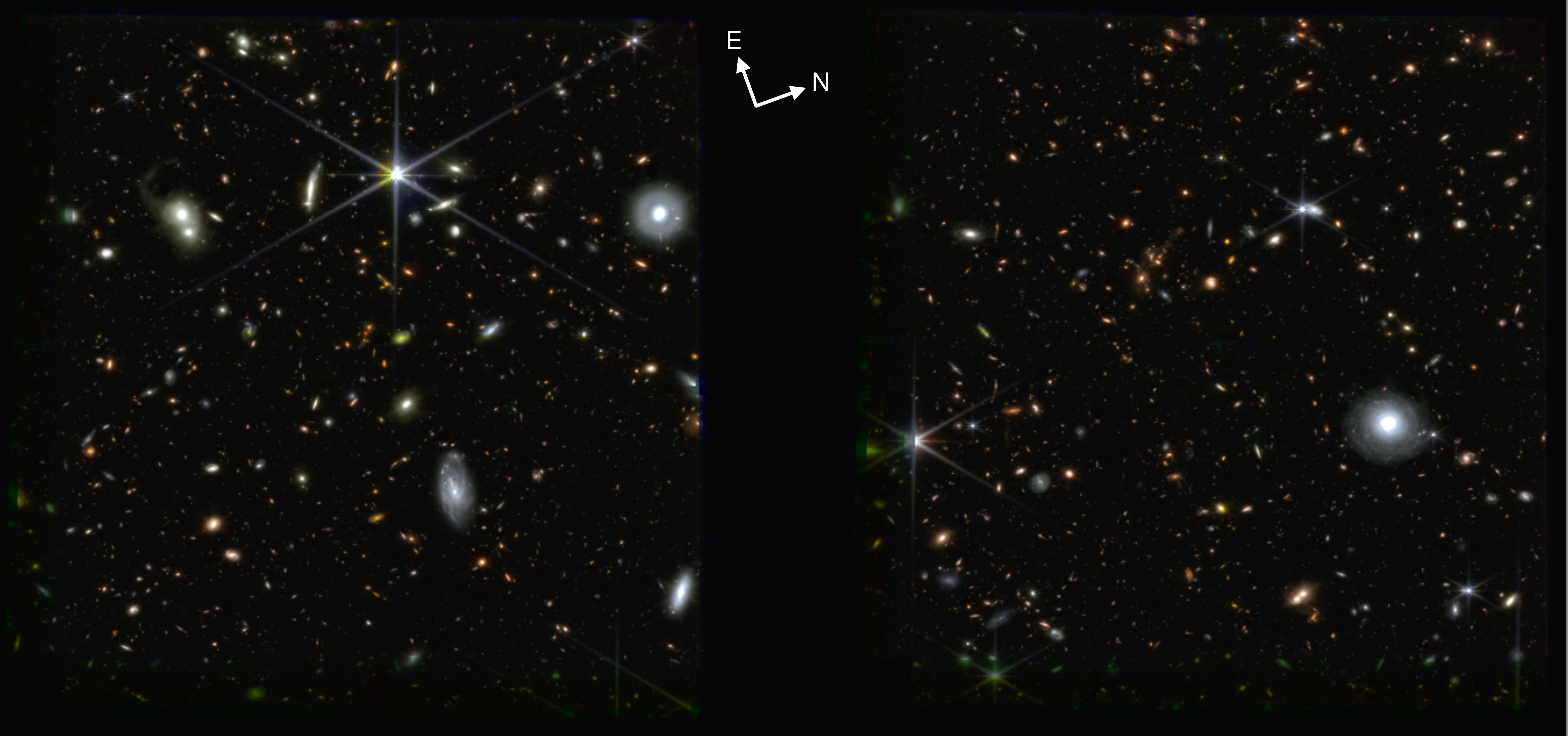}
		\caption{Color composite (b=F115W+F150W, g=F200W+F277W, r=F277W+F356W) image of the NGDEEP field from our NIRCam mosaics. These fully reduced mosaics will be made publicly available following the NGDEEP release schedule. The NIRCam imaging reaches 29.9--30.4 AB mag, making it the deepest public {\it JWST} imaging dataset at the time of writing.}
		\label{fig:mosaic}
\end{figure*}

\subsection{JWST Imaging}

The Epoch 1 NGDEEP NIRCam data obtained in 2023 alone represent the deepest public {\it JWST} imaging dataset taken in the first year of its operations. Observations were taken in six filters: F115W, F150W, and F200W with the short-wavelength detector, and F277W, F356W, and F444W with the long-wavelength detector. We use a combination of DEEP8 and SHALLOW4 readout patterns to achieve exposure times of 97 ks (F115W), 93 ks (F444W), and 32 -- 45 ks (F150W, F200W, F270W, and F356W). The exposure times are allocated to reach approximately uniform depth in all filters except F115W, where we increase the exposure time to improve detection of the \lya ~break at $z \gtrsim 9$.

We reduce the raw NIRCam imaging data using the {\it JWST} pipeline with custom modifications to correct for additional features in the data. The custom modifications were first employed by the CEERS survey, and are described in detail in \citet{bag22}. For NGDEEP, we include an additional custom procedure to remove residual flat field features from the images. Here we highlight some key aspects of the data reduction.

\subsubsection{Custom Flat Field Correction}\label{sec:flats}

For the first year of the commissioning of {\it JWST}, the reference flat images available through the CRDS were produced prior to the launch of using pre-flight data, and likely do not fully capture the most up to date in-flight flat field patterns of the NIRCam detector. Therefore, we apply an additional custom flat field correction to remove residual flat field features. The custom flat images are constructed as follows using {\it JWST} pipeline version 1.9.2 and CRDS context 1045.

We downloaded available public NIRCam imaging data from 14 extragalactic programs on the MAST archive in the six NGDEEP filters. This resulted in $\sim 200-400$ exposures per detector per filter. We reduced these exposures following the same procedures as our science data up to and including flat field correction using the reference flat images. Therefore, any flat field features left in these images will be the result of the residuals of the reference flat images. We then detect and mask sources in each image before median combining all the images to obtain a source-free sky flat image per filter per detector.
These custom sky flats are applied to the short-wavelength filters in our final reduction during Stage 2 processing as described in the following section.

\subsubsection{Data Reduction}

At the time of data acquisition, we initially reduced the science images using {\it JWST} pipeline version 1.9.2 and CRDS context 1045, combining with our custom sky flats. In May 2023, in-flight reference flats for the long-wavelength channels were released with CRDS context 1084 shortly before the submission of this paper. The reference flats for the short-wavelength channels were not updated in this context. Therefore, we re-reduced the images for our long-wavelength channels (F277W, F356W and F444W) using pipeline version 1.10.2 and CRDS context 1084. We follow the same procedures as the initial reduction, except that we do not include our custom sky flats in Stage 2, and use only the updated reference flats. 
In this section, we describe our data reduction process including the different flat-field treatments for the short and long-wavelength channels.

We reduce the science images using the following procedures. We first process the raw images through detector-level calibration by Stage 1 of the calibration pipeline using default parameters. We then perform custom corrections to flag and remove snowballs from all exposures, remove wisps from the F115W, F150W and F200W exposures using the wisp templates produced by the JADES collaboration in October 2022 \citep[S. Tacchella, priv. comm.;][]{tacchella23,rieke23}, 
and remove $1/f$ noise from all exposures. We find that the available wisp templates are not able to fit all the wisp features, introducing additional noise in the images.

We then process the exposures through Stage 2 of the calibration pipeline. In our initial reduction (pipeline v1.9.2, CRDS context 1045), we run Stage 2 in two steps to incorporate our custom sky flats. In the first step, we run Stage 2 up to and including flat field correction using default parameters and the reference flat images. 
In the second step, we resume Stage 2 by performing flat field correction using the custom sky flats (Section~\ref{sec:flats}) and the default flux calibration. This produces images in units of MJy/sr. We find that the inclusion of the custom sky flats has improved the $5\sigma$ photometric depths in the final mosaics by $0.28-0.64$ mag in the long-wavelength channels and $0.07-0.12$ mag in the short-wavelength channels (all compared to a reduction with CRDS context 1045 without this custom correction; see Section \ref{sec:depth} for the depth estimation procedures). 

In the reduction using the updated reference flats for the long-wavelength channels (pipeline v1.10.2, CRDS context 1084), we perform Stage 2 in one step, without including the custom sky flats. We find that the new reference flats have slightly improved the quality of the F277W and F444W mosaics, leading to an additional $\sim 0.1$ mag increase in $5\sigma$ depths over the version with our custom flat correction. The depth for F356W has decreased by $<0.1$ mag with the new reference flats. We have also tested the reduction using the new CRDS in a short wavelength channel, F200W. We found that our previous reduction using our custom sky flats provides higher image quality than the new CRDS alone, with the former reaching $\sim 0.1$ mag deeper in this short-wavelength channel. For the reasons stated above, we use our reduction with custom sky flats using CRDS context 1045 for the short-wavelength channels, and the reduction with in-flight reference flats using CRDS context 1084 for the long-wavelength channels.


Before processing the images through Stage 3 of the pipeline, we align the images using a custom version of the \texttt{TweakReg} routine of the calibration pipeline. Our modified approach uses \texttt{Source Extractor} \citep{ber96} to create catalogs for each individual image to provide improved source detection, deblending and centroiding over the pipeline. We align the images using a reference catalog constructed from a {\it HST}/ACS F814W 30 mas/pixel mosaic in the HUDF-Par2 field with astrometry tied to Gaia DR3. We remove stars from the reference catalog by excluding sources with a stellarity $\ge 0.8$ and FWHM $\le 5$ pixels, since the proper motion of stars over the large observation time difference of the {\it HST} and {\it JWST} imaging can adversely affect astrometric alignment. 

We obtain excellent relative alignment between NIRCam filters, with median offsets of $\lesssim 3$ mas, and a median absolute deviation (MAD) of $5-8$ mas. Absolute alignment between ACS and NIRCam is achieved within a median offset of $\lesssim 10$ mas and MAD of $10-14$ mas. A small systematic offset is observed with the NIRCam coordinates shifted in the positive declination direction by a median of $5-10$ mas across all NIRCam filters. In the F115W filter, we do not obtain satisfactory astrometric alignment for the SHALLOW4 exposures, where few sources are available for alignment due to a combination of the short exposure time and lower filter transmission. We have excluded these exposure from our analysis, which comprises 3.6 ks of exposure time out of a total of 97 ks in F115W. 

We then perform outlier detection on the aligned images using the calibration pipeline using \texttt{maskpt} = 0.5, \texttt{nhigh} = 1, \texttt{good\_bits} = \texttt{$\sim$ DO\_NOT\_USE+UNRELIABLE\_SLOPE}, and the default values for all other parameters. Next, we subtract a pedestal background value from each image, robustly measure the sky variance, and scale the read noise variance maps to match the measured values. Thereafter, we create a mosaic for each filter using the \texttt{Resample} routine in the calibration pipeline to drizzle the images onto a pixel scale of 30 mas/pixel and to the same WCS of the {\it HST}/ACS F814W reference image, so that the mosaics in all filters in NIRCam and ACS are pixel-aligned. Finally, we estimate and subtract any remaining background in the mosaics using a custom procedure that masks sources in all filters to create a combined source mask before fitting a two-dimensional model to the global background.


\subsection{HST Imaging}

The existing archival public {\it HST} imaging across the NGDEEP region was retrieved from the MAST Archive\footnote{\url{https://archive.stsci.edu}} and processed into mosaic combination, incorporating improvements in astrometry to align these images directly to Gaia DR3\footnote{\url{https://www.cosmos.esa.int/web/gaia/dr3}}, following approaches first described in \citet{koe11} where more specific details are presented. Briefly, the HST ACS/WFC imaging data in F814W were first processed for each individual visit, using the \texttt{DrizzlePac}\footnote{\url{https://github.com/spacetelescope/drizzlepac}} \texttt{TweakReg} routine to align all the exposures to one another within each visit, and subsequently align each full visit directly to Gaia DR3, taking account the proper motions of Gaia stars by applying the proper motion corrections to the epoch of observation in all cases. Within each visit, excellent alignment was achieved between all the individual exposures, with median absolute deviation (MAD) generally $\lesssim 3 - 5$ mas. Exposures within overlapping visits were similarly aligned with each other, to a similar level of accuracy, across the entire field. For absolute astrometry, the visits were all aligned directly to Gaia DR3, reaching an overall level of absolute astrometric alignment accuracy across the entire field to MAD values $\lesssim 7 - 9$ mas, generally limited by uncertainties in the proper motion values across the long time baselines spanning up to two decades.

The full aligned dataset was then processed through \texttt{DrizzlePac} \texttt{AstroDrizzle} to produce a combined mosaic at 30 mas/pixel, a sufficiently small pixel scale to provide Nyquist sampling of the {\it HST} ACS PSF, and with the drizzle weighting using custom inverse variance images (IVM) that were created for each individual exposure taking into account all the background noise terms for that exposure, including the sky emission which can vary throughout an orbit for {\it HST}. The resulting ACS F814W mosaic was then used to produce a catalog with \texttt{Source Extractor}, which was subsequently trimmed to remove stars by excluding all sources with stellarity $\geq 0.8$ and FWHM $< 5$ pixels (0$\farcs$15), since most stars are too faint to be included in the Gaia catalog and therefore their proper motions are unknown, which can adversely impact the quality of the astrometric alignment when this catalog is used to align the {\it JWST} data, given the significant time baseline since the {\it HST} data were obtained.

\subsection{Photometric Catalog}

The photometric catalog procedure is very similar to Finkelstein et al.\ 2023 (in prep), which we summarize here.  Photometry was measured with \textsc{Source Extractor} (\textsc{SE}; v.2.25.0) in dual-image mode, using a weighted mean of the F277W and F356W as the detection image.  Photometry was then performed on each of the six NIRCam images and the F814W {\it HST}/ACS image.  Colors were measured in Kron apertures, using PHOT\_AUTOPARAMS $=$ 1.1, 1.6 (colors were also measured in circular apertures with a range of radii for later use). For images with PSF FWHM smaller than F277W, we derived convolution kernels with \textsc{Pypher} to convolve their PSFs to match that of the F277W image.  For the F356W and F444W images, which had larger PSFs, a source-specific correction factor was derived as the ratio between the flux in a F277W image convolved to match the PSF of a given image, and the native F277W image.  To derive estimates of the total flux, an aperture correction was derived as the ratio of the flux in the larger default Kron aperture (PHOT\_AUTOPARAMS $=$ 2.5, 3.5).  Finally, these total flux corrections were validated via source-injection simulations, where mock sources with a range of magnitudes were added to the images, with fluxes measured in the same way as real sources.  An additional magnitude-dependent correction was found to be needed to correctly estimate total fluxes, ranging from $\sim$2\% at m$ =$ 25, to 8\% at $m =$ 29.   

Flux uncertainties were measured following \citet{pap16} and \citet{fin23}.  We first measured the fluxes in each image at a range of random positions with 30 different circular apertures with diameters ranging from 3 to 100 pixels.  The noise in each aperture diameter was calculated as the normalized median absolute deviation of the measured fluxes.  Then a functional form was fit to the measured noise as a function of pixels $N$ enclosed in an aperture, using a function of the form:
\begin{equation}
    \sigma_N = \sigma_1 (\alpha N^{\beta} + \gamma N^{\delta}), 
\end{equation}
using MCMC to derive posterior constraints on the free parameters $\alpha$, $\beta$, $\gamma$, and $\delta$, where the pixel-to-pixel rms $\sigma_1$ was measured directly from the images.  This equation was then used to estimate the noise for a given object scaling to the size of its aperture radius (normalized by the ERR map value at the position of an object).
All fluxes and uncertainties were corrected for Galactic extinction assuming E(B-V) $=$ 0.008 (for the GOODS-S field) with a \citet{cardelli89} Milky Way attenuation curve.

\begin{deluxetable}{cccc}
\vspace{2mm}
\tablecaption{Imaging Summary}
\tablewidth{\textwidth}
\tablehead{\multicolumn{1}{c}{Filter} & \multicolumn{1}{c}{FWHM} & \multicolumn{1}{c}{PSF Enclosed} & \multicolumn{1}{c}{Point-Source Limiting}\\
\multicolumn{1}{c}{$ $} & \multicolumn{1}{c}{$ $} & \multicolumn{1}{c}{Flux (d$=$0\dotarcsec2)} & \multicolumn{1}{c}{Magnitude (5$\sigma$)}}
\startdata
F814W&0.114$^{\prime\prime}$&0.60&29.9\\
F115W&0.067$^{\prime\prime}$&0.76&30.3\\
F150W&0.071$^{\prime\prime}$&0.76&30.2\\
F200W&0.077$^{\prime\prime}$&0.73&29.9\\
F277W&0.125$^{\prime\prime}$&0.64&30.4\\
F356W&0.142$^{\prime\prime}$&0.57&30.0\\
F444W&0.162$^{\prime\prime}$&0.51&30.4\\
\enddata
\tablecomments{The limiting magnitude is that measured in a 0.2$^{\prime\prime}$ diameter aperture on the unmatched images, corrected to total based on the PSF flux enclosed in that aperture size.}
\label{tab:tab1}
\vspace{-8mm}
\end{deluxetable}

\subsection{Photometric Depths} \label{sec:depth}

We estimated the point-source depth of our mosaics using the empirical noise function described above.  We first used these functions to derive the 1$\sigma$ flux error in a 0.2$^{\prime\prime}$ diameter aperture.  We then corrected these measurements to total flux uncertainties using an aperture correction derived from the curves-of-growth of stars used to create the PSFs. We summarize our depth measurements in Table \ref{tab:tab1}. These depths are shallower than those predicted by the {\it JWST} Exposure Time Calculator using our exposure time and setup by $0.1-0.5$ mag, despite the $0.1-0.6$ mag improvement resulting from the custom  flats in the short-wavelength filters and updated reference flats in the long-wavelength filters. A potential cause can be intrinsic noise in any calibration images that get propagated and amplified through the long exposure time of our program. We are continuing to investigate this issue for future reductions of our data, as well as to potentially alter the observing strategy for the second epoch.

\begin{figure*}[thbp]
	\centering
		\includegraphics[width=0.75\textwidth]{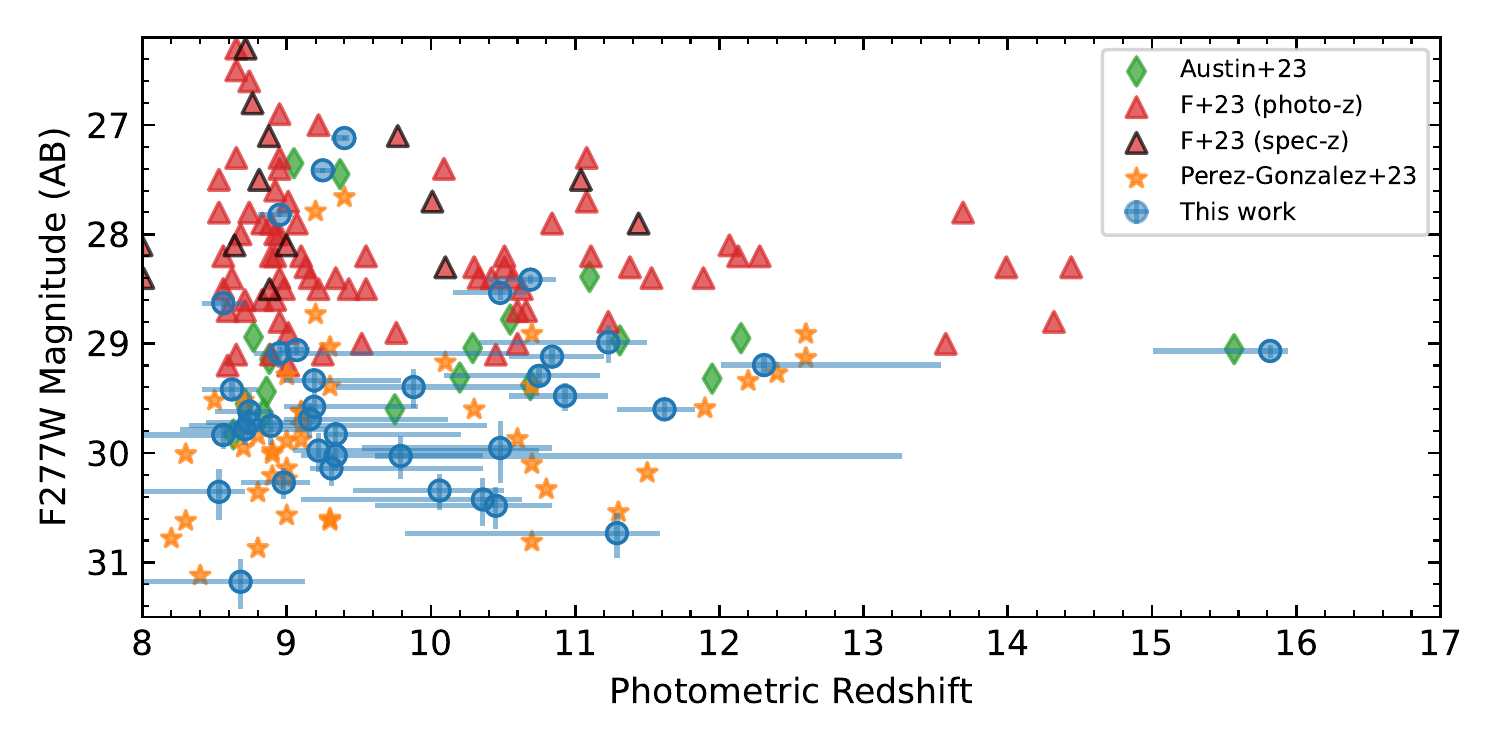}
		\caption{Distribution of F277W magnitude and photometric redshift of our sample. We show objects in our sample in blue circles, along with photometric redshift samples in MDS \citep[][orange asterisks]{pg23} and \citet[][green diamonds]{aus23}. We also show the sample from the completed CEERS NIRCam dataset (\citealt{fin23}, Finkelstein et al. in prep), denoting sources spectroscopic redshift with red triangles and photometric redshift with red triangles with black outlines. Even this first epoch of NGDEEP NIRCam imaging allows us to reach fainter magnitudes than CEERS by $\sim 1.5$ mag. While MDS reaches $\sim 0.5$ mag deeper than NGDEEP, we probe $\sim 1$ mag fainter than the very early analysis of NGDEEP data by \citet{aus23}, primarily due to our custom flat field correction. }
		\label{fig:zmag}
\end{figure*}

\subsection{Photometric Redshifts}

We measured photometric redshifts with \textsc{EAZY} \citep{bra08}.  Following \citet{fin23} we use the default set of 12 ``tweak FSPS" templates in combination with six additional templates constructed by \cite{larson22} inclusive of the blue colors expected at such high redshifts.  A flat redshift prior with respect to luminosity was assumed (given the lack of knowledge about the bright-end of the luminosity function at early times), and redshifts from $z =$ 0--20 were considered.  EAZY was run three times -- a fiducial run with our Kron-measured colors, a ``circular" run using colors measured on the PSF-matched images with a 0.3$^{\prime\prime}$ diameter circular radius, and a ``low-redshift" run, with the maximum redshift set to $z =$ 7 (to allow visualization of the best-fitting low-redshift model).

\section{Selection of Redshift $\gtrsim 9$ Galaxy Candidates}
\label{sec:method}

\begin{figure*}
	\includegraphics[width=0.5\textwidth]{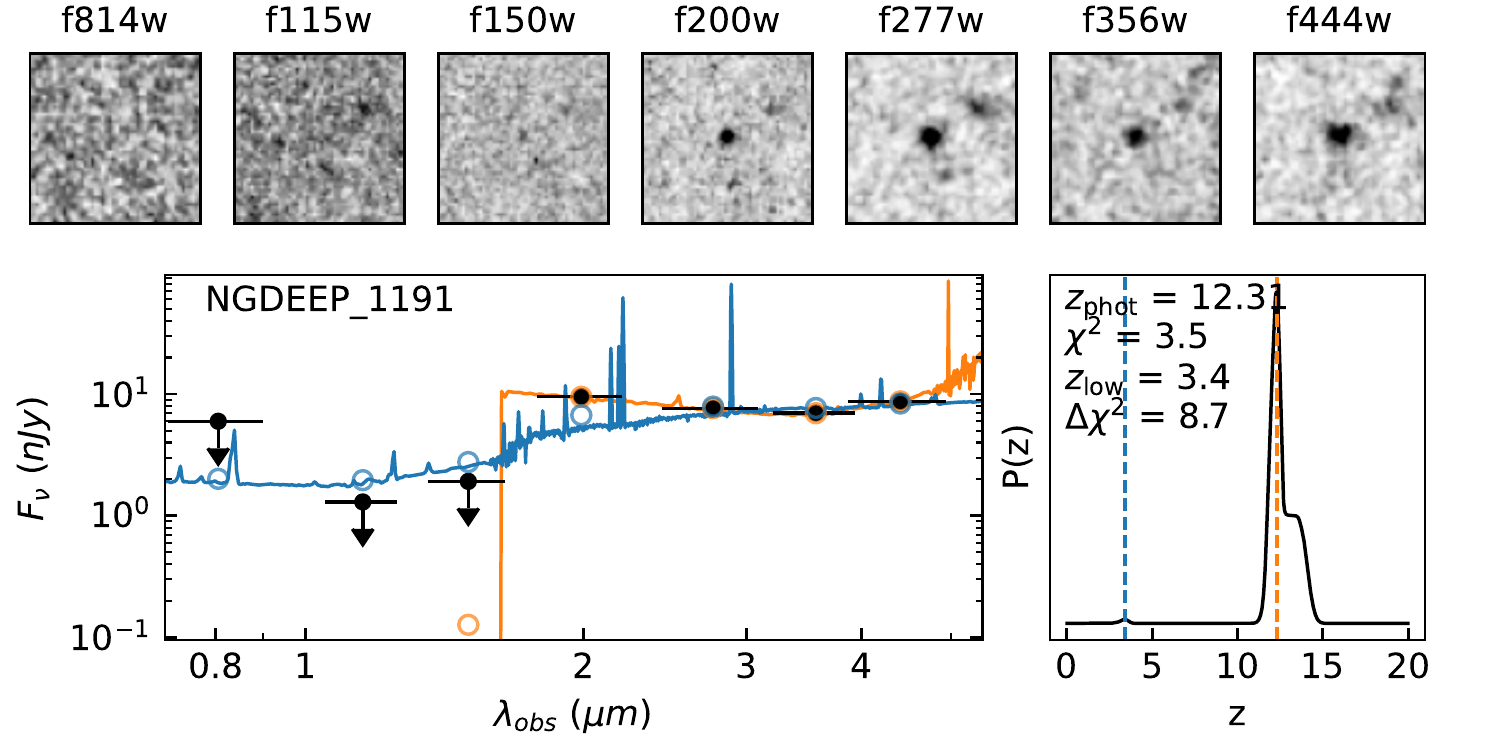}
	\includegraphics[width=0.5\textwidth]{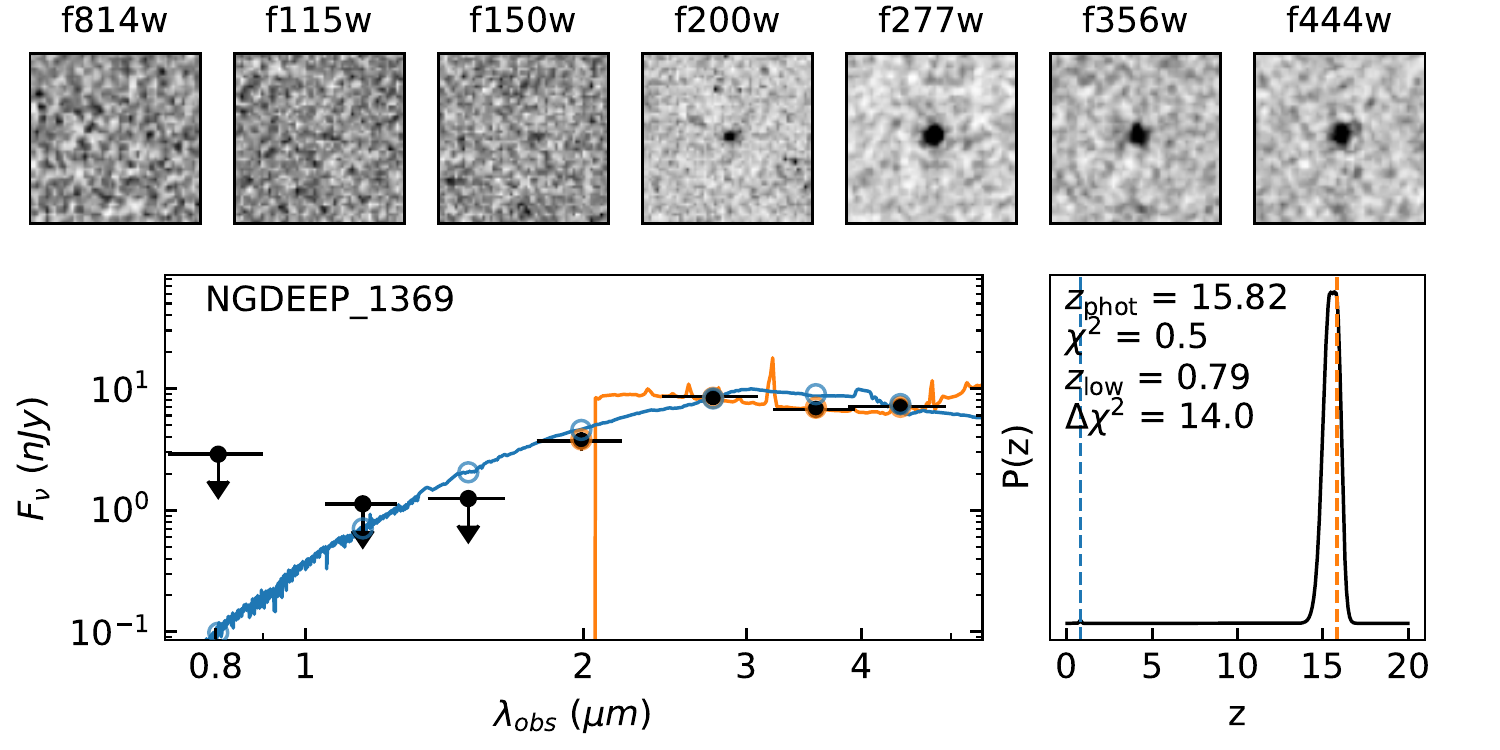}
    \caption{The two sources with photometric redshifts at $z \ge 12$ in our sample. The top panel shows $1\dotarcsec5$ stamp images. The bottom left panel shows the observed photometry in black points, the best-fit EAZY model spectrum (photometry) in orange curves (open circles) and the best-fit $z < 7$ model spectrum (photometry) in blue curves (open circles). Observed fluxes with S/N$<2$ are shown as $2\sigma$ upper limits. The bottom right panel shows the EAZY photometric redshift probability density function in the black curve. The best-fit and secondary redshifts are denoted by the orange and blue dashed vertical lines, respectively. The full sample in shown in the Appendix.}\label{fig:bio_z15}
\end{figure*}

\subsection{Sample Selection}

To select our sample of $z \gtrsim 9$ galaxy candidates, we use selection criteria based on a combination of flux detection significance values and quantities derived from photometric redshift fitting. We denote the probability density function of the photometric redshift as $P(z)$. 
Signal-to-noise ratios (S/N) below are measured in $0\dotarcsec2$ diameter apertures in the non-PSF matched images. Our primary selection criteria are:
\begin{enumerate}
    \item Best-fit photometric redshift ($z_a$) $> 8.5$.
	\item S/N $> 5.5$ in at least two bands or S/N $> 4.5$ in at least three bands to reduce spurious detections.
	\item S/N $< 3$ in all bands blueward of the Lyman break. This includes F814W for $z \gtrsim 9$, F115W for $z \ge 11$, F150W for $z \ge 14$, and F200W for $z \ge 19$.
	\item The $\chi^2$ of the best-fit model $< 60$ to ensure a good fit to the photometry.
	\item $\Delta \chi^2 > 4$ calculated as the difference between the best-fit $\chi^2$ for the low-redshift ($z < 7$) and high redshift models, corresponding to a $2\sigma$ significance \citep{bow20}.
	\item $\int P(z>7) > 0.95$ such that the high-redshift probability density peak must include at least $95\%$ of the total probability.
\end{enumerate}

We first perform a selection using the photometric redshift quantities (criteria 1, 4 - 6) derived from Kron apertures, as they are expected to yield optimal flux extraction and thus photometric redshift measurements. However, in the presence of close neighboring sources, Kron apertures in \texttt{SourceExtractor} are occasionally skewed to produce large apertures that include blended emission from both sources.  This effect is enhanced with these very deep NIRCam images, where brighter low-redshift galaxies have detectable emission to larger radii. 
As the apparent sizes of $z \gtrsim 9$ galaxies are expected to be small, when the area of the Kron aperture is $>3$ times that of a $0\dotarcsec3$ diameter circular aperture, we use photometric redshift quantities derived from the circular aperture and disregard the Kron aperture results. This results in an initial sample of 69 $z \gtrsim 9$ candidates, where 66 are based on Kron apertures, and 3 are based on circular apertures.

This initial sample was then visually vetted by authors GL and SLF. Upon inspection of the image cutouts, we found 31 spurious or unreliable sources. The majority of these sources are spurious detection at image edges, constituting 22 ($71\%$) of the removed sources. These represent spurious detections that do not originate from astrophysical objects. We also find two sources located on a diffraction spike of a nearby bright star, three sources potentially blended with a bright extended neighbor, and one source located in a region with an elevated diffuse background. We find three more sources which show visible flux in dropout F115W band image although the extracted photometry shows a non-detection, likely driven by outlier negative pixels within the aperture. These nine sources are true astrophysical objects whose photometry is considered unreliable due to their projected locations in the sky by chance. We show the image cutouts of all 31 sources identified as spurious or unreliable in the Appendix. These 31 spurious or unreliable sources are removed from the sample, resulting in a final sample of 38 $z\gtrsim 9$ galaxies. We note that the majority ($71\%$) of the removed sources are spurious detections not corresponding to any astrophysical objects, and therefore does not affect the completeness of our sample.

\subsection{Sample Completeness}
\label{sec:compl}

We estimate the completeness of our photometric selection via source injection simulations.  Full details of this process are described in Finkelstein et al.\ (in prep), but we summarize briefly here.  Sources are created with a range of F277W magnitudes, colors, and surface brightness profiles.  The latter are modeled with Galfit \citep{peng02} as Sersic profiles with a log-normal distribution of the sersic index peaked near $n =$ 1, and a range of input half-light radii from 1--8 pixels (encompassing the observed sizes of galaxies in the sample).  After convolution with the PSF of a given image, these mock objects are added to a given image.  We simulate 50,000 such objects (injecting 1000 at a time), performing photometry and measuring photometric redshifts in the same manner as was done on the real images.  The completeness is then defined as the number of galaxies recovered both photometrically, and by our high-$z$ sample selection criteria, in bins of apparent or absolute magnitude.  

As the size of a galaxy can impact the completeness, we include it as a parameter in our completeness estimation.  We correct for any bias in the measured sizes using these simulations, where we find that the input half-light radius (measured by \textsc{SE} on the model, noise-less, images) compared to the recovered half-light radius (measured by \textsc{SE} from the real images with injected sources) were $\sim$1.5$\times$ larger (with no significant dependence on source brightness).  We thus multiply the \textsc{SE} measured half-light radii for the recovered sources by 1.5 prior to calculating the completeness. 

For the results below we use the completeness in two different ways.  First, for the surface density calculation, we calculate an individual completeness for each object, given its observed magnitude, best-fitting photometric redshift, and \textsc{SE}-measured half-light radius (corrected for this scale factor of 1.5).  For the rest-UV luminosity function we estimate the completeness in bins of UV absolute magnitude.  In a given magnitude bin, we calculate the completeness in bins of half-light radius.  We calculate a weight for these completeness values as the number of real objects observed in this magnitude bin in each bin of half-light radius, providing a single weighted volume per magnitude bin.  The effective volume in a given magnitude bin is then calculated as:
\begin{equation}
V(M_{UV})_\mathrm{eff} = \int \frac{dV}{dz} C(M,z) dz,
\end{equation}
where $C(M,z)$ is the completeness in a given bin of absolute magnitude and redshift after the half-light radius weighting.  

\begin{figure*}[thbp]
	\centering
		\includegraphics[width=0.75\textwidth]{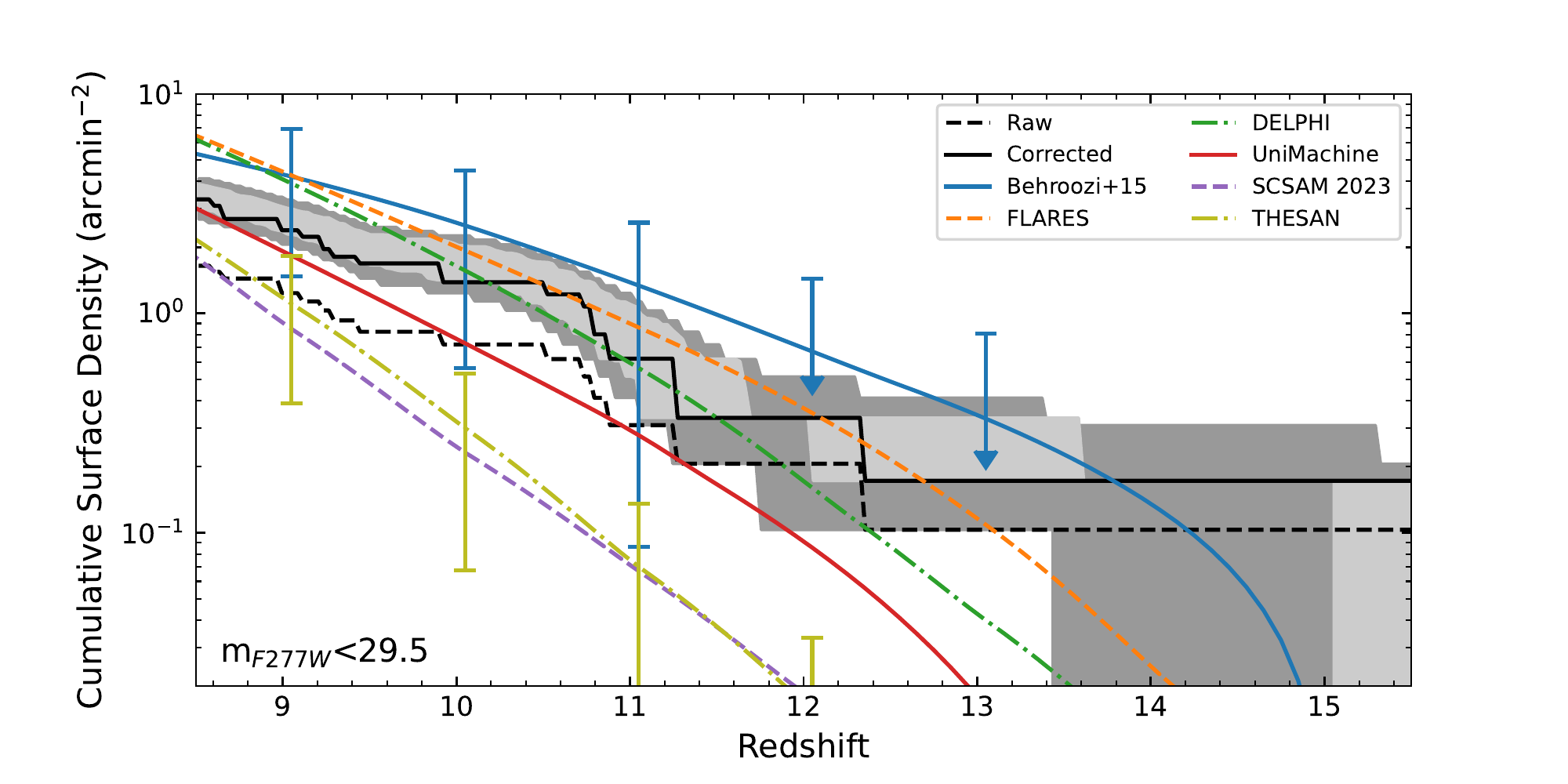}
		\caption{Galaxy cumulative surface density of sources with $\m < 29.5$ as a function of redshift. The black dashed line shows the completeness uncorrected values, while the black solid line shows the completeness corrected values. The light shaded region shows the $68\%$ confidence intervals derived from Monte Carlo simulations sampling the F277W flux and photometric redshift distributions. The darker shaded region includes Poisson uncertainty in the simulations. For the highest- and lowest-predicting models, we denote the $1\sigma$ spread in the predicted surface density due to cosmic variance with vertical error bars. We compare the observed results with predictions from theoretical models (see main text for details). Accounting for cosmic variance, our observed values are higher than those predicted by THESAN and the Santa Cruz SAM at all redshifts.}
		\label{fig:sur}
\end{figure*}

\section{Results}
\label{sec:results}

\subsection{NGDEEP $z \gtrsim 9$ Galaxy Sample}

Using the selection procedures described in the previous section, we arrive in a sample of 38 galaxies at $z \gtrsim 9$. We tabulate the sample in the Appendix (Table \ref{tab:samp}). We plot the distribution of apparent magnitude and redshift of our final sample in Figure \ref{fig:zmag}. We also show samples reported at similar redshift ranges from CEERS \citep[][Finkelstein et al. in prep]{fin23}, the NIRCam program of the MIRI Deep Survey (MDS, PID: 1283, PI: H.U. Norgaard-Nielsen, G. Oestlin, \citealt{pg23}), and an early NGDEEP analysis \citep{aus23}. Our sample spans a magnitude range of $\m \sim 27-30.5$ and photometric redshifts up to $z \approx 16$. Our sample reaches $\sim 1.5$ mag fainter than CEERS due to our deeper NIRCam imaging. Our sample also reaches $\sim 1$ mag deeper than \citet{aus23}, likely due to the use of our custom flat correction and the updated reference flats, while our data is $\sim 0.5$ mag shallower than the MDS. We will compare our findings with MDS and \citet{aus23} in detail in Section \ref{sec:comp}. We show the SEDs and image cutouts of the two $z>12$ galaxy candidates in Figure \ref{fig:bio_z15}. The plots for the remaining sources are shown in the Appendix.

The highest-redshift source in our sample has a best-fit photometric redshift of 15.8. It shows a secondary photometric redshift solution at $z=3.6$. While it has a $\Delta \chi^2$ of 6.8 and an integrated $P(z>7)$ of 0.98, the recent spectroscopic identification of $z = 4.9$ for the bright $z \sim$ 16 candidate introduced by \citet{don23a} implies we should treat this candidate with caution \citep{arrabalharo23}. A similar effect, where bright [OIII] increases the F277W flux, and bright H$\alpha$ increases the flux in both F356W and F444W in a narrow redshift range around $z \sim 5$ is also possible here. Deep spectroscopic followup will be needed to validate this object.

\subsection{Cumulative Galaxy Surface Density}

A very useful way to compare the observed galaxy population to model predictions across a wide range of redshift is the cumulative surface density of galaxies as a function of redshift. In Figure \ref{fig:sur}, we plot the observed cumulative surface density for sources in our sample.  We show the cumulative surface density down to a limiting magnitude of $m_{F277W} < 29.5$ where our completeness is high. 
To correct for completeness, we count each galaxy as one divided by the completeness estimated using its magnitude, redshift, and size (\S \ref{sec:compl}). We show the completeness corrected surface density in the solid black line and the raw surface density in the dashed black line. Across the magnitude and redshift range of our galaxy sample, the completeness correction typically ranges between $\sim \times 2-5$.

We use Monte Carlo simulations to estimate the uncertainty in the cumulative surface density taking into account the flux errors, photometric redshift uncertainty, and Poisson noise. In $10^4$ simulations, we randomly sample the F277W flux using a normal distribution with a standard deviation equal to the flux error, the redshift from the $P(z)$, and perturb the number of galaxies by a Poisson distribution, and recalculate the cumulative surface density. We plot the $68 \%$ spread in the calculated cumulative surface density as the $68\%$ confidence interval as the gray shaded region.

We find that at $\m < 29.5$, the surface density reaches 5 galaxies per arcmin$^2$ at $z = 8.5$ and declines steadily to 0.2 arcmin$^{-2}$ at $z \sim 12$. We compare our observations to six recent model predictions, including the Santa Cruz semi-analytic model (SAM) with \textsc{gureft} merger trees \citep{yun19a,yun23}, the DELPHI SAM \citep{day17}, empirical models from \citet{beh15} and the \textsc{UniverseMachine} \citep{beh19, beh20}, and cosmological hydrodynamical simulations FLARES \citep{lov21, vij21, wil23} and THESAN \citep{kan22}. Note that the Santa Cruz SAM predictions do not include dust attenuation, while the other models assumes dust attenuation.

We estimate the effect of cosmic variance on the predicted surface densities based on the measured cosmic variance in the \textsc{BlueTides} simulations \citep{bho20}. We show the $1\sigma$ uncertainty in the predicted surface densities due to cosmic variance for the
models that predict the highest and lowest number densities in our comparison,
\citet{beh15} and THESAN. For cosmic variance fractional uncertainties greater than unity, we show an upper limit for the predicted surface density. Accounting for cosmic variance, our results are consistent with predictions from the empirical model of \citet{beh15}, the DELPHI SAM and the FLARES hydrodynamical simulation.
Our results are substantially higher than the predictions from THESAN hydrodynamical simulation and the Santa Cruz SAM by a factor of 2 at $z=8.5$ to over an order of magnitude at $z \gtrsim 12$. We note that the measured surface density at $z\gtrsim 12.5$ is  driven by one source at $z=15.6$. A similar analysis of the cumulative surface density of sources with $\m < 28.5$ in the first epoch of the CEERS survey, along with a comparison with many of the same theoretical models, is shown in Figure 14 of \citet{fin23}, which shows a more significant discrepancy between observations and models. This shows that some models give relatively more accurate predictions for the faint galaxy population probed by this study than the brighter galaxies probed by CEERS.

\begin{figure*}[thbp]
	\centering
		\includegraphics[width=\textwidth]{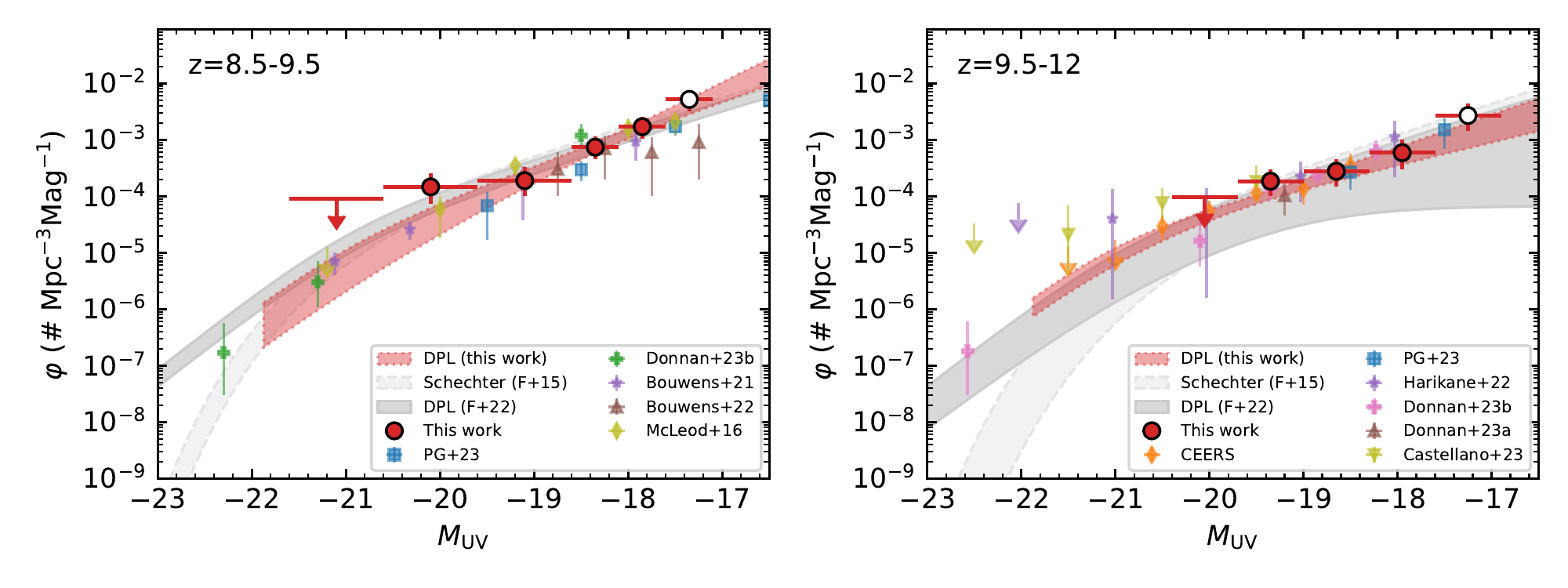}
		\caption{{\it Left}: The rest-frame UV luminosity function at $z=8.5-9.5$. We show our luminosity function in red circles (open circles denotes bins where the completeness is $< 30\%$). We also show literature values from \citet{bou21, bou22, cas23, don23a, don23b, har22, mcl16} and \citet{pg23}. We show our best-fit DPL function to the NGDEEP and CEERS data points in the red shaded regions. The dark (light) grey shaded regions show the DPL (Schechter) functions from \citet{fin22} \citep{fin15}, extrapolated to $z=8.5$ (upper bound) and 9.5 (lower bound). {\it Right}: Same as the left panel but for $z=9.5-12.0$. We also show results from the two-epoch CEERS sample (Finkelstein et al. in prep). The observed luminosity function at $z=9.5-12$ is consistent with the upper end of the extrapolation, suggesting that a slower evolution of the luminosity function at $z\gtrsim10$. We measure a faint end slope of $-2.4\pm0.4$ at $z\sim 9$ and $-2.5 \pm 0.2$ at $z\sim 11$, finding no significant evolution at the faint-end slope of the luminosity function.}
		\label{fig:uvlf}
\end{figure*}

\begin{deluxetable}{CCccccc}
\tablecaption{UV Luminosity Function}\label{tab:lf}
\tablehead{\colhead{$M_{1500}$} & \colhead{$\Delta M$} & \colhead{$\phi \times 10^{-5}$} & \colhead{Number of} & \colhead{$V_\mathrm{eff}$} \\
\colhead{(mag)} & \colhead{(mag)} & \colhead{(Mpc$^{-3}$mag$^{-1}$)} & \colhead{Galaxies} & \colhead{(Mpc$^{3}$)}}

\startdata
\multicolumn{5}{c}{$z \sim 9$} \\
\hline
-21.1 & 1.0 & $<8.9$ & 0 & 18700 \\
-20.1 & 1.0 & $14.7_{-7.2}^{+11.1}$ & 2 & 18500 \\
-19.1 & 1.0 & $18.9_{-8.9}^{+13.8}$ & 2 & 15800 \\
-18.35 & 0.5 & $74.0_{-29.0}^{+41.4}$ & 5 & 13100 \\
-17.85 & 0.5 & $170_{-65}^{+85}$ & 5 & 7770 \\
-17.35 & 0.5 & $519_{-198}^{+248}$ & 7 & 2520 \\
\hline
\multicolumn{5}{c}{$z \sim 11$} \\
\hline
-20.05 & 0.7 & $<9.7$ & 0 & 27900 \\
-19.35 & 0.7 & $18.5_{-8.3}^{+11.9}$ & 3 & 26100 \\
-18.65 & 0.7 & $27.7_{-13.0}^{+18.3}$ & 3 & 20800 \\
-17.95 & 0.7 & $59.1_{-29.3}^{+41.9}$ & 3 & 9840 \\
-17.25 & 0.7 & $269_{-124}^{+166}$ & 4 & 2210
\enddata
\end{deluxetable}

\subsection{UV Luminosity Function}

A key observational diagnostic to the evolution and assembly history of galaxies in the first 500 Myr of cosmic time is the UV luminosity function. We calculate the UV luminosity function in two redshift bins, $z=8.5-9.5$ ($z \sim 9$) and $9.5-12.0$ ($z \sim 11$). Galaxies are assigned to redshift bins using the best-fit photometric redshift. To measure the rest-frame UV absolute magnitude ($M_{1500}$), we perform SED fitting using \texttt{Bagpipes} \citep{car18}. The procedures and results of the SED fitting is presented in Morales et al. (in prep). We measure $M_{1500}$ by averaging the \texttt{Bagpipes} posterior model spectrum from rest-frame 1450 \AA\ to 1550 \AA. There are two sources in our sample with remarkably red colors driven primarily by high F444W fluxes (see Section \ref{sec:red} for detailed discussion), leading to model spectra that poorly fit the photometry at the bluer wavelengths near rest-frame 1500 \AA. For these sources, we exclude F444W from the SED fitting.   We calculate the luminosity function following the methodology of \citet{fin15} and \citet{fin23}. We calculate a non-parametric step-wise maximum likelihood number density in each magnitude bin assuming a Poisson likelihood function. For the $z \sim 9$ redshift bin, we use magnitude bins of 1 mag from $M_{1500}=-21.6$ to $-18.6$, and 0.5 mag from $-18.6$ to $-17.1$ mag. For $z \sim 11$, we used 0.7 mag bins spanning $-20.4$ to $-16.9$ mag. We estimate the uncertainty of the number density with a Markov chain Monte Carlo (MCMC) technique with no prior on the number densities. In each step of the MCMC chain, we randomly sample $M_{1500}$ for each galaxy from 500 \texttt{Bagpipes} posterior model spectra, allowing a given galaxy to move between bins in each MCMC step. This accounts for both Poisson statistics and the uncertainty in $M_{1500}$. We take the median posterior number density as the luminosity function, and the 16- and 84-percentiles as the uncertainty. We tabulate our luminosity functions in Table \ref{tab:lf}.

\begin{figure}[thbp]
	\centering
        \includegraphics[width=0.5\textwidth]{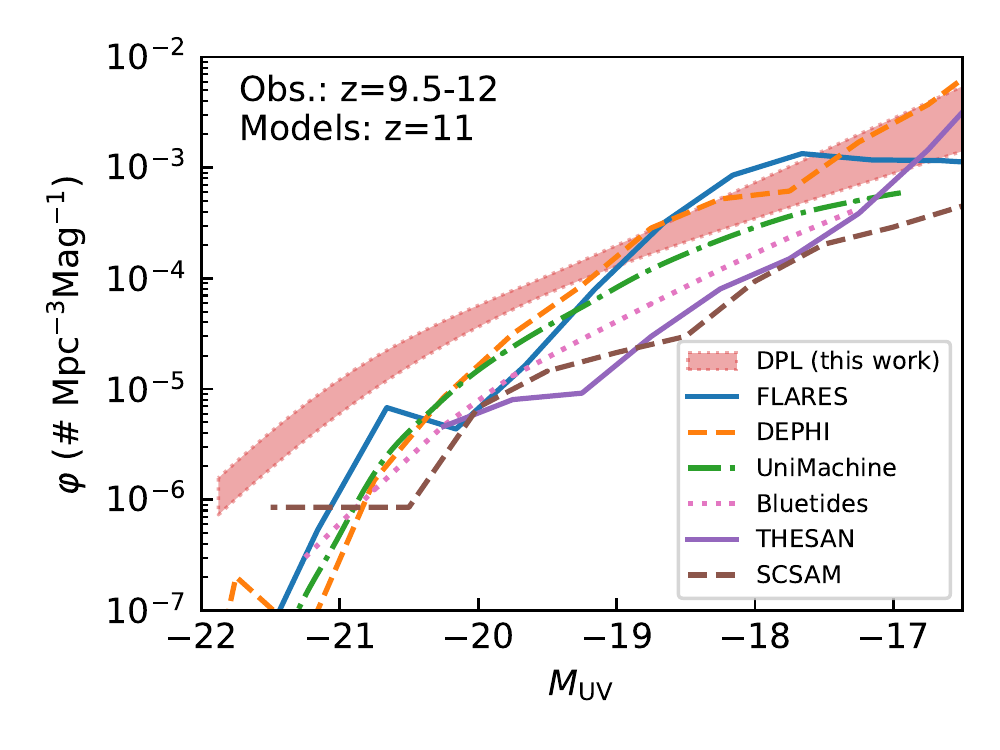}
		\caption{Comparison between our $z=9.5-12$ rest-frame UV luminosity function with model predictions at $z=11$, close to the volume weighted average redshift of the observed sample of $z=10.7$. Our best-fit DPL is shown in the red shaded region. Some models underpredict the number density of galaxies across the full studied range of luminosities, while others underpredict bright galaxies but reproduce the number of fainter galaxies, implying a different predicted LF slope.
  }
		\label{fig:lf_comp}
\end{figure}

In Figure \ref{fig:uvlf}, we plot the measured binned luminosity functions, showing the median and $68\%$ error from the MCMC analysis, in the two redshift bins along with measured luminosity function values in the literature at comparable redshifts. We compare these results to pre-{\it JWST} empirical extrapolations of the luminosity function, including a Schechter function from \citet{fin16}, who fit the luminosity function using observations at $z=4-8$, and a double power-law (DPL) from \citet{fin22}, who incorporated observations at $z=3-9$. These models are parameterized by $(1+z)$, and we extrapolate them to the lower and upper limits of each redshift bin to show the range of predicted values encompassing the respective redshift interval, under the assumption that the observed smooth evolution at these lower redshifts continues into this epoch. We find that the measured $z\sim 9$ luminosity function is consistent with the range of values predicted by the extrapolated functions, while the measured $z\sim 11$ luminosity function is more consistent with upper end ($z=9.5$) of the extrapolated values. This suggests a slower evolution of the luminosity function at $z\gtrsim10$ than at $z=3-9$, where the extrapolated functions were fitted.

To quantify the evolution of the luminosity function at $z > 8.5$, we fit a DPL to our observed sources in the two redshift bins above. At $z\sim 9$, the NGDEEP sample spans absolute magnitudes from $-21$ to $-17$ mag. At $z \sim$ 11 we extend the dynamic range in absolute magnitude by supplementing our analysis with the calculated number densities from the full CEERS survey from Finkelstein et al. (2023, in prep). We included the CEERS sample because of the similarities in the galaxy selection methodology and filter coverage with NGDEEP, resulting in a relatively homogeneous sample. The NGDEEP sample spans magnitudes from $-20$ to $-17$ mag, while the CEERS samples covers magnitudes from $-21$ to $-19$. 

\begin{deluxetable}{cCC}
\tablecaption{Double Power-Law Parameters}\label{tab:dpl}
\tablewidth{\textwidth}
\tablehead{
\colhead{Parameter} & \colhead{Prior} & \colhead{Posterior}
}
\startdata
\multicolumn{3}{c}{$z\sim 9$} \\
\hline
$\log(\phi^*[\mathrm{Mpc}^{-3}\mathrm{mag}^{-1}])$  & [-10, -1] & $-4.76^{+0.38}_{-0.43}$\\
$\alpha$ & [-4, 0] &  $-2.45^{+0.38}_{-0.41}$ \\
$\beta$ & -4.28\tablenotemark{a} & -\\
$M^*$ (mag) & -21.03\tablenotemark{a} &  - \\
\hline
\multicolumn{3}{c}{$z\sim 11$} \\
\hline
$\log(\phi^*[\mathrm{Mpc}^{-3}\mathrm{mag}^{-1}])$ & [-10, -1] & $-4.75^{+0.15}_{-0.16}$\\
$\alpha$ & [-4, 0] &  $-2.22^{+0.23}_{-0.23}$ \\
$\beta$ & -4.19\tablenotemark{a} & - \\
$M^*$ (mag) & -20.99\tablenotemark{a} &  - \\
\enddata
\tablenotetext{a}{We fix the values of $\beta$ and $M^*$ to the extrapolated values from \citet{fin22} since our data do not probe the bright end of the luminosity function.}
\end{deluxetable}

We parameterize luminosity function using a DPL given in the following form:
\begin{equation}
    \phi(M) = \phi^* \left[10^{0.4(\alpha+1)(M-M*)} + 10^{0.4(\beta+1)(M-M*)}\right]^{-1},
\end{equation}
where $\phi^*$ is the characteristic number density, $M^*$ is the characteristic magnitude, $\alpha$ is the faint-end slope, and $\beta$ is the bright-end slope.  We use MCMC to estimate the values of $\phi^*$ and $\alpha$ assuming a Poisson likelihood and a flat prior in the parameters. Since the combined NGDEEP and CEERS sample does not cover the bright end of the luminosity function, we fix $M^*$ and $\beta$ to the calculated or extrapolated values at $z=9$ and $z=11$, respectively, using the empirical function in \citet{fin22}. We show the results of our DPL fit and the priors in Table \ref{tab:dpl}. We find a faint-end slope of $\alpha=-2.5\pm 0.4$ at $z\sim 9$ and $\alpha=-2.2\pm 0.2$ at $z \sim 11$. These results at $z \sim$ 9 are consistent with pre-{\it JWST} observations of $-2.4$ to $-2.0$ \citep[e.g.][]{bou15, mcl16, fin22}, while we observe no significant evolution in the faint-end slope from $z\sim9$ to $z \sim 11$. 

In Figure \ref{fig:lf_comp}, we compare our $z\sim 11$ luminosity function with predictions from the same theoretical models shown in the previous section. We also include predictions from the \textsc{BlueTides} simulation \citep{feng16, wil17}. We show comparisons of our measured luminosity function with model predictions at $z=11$, which is close to the equal volume midpoint of the $z\sim 11$ redshift bin of $z=10.7$. All of the models predict lower number densities than the observations at bright magnitudes ($M_{\rm UV} \lesssim -19.5$) by a significant factor (up to an order of magnitude or more), while FLARES and DELPHI are consistent with our new observational constraints at fainter luminosities ($M_{\rm UV} > -19.5$).  The Santa Cruz SAM, THESAN, and BlueTides all significantly underpredict the number density of galaxies in this luminosity range, and UniverseMachine underpredicts by a smaller amount. It is notable that FLARES predicts a somewhat steeper faint end LF slope than the observed one, while the other models mostly predict a slope that appears consistent with the observed one even if the amplitude of the luminosity function is too low.

\section{Discussion}
\label{sec:discuss}

\subsection{Other High-redshift Galaxy Samples in HUDF-Par2}
\label{sec:comp}

The HUDF-Par2 field is also targeted by the MDS. The MDS has conducted NIRCam observations in four filters down to $30.2-30.8$ mag in a region partially overlapping with the NGDEEP footprint. While we have observed in two additional filters, F200W and F444W, the MDS NIRCam imaging in the remaining filters are $0.1-0.7$ mag deeper than ours. Using the NIRCam photometry, they report 45 galaxy candidates at $z>8$ \citep{pg23}, 22 of which are within our F277W footprint. We find counterparts for 19 of these sources in our catalog, with the remaining three undetected due to our shallower depth. None of these 19 sources is selected in our sample. 

Here, we examine the 19 sources not selected. One source has best-fit photometric redshifts of $8 < z < 8.5$, marginally falling outside of our redshift selection threshold. Two sources have a best-fit photometric redshift $>8.5$, but did not pass one or both of our $\Delta \chi^2$ and $P(z)$ requirements. One source has been selected in our initial sample, but is removed after visual inspection because of visible flux in the F115W dropout filter. Seven sources are better fit in our analysis with a Balmer break at $z\sim 2$, but display a secondary $P(z)$ peak at $z > 8.5$. The remaining eight sources are either faint sources undetected in multiple bands or located near image edges with elevated noise, leading to poorly constrained $P(z)$ in our data. These differences are likely due to the fainter limiting magnitudes reached by the MDS observations. Many of the $z>8$ galaxy candidates reported in \citet{pg23} are fainter than 30 mag (see Figure \ref{fig:zmag}), where our completeness is $\sim 20\%$.

An early analysis using the NGDEEP dataset is also presented by \citet{aus23}, where they identify 18 $z > 8$ galaxy candidates. Of these, 6 sources are selected in our final sample. A key difference between this study and \citet{aus23} is that we incorporate {\it HST}/ACS F814W imaging, which is useful for identifying dropouts at $z \gtrsim 9$ and rejecting low-redshift contaminants whose NIR photometry alone resembles a high-redshift source. Out of the 12 sources not recovered in our sample, we find that two sources have $z \approx 8.4$, thus marginally falling outside of our selection threshold. One source has a best-fit redshift of $z > 8.5$, but do not pass our $\Delta \chi^2$ and/or $P(z)$ requirement. Five sources are better fit with a Balmer break at $z \sim 2$, while a secondary $P(z)$ peak is located at $z > 8.5$. Three sources are better fit with a $z\sim 2$ solution, and do not show a substantial $P(z)$ opeak at $z>8.5$. One source among these three is located in a region with elevated background noise in our data. One more source passes all of our automated selection criteria, but is rejected after visual inspection suggests potential blending with a nearby bright source. 

Our sample includes 32 galaxy candidates that are not selected by \citet{aus23}. While some differences can be expected due to different source extraction and photometric redshift configurations, this relatively large discrepancy is likely because of the improvement in our photometric depths resulting from the custom flat correction. \citet{aus23} use F444W as the detection band, which is the filter the most severely affected by residual flat features. The inclusion of the updated reference flats in our data reduction has improved the $5\sigma$ depth in F444W from 29.7 mag to 30.4 mag, a 0.7 mag gain. In our analysis, we use F277W as the detection band, which is also 0.7 mag deeper than the pre-update F444W. In fact, 20 of our galaxy candidates have $\m > 29.5$, compared to only 4 in \citet{aus23}.

\begin{figure*}
	\includegraphics[width=0.5\textwidth]{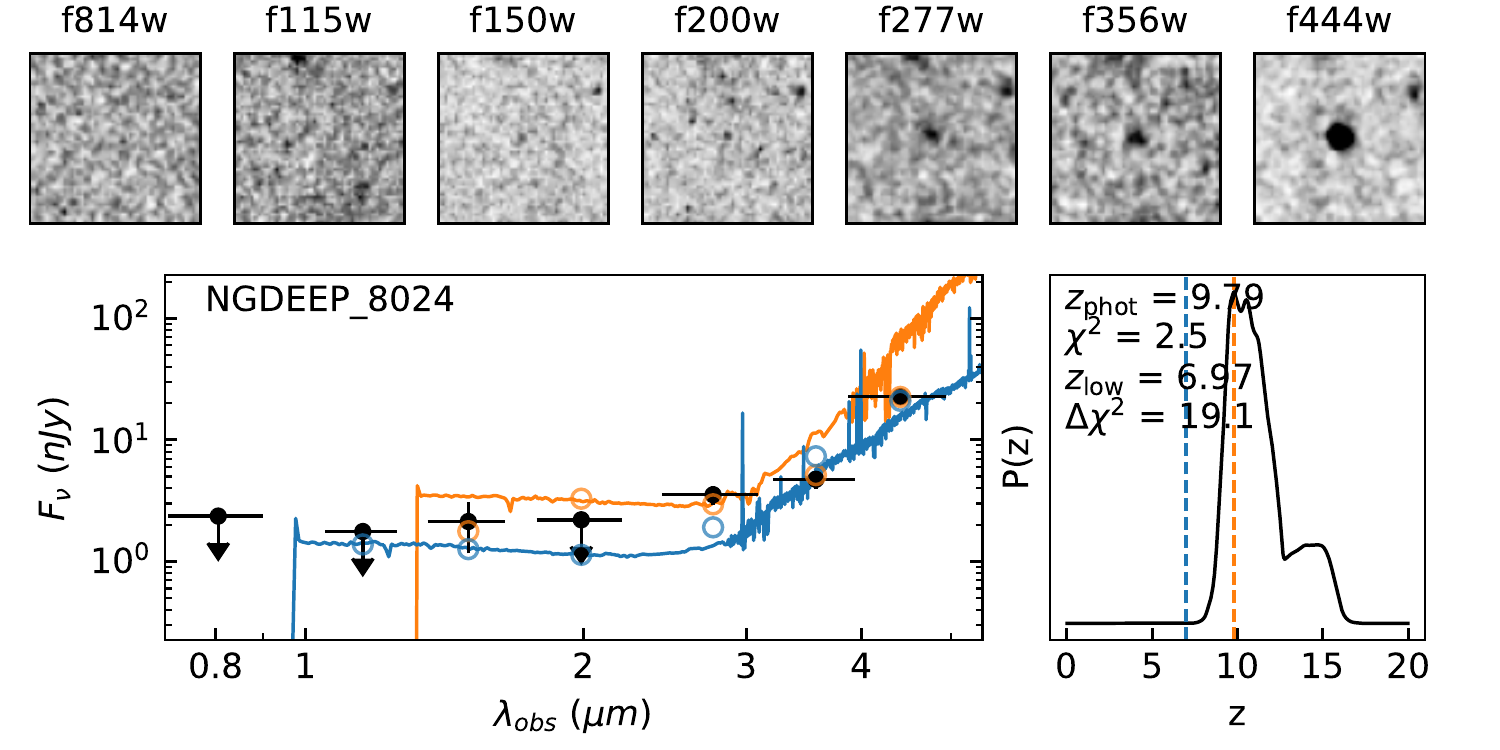}
	\includegraphics[width=0.5\textwidth]{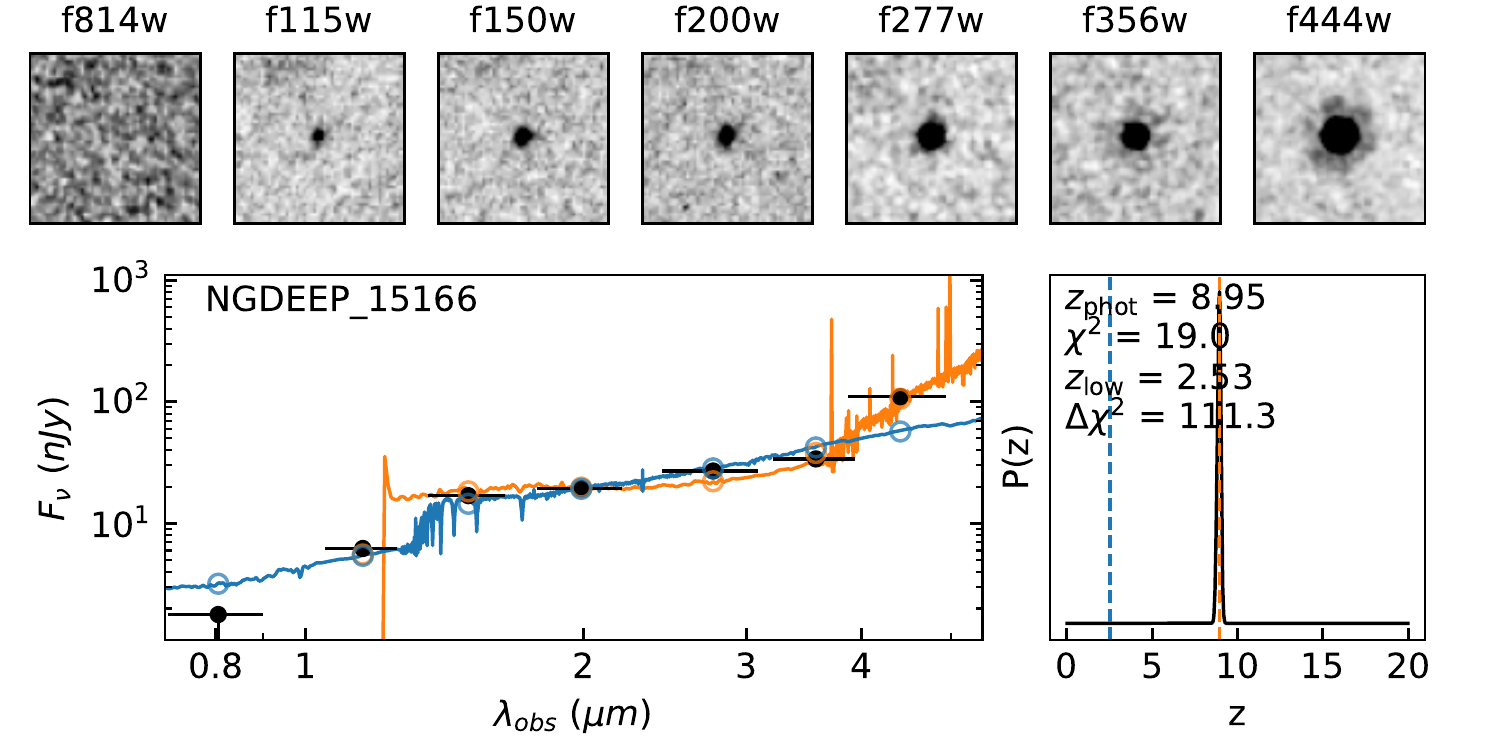}
    \caption{Same as Figure \ref{fig:bio_z15}, but for the two sources are found in our sample with a remarkably red color at $\gtrsim 3 \mu$m. The morphology is compact to point-like. These properties are similar to a class of sources recently discovered with {\it JWST} \citep[e.g.][]{aki23, fur22, koc23}, whose origins have been attributed to dust-obscured galaxies or SMBH accretion. (Note that the model spectra are the best-fit combinations of EAZY templates considering only stellar emission.) The objects here are the first to be observed at $z \sim 9$ and $m_\mathrm{F444W} \sim 28$.}\label{fig:bio_red}
\end{figure*}

\subsection{Red and Compact $z > 9$ Galaxies}
\label{sec:red}

In our $z > 9$ sample, we find two galaxy candidates displaying remarkably red F356W - F444W colors. Their IDs are NGDEEP 8024 and NGDEEP 15166, and their SEDs and image cutouts are shown in Figure \ref{fig:bio_red}. The SEDs of these sources show a flat to blue continuum below $\sim 3 \mu$m, with F150W - F200W colors of $\lesssim 0.2$, before steeply turning red to F277W - F444W colors of $>1.5$. In the images, these sources appear to be compact to point-like in all filters. We note that the morphology of NGDEEP 20351 in the F444W filter appears to resemble diffraction spikes produced by point sources. A class of sources with similar properties has been discovered by a number of studies at $z=5-8$ using {\it JWST} imaging \citep{fur22, koc23, aki23, bar23}. The origin of these sources have been attributed to dust-obscured galaxies \citep{aki23} and/or SMBH accretion \citep{fur22, koc23}. These objects, however, are the first to be observed at $z\sim 9$ and $m_\mathrm{F444W} \sim 28$. This shows that these objects exists in fainter systems in addition to the bright initial discoveries. A full analysis of these objects will be presented in a coming paper. Followup spectroscopic or deep radio observations of these objects will help determine their nature conclusively.

\subsection{Theoretical Implications of our Results}
\label{sec:theoryimp}

We have shown a comparison between the observed cumulative number density of galaxies at $z>9$ and luminosity functions at $9.5 < z < 12$, and the corresponding predictions from a range of theoretical models. Several papers have already pointed out that most published theoretical models underpredict the number density of luminous galaxies at $z\gtrsim 10$ \citep{har22,fin23,yun23,adams23}. Here we find the interesting result that some of these models (FLARES, DELPHI) are in reasonable agreement with the fainter galaxy population probed by NGDEEP at $z\sim 11$, while others (THESAN, Santa Cruz SAM) also significantly underpredict the faint population. This implies that models predict rather different faint-end \emph{slopes} for the UV LF at $z\sim 11$. 

Empirical models, semi-analytic models, and numerical hydrodynamic simulations contain different ingredients and therefore can lead to different kinds of insights and conclusions. Empirical models like UniverseMachine and the model of \citet{beh15} do not attempt to model or characterize physical processes, but rather obtain empirical constraints on the mapping between an observable galaxy property, such as stellar mass or star formation rate, and a dark matter halo property such as mass or mass accretion rate. The mappings obtained from pre-JWST observations at lower redshifts ($z \lesssim 8$) are extrapolated to obtain predictions at higher redshifts. The \citet{beh15} model predicts a strong increase in the stellar-to-halo mass ratio from $z\sim 8$--15. This model appears promising at matching the observed number densities of galaxies at $z \gtrsim 10$, however, it does not explain in detail how physical processes could achieve this increase.  At least some physics-based models, such as the Santa Cruz SAM, predict a stellar-to-halo mass ratio that does not evolve significantly across this redshift interval \citep{yun23}.

Semi-analytic models solve systems of ordinary differential equations that track flows of mass and metals between different reservoirs (intergalactic medium, circumgalactic medium, interstellar medium). They contain simple parameterized recipes that describe physical processes such as cooling, star formation, stellar driven winds, etc., which are typically tuned to match global galaxy observations. Numerical hydrodynamical simulations solve systems of partial differential equations for particles or grid cells, but still contain phenomenological ``sub-grid'' recipes that describe physics occurring on scales below those that the simulation can resolve explicitly (such as star formation, stellar feedback, and black hole growth and feedback). These sub-grid recipes contain tuned parameters which are also typically calibrated to match observations.

In both semi-analytic models and numerical hydrodynamic simulations, to first order the \emph{amplitude} of the UV LF is driven by the normalization of the mass and energy loading (mass outflow rate or energy outflow rate divided by star formation rate) of stellar driven winds, while the \emph{slope} is determined by the (input or emergent) dependence of mass and/or energy loading on global galaxy properties such as velocity dispersion or halo mass. 
In addition, the \emph{scatter} in the predicted $M_{\rm UV}$ (e.g. in the $M_{\rm UV}$--$M_{\rm halo}$ relation), due to very different implementation of physical processes across these simulations, can also have an impact in the predicted number density of galaxies, especially at the bright-end of the UV LF because of Eddington bias.
We note that the predicted $M_{\rm UV}$ from simulations is idealized and does not account for various noise and flux uncertainties that are faced by observations.
It is intriguing that the UV LF predicted by the DELPHI SAM and the FLARES hydro simulations agree quite well, while the Santa Cruz SAM and THESAN hydro simulations also agree well with one another, but these two sets of models predict number densities of faint galaxies that differ by as much as an order of magnitude at $z=11$.  
It is clear that measuring the properties of galaxies over a wide range of luminosity/mass and redshift will be invaluable for discriminating between models and constraining the physical processes that shape galaxy formation.

\section{Conclusions}
\label{sec:conclusions}

We present results from a study of galaxies at $z \gtrsim 9$ using $\simeq 50$ hours of ultra-deep NIRCam observations from the first half of the NGDEEP survey. The imaging reaches $5\sigma$ depths of 29.9--30.4 mag, making it the deepest public {\it JWST} imaging dataset to date. We perform a detailed data reduction process including a number of custom procedures. 
We have identifed a robust sample of 38 galaxies at $z \gtrsim 9$ using photometric redshift selection. Our main findings are summarized below:

\begin{itemize}
    \item We measure the cumulative surface density of galaxies as a function of redshift. We find that our results are in agreement with the higher end of theoretical predictions, suggesting that some models give relatively more accurate predictions for the faint galaxy population than for brighter galaxies probed by previous studies. 
    \item We present the rest-frame UV luminosity function at $z=8.5-9.5$ and $z=9.5-12.0$ using our sample. We fit a DPL function to quantify the evolution of the faint-end slope and number density at $z \gtrsim 9$. We find a faint-end slope of $\alpha = -2.5 \pm 0.4$ and $-2.2 \pm 0.2$ at $z\sim 9$ and $z\sim 11$, respectively. This shows no significant evolution of the faint-end slope from $z=9$ to 11.
    \item We compare our luminosity function at $z\sim 11$ with empirical extrapolations and theoretical predictions. All of the physics-based models under-predict the number of luminous galaxies significantly. Some models reproduce the number density of fainter galaxies, while others under-predict these as well,  implying a different predicted faint end slope for the UV LF. This likely arises from differences in the modeling of stellar feedback.
    \item We have discovered two objects with remarkably red colors at $\gtrsim3 \mu$m and compact to point-like morphology. These sources show similarities to a class objects recently discovered with {\it JWST} observations. The origins have been attributed to dust-obscured galaxies and/or quasar activity.
\end{itemize}

Using only half of the NGDEEP dataset, our study has demonstrated the value of deep field observations in the study of galaxies at the earliest times. The second half of the NGDEEP Survey will be completed in early 2024. The new data is expected to increase the detection limits by $\sim 0.7$ mag. The full NGDEEP dataset will allow us to robustly probe the faint galaxy population down to $M_\mathrm{UV} \sim -17$. By combining the full NGDEEP dataset with the JADES and MDS programs in the HUDF-Par2 field, these legacy data will transform our understanding of formation and evolution of galaxies at the earliest epochs.

\begin{acknowledgments}

We thank Sandro Tacchella and the NIRCam team for sharing the wisps templates. GL thanks Max Franco for useful discussions on NIRCam data reduction. GL, MB, SLF, SF, RL, DB, OCO, AM acknowledge that the location where most of this work took place, the University of Texas at Austin, sits on the Indigenous lands of Turtle Island, the ancestral name for what now is called North America. Moreover, we would like to acknowledge the Alabama-Coushatta, Caddo, Carrizo/Comecrudo, Coahuiltecan, Comanche, Kickapoo, Lipan Apache, Tonkawa and Ysleta Del Sur Pueblo, and all the American Indian and Indigenous Peoples and communities who have been or have become a part of these lands and territories in Texas.

GL, MB and SLF acknowledge support from NASA through STScI award JWST-GO-2079. MC acknowledges support from INAF Minigrant ``Reionization and fundamental cosmology with high-redshift galaxies". This work is based on observations made with the NASA/ESA/CSA {\it JWST}. The data were obtained from the Mikulski Archive for Space Telescopes at the Space Telescope Science Institute, which is operated by the Association of Universities for Research in Astronomy, Inc., under NASA contract NAS 5-03127 for JWST. These observations are associated with program \#2079.
The authors acknowledge the Texas Advanced Computing Center (TACC) at The University of Texas at Austin for providing HPC resources that have contributed to the research results reported within this paper.

\end{acknowledgments}

\vspace{5mm}
\facilities{JWST(NIRCam), HST(ACS)}

\software{astropy \citep{ast13, ast18}, Bagpipes \citep{car18}, EAZY \citep{bra08}, Source Extractor \citep{ber96}}

\bibliography{ms}{}
\bibliographystyle{aasjournal}

\appendix

In this Appendix, we present tabulated data for the full $z \gtrsim 9$ sample, as well as figures for sources in the sample that have not appeared in the previous sections. We also show the image cutouts for the 31 sources rejected by visual inspection.

\begin{figure*}
	\includegraphics[width=0.33\textwidth]{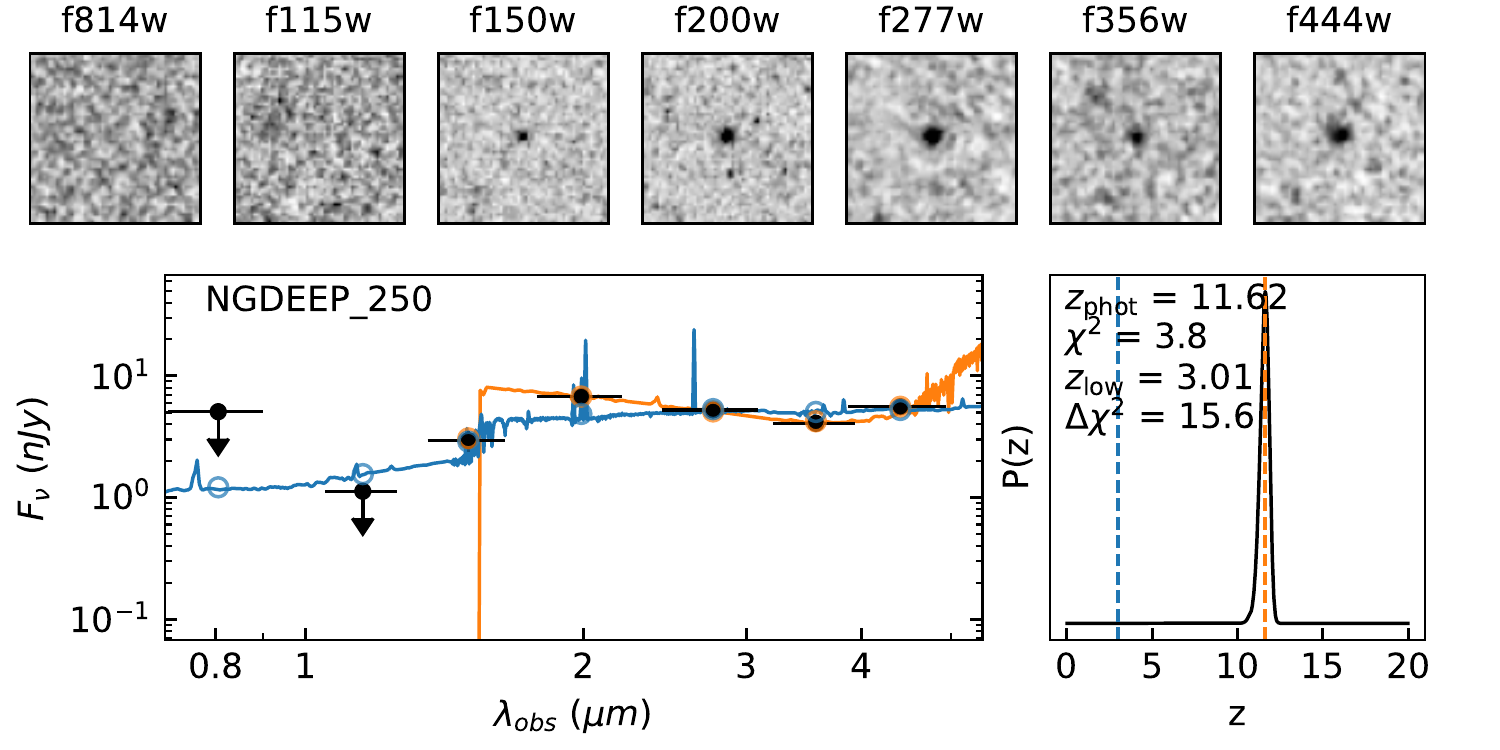}
	\includegraphics[width=0.33\textwidth]{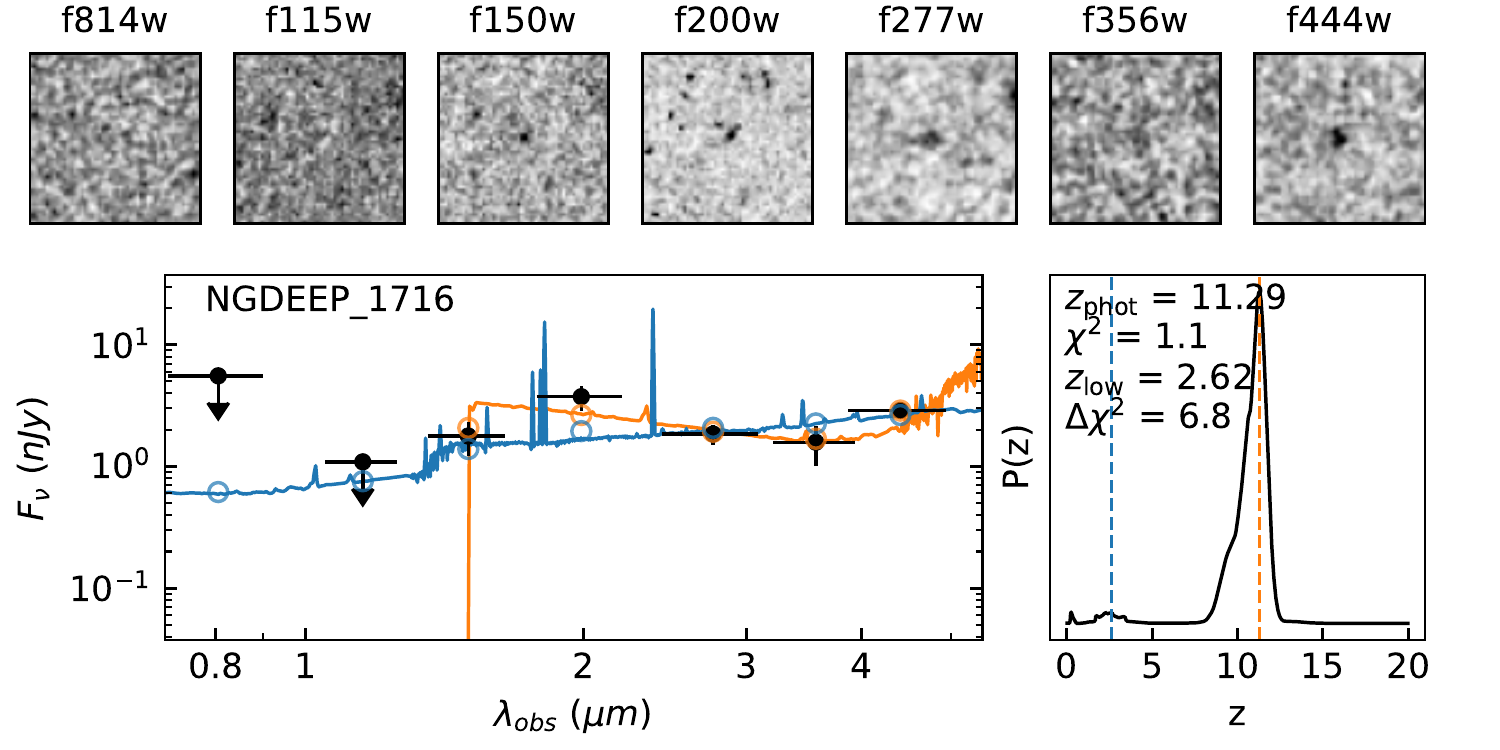}
	\includegraphics[width=0.33\textwidth]{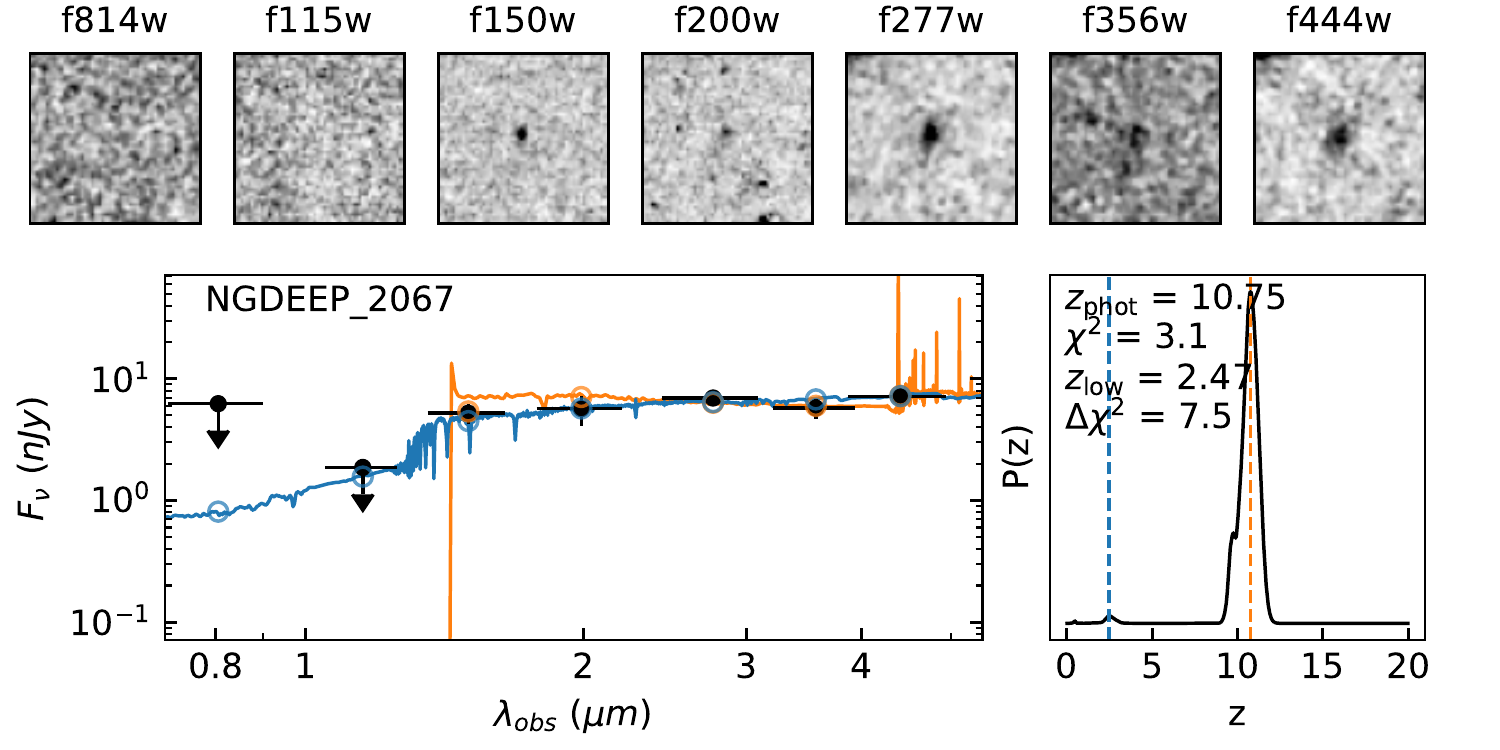}
	\\
	\includegraphics[width=0.33\textwidth]{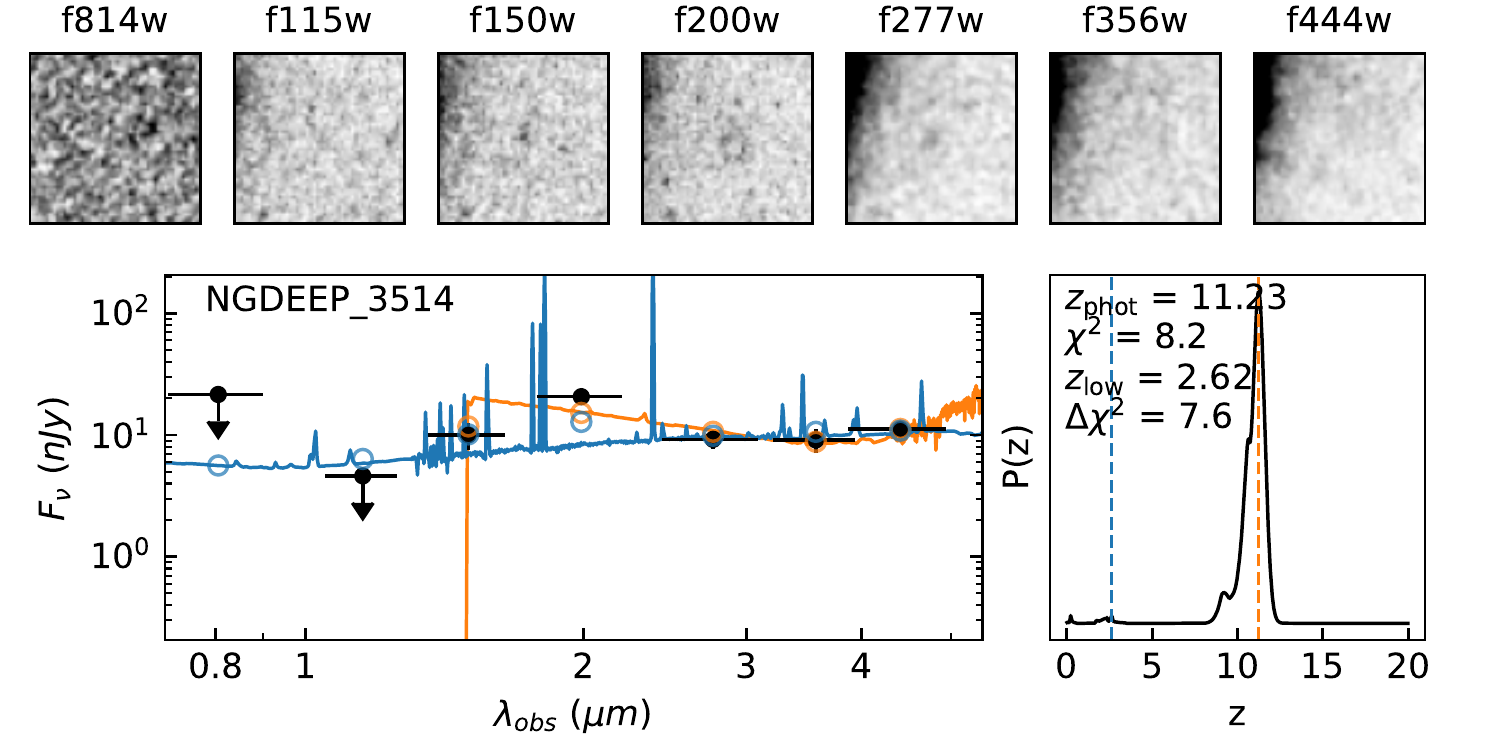}
	\includegraphics[width=0.33\textwidth]{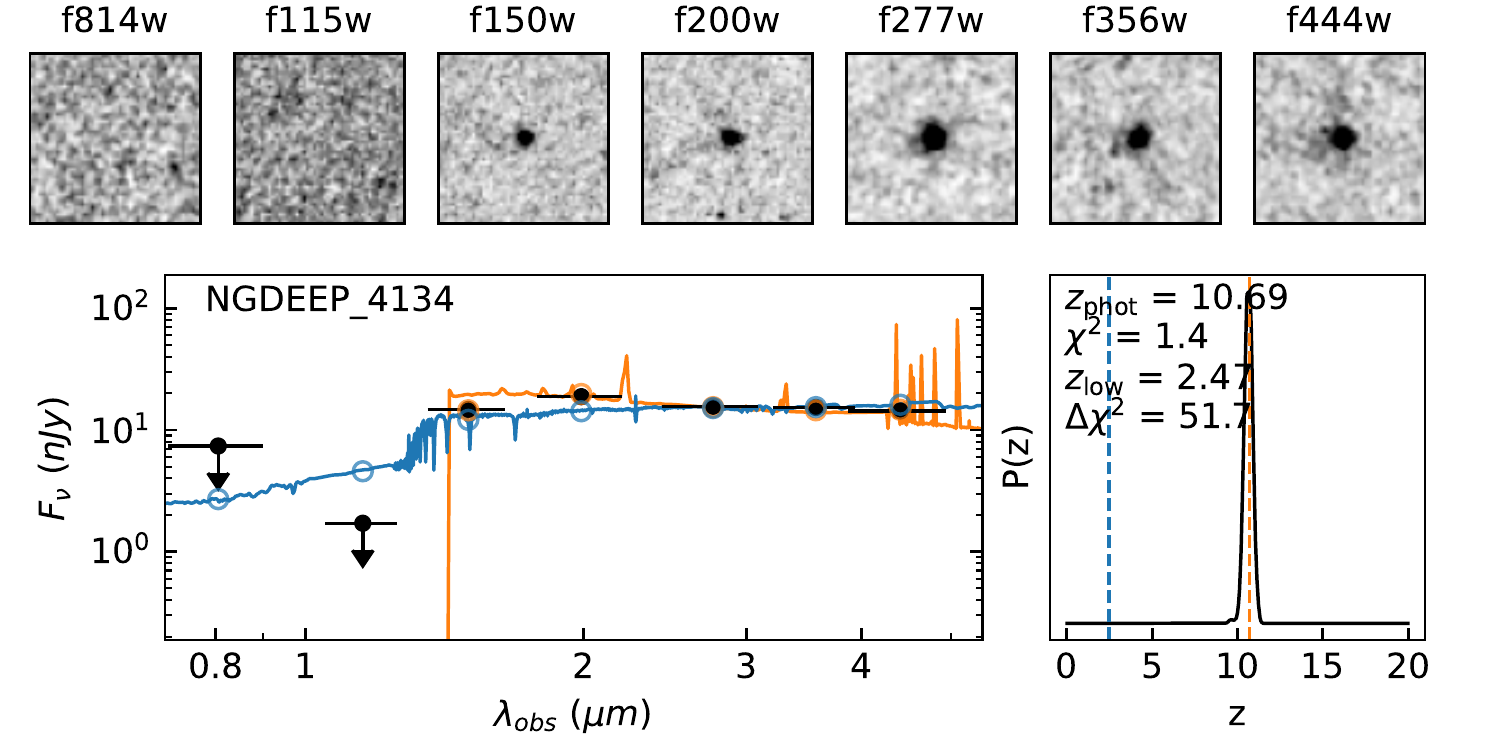}
	\includegraphics[width=0.33\textwidth]{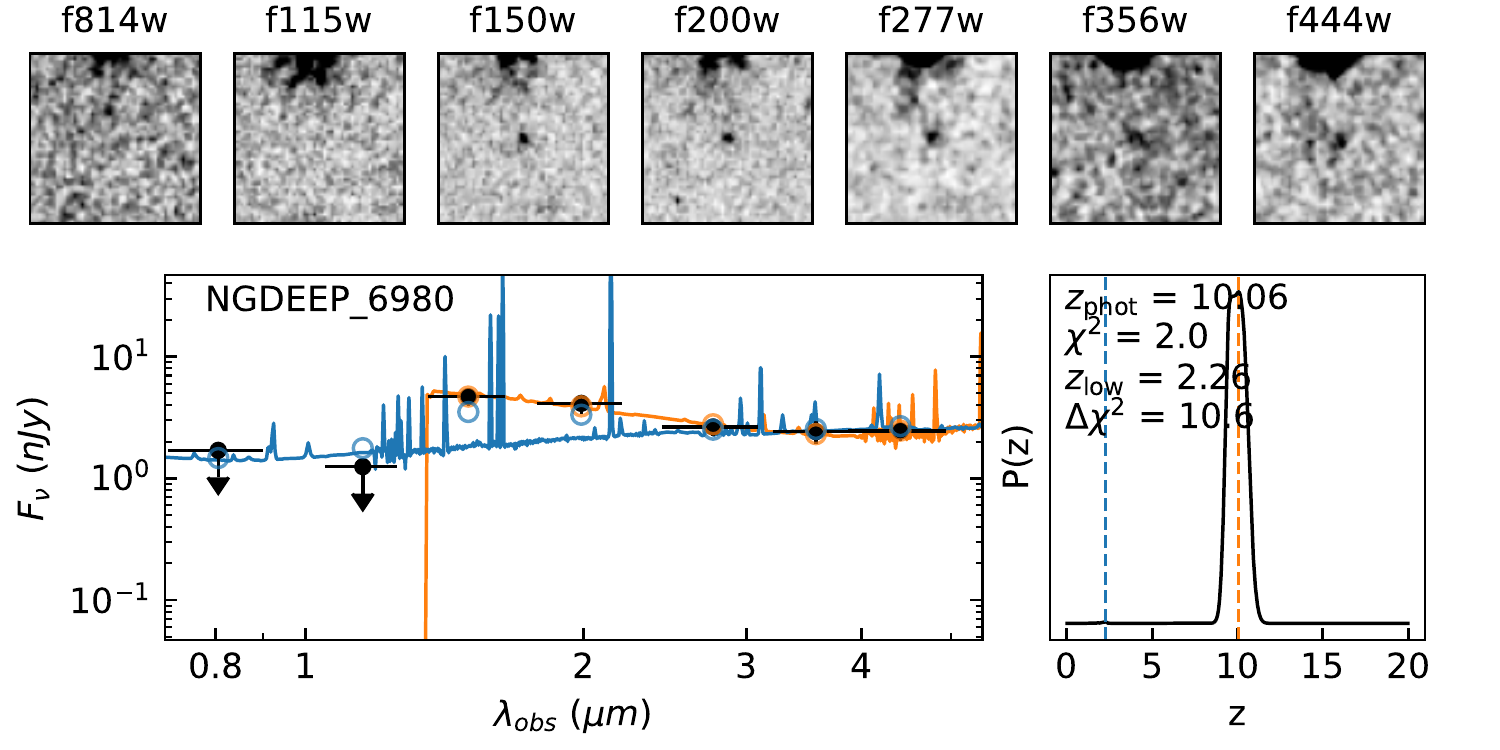}
	\\
	\includegraphics[width=0.33\textwidth]{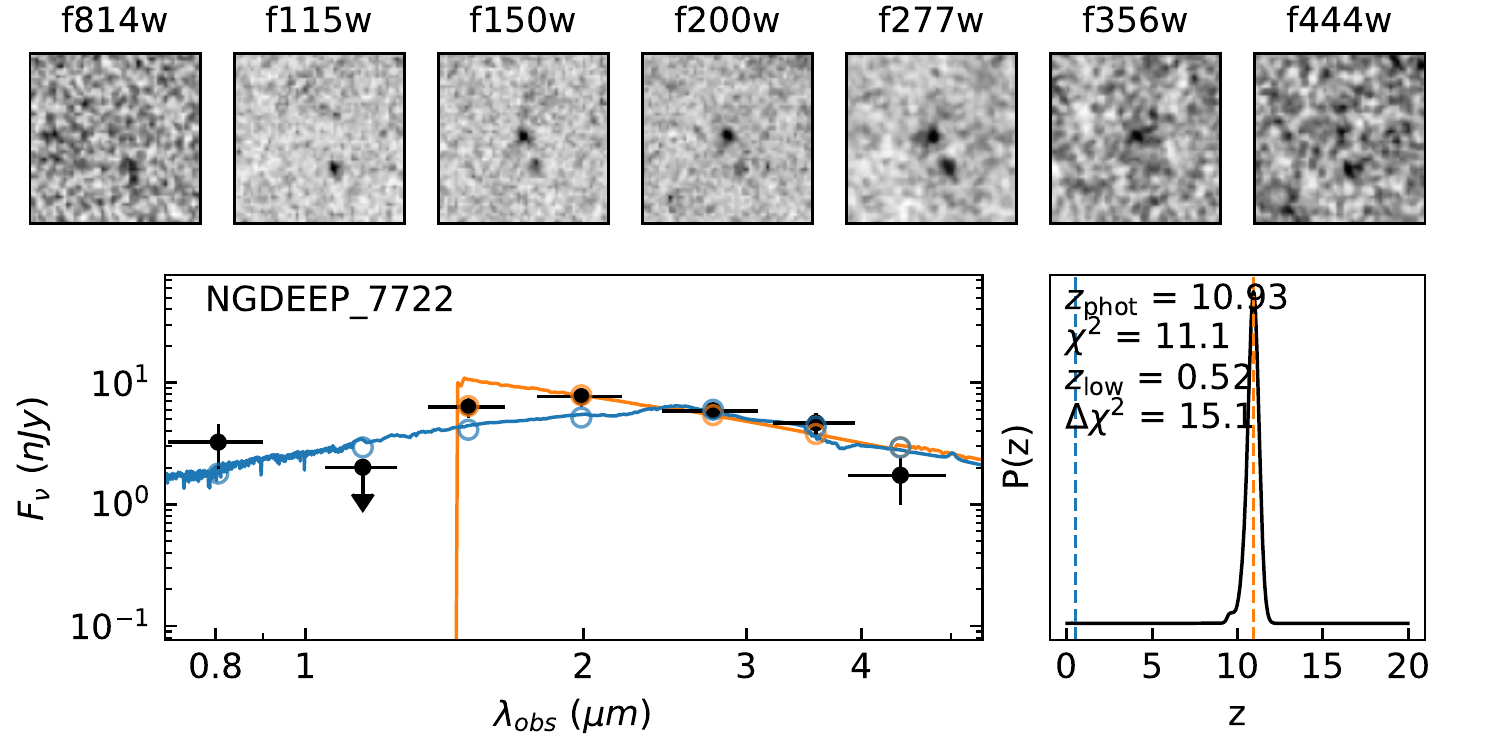}
	\includegraphics[width=0.33\textwidth]{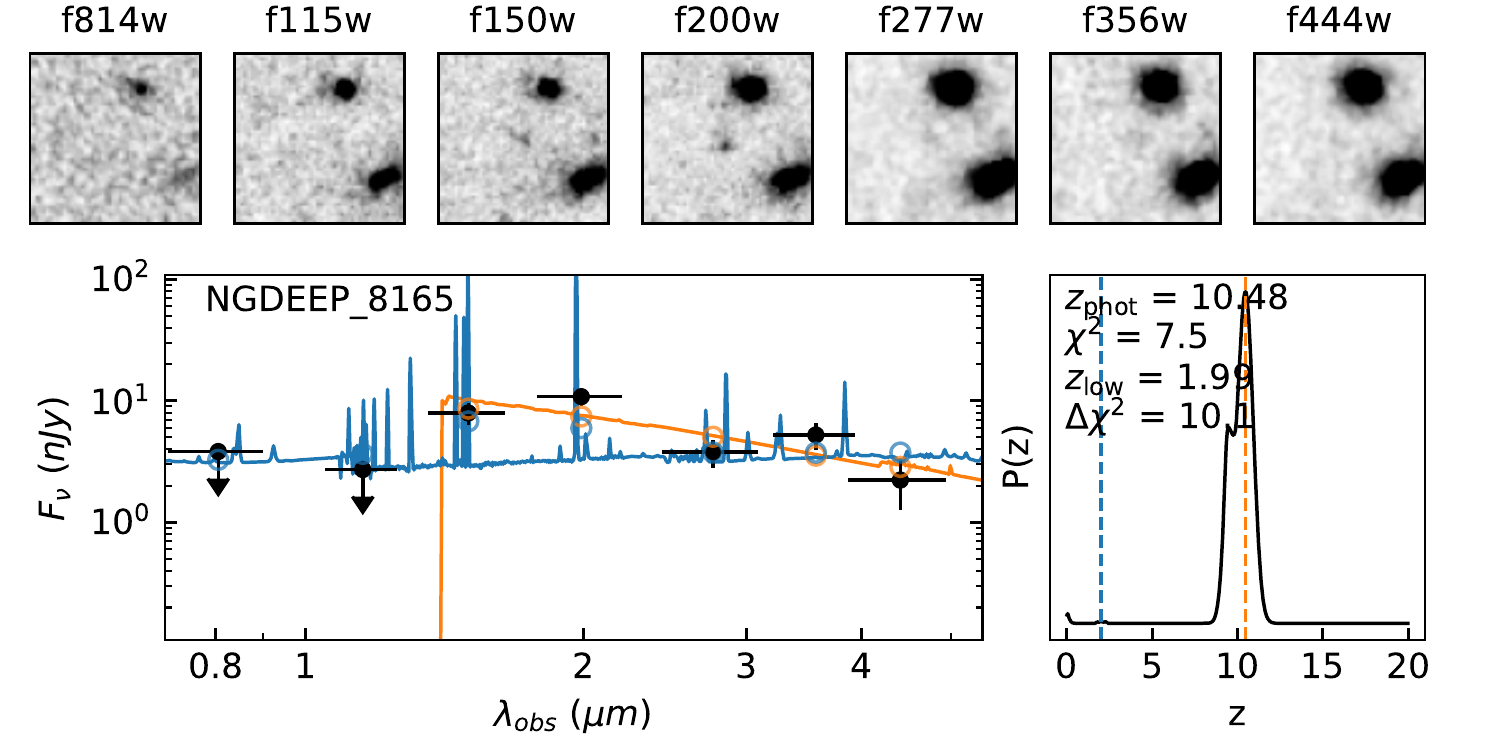}
	\includegraphics[width=0.33\textwidth]{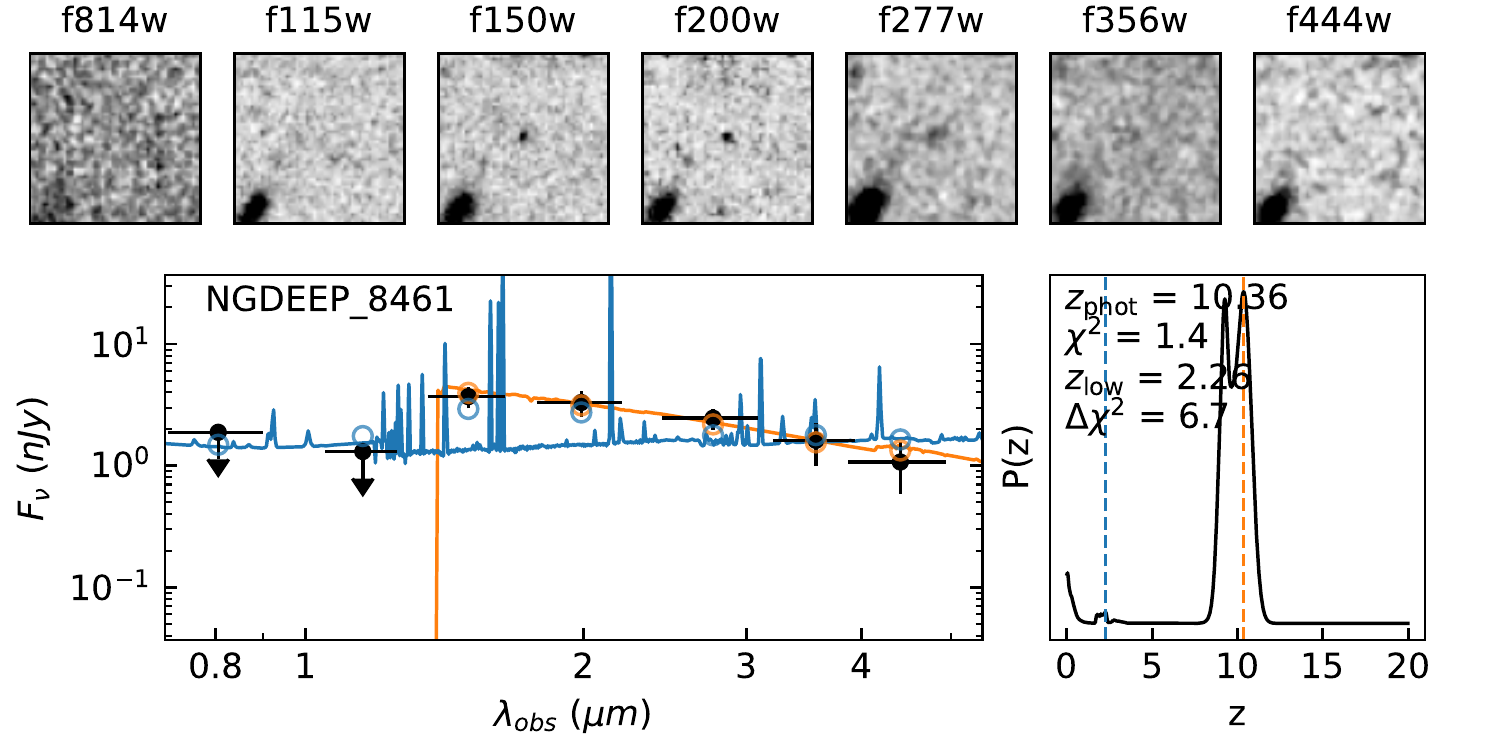}
	\\
	\includegraphics[width=0.33\textwidth]{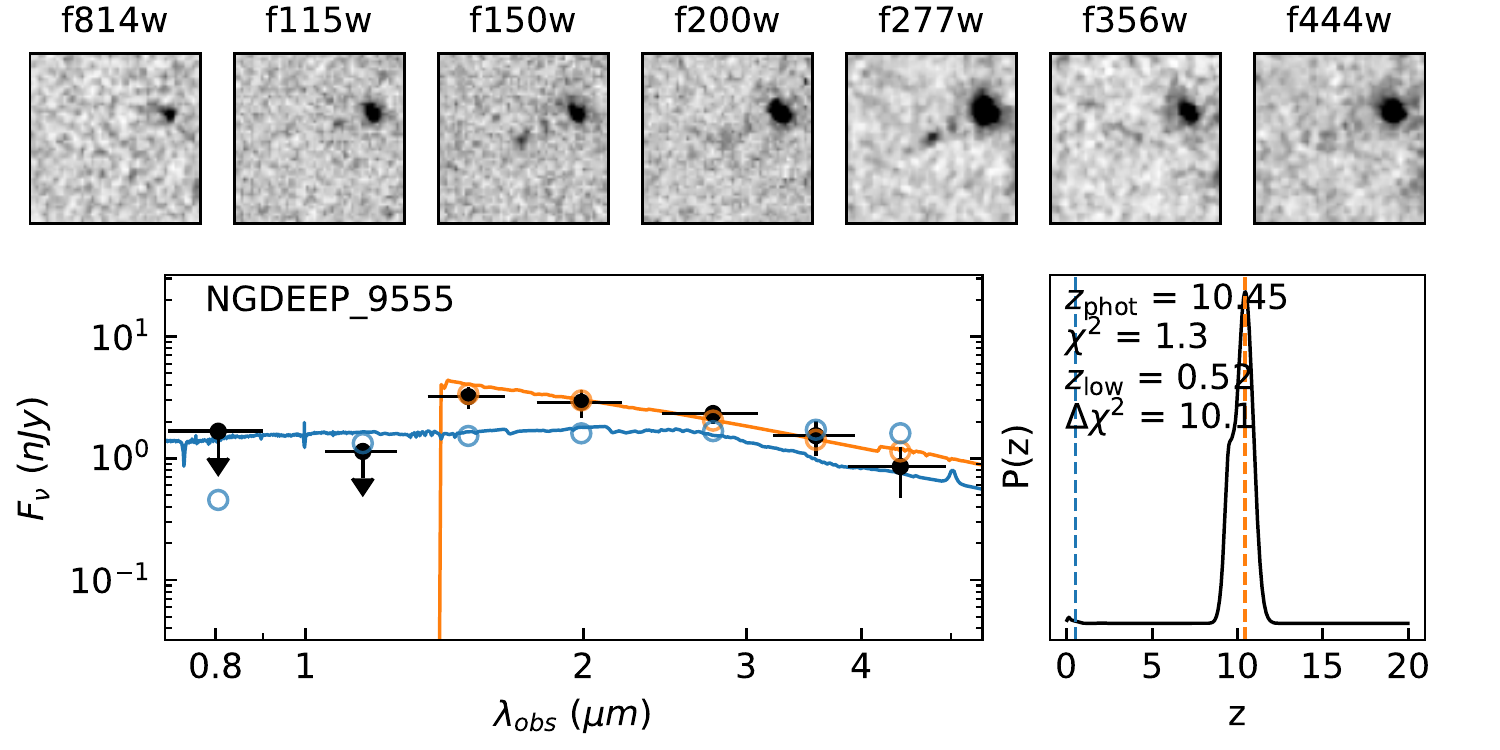}
	\includegraphics[width=0.33\textwidth]{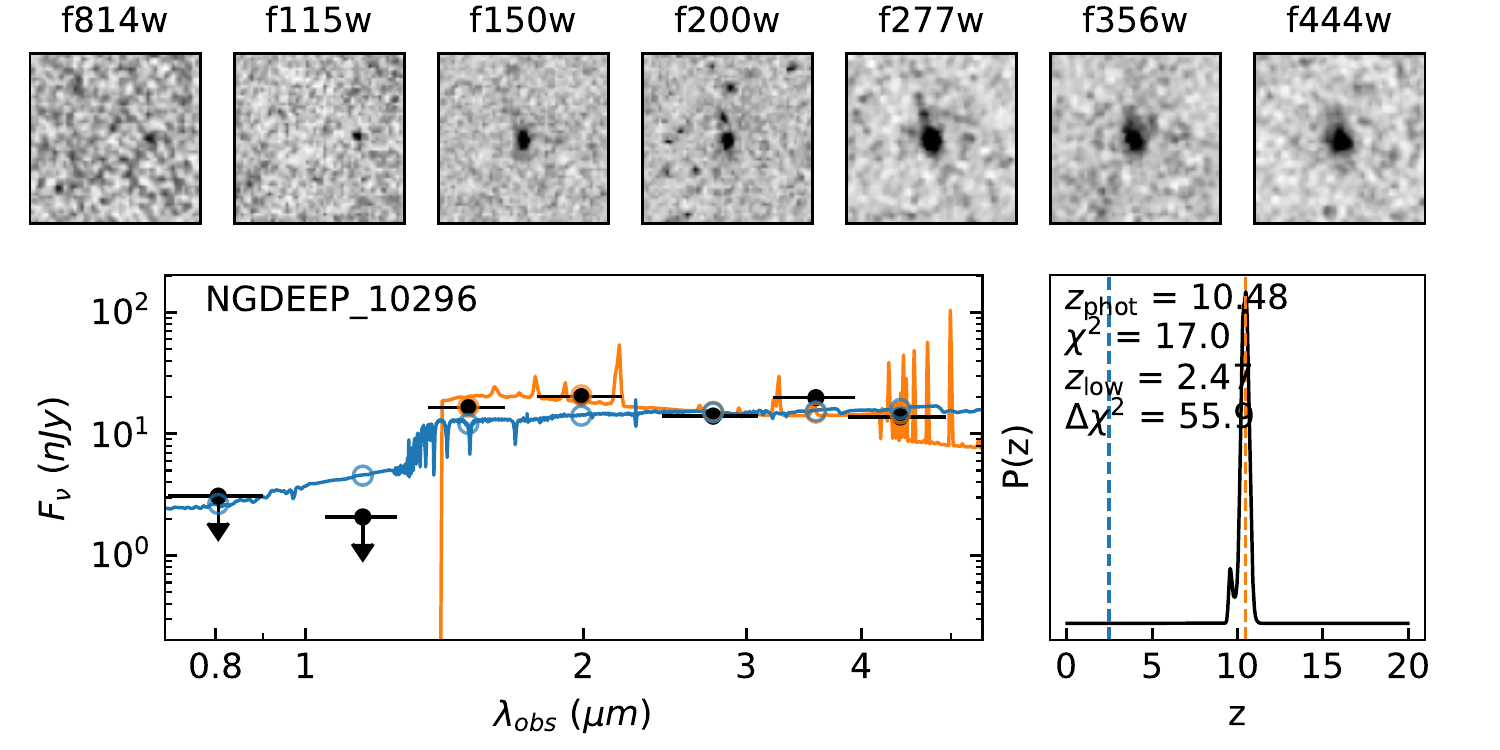}
	\includegraphics[width=0.33\textwidth]{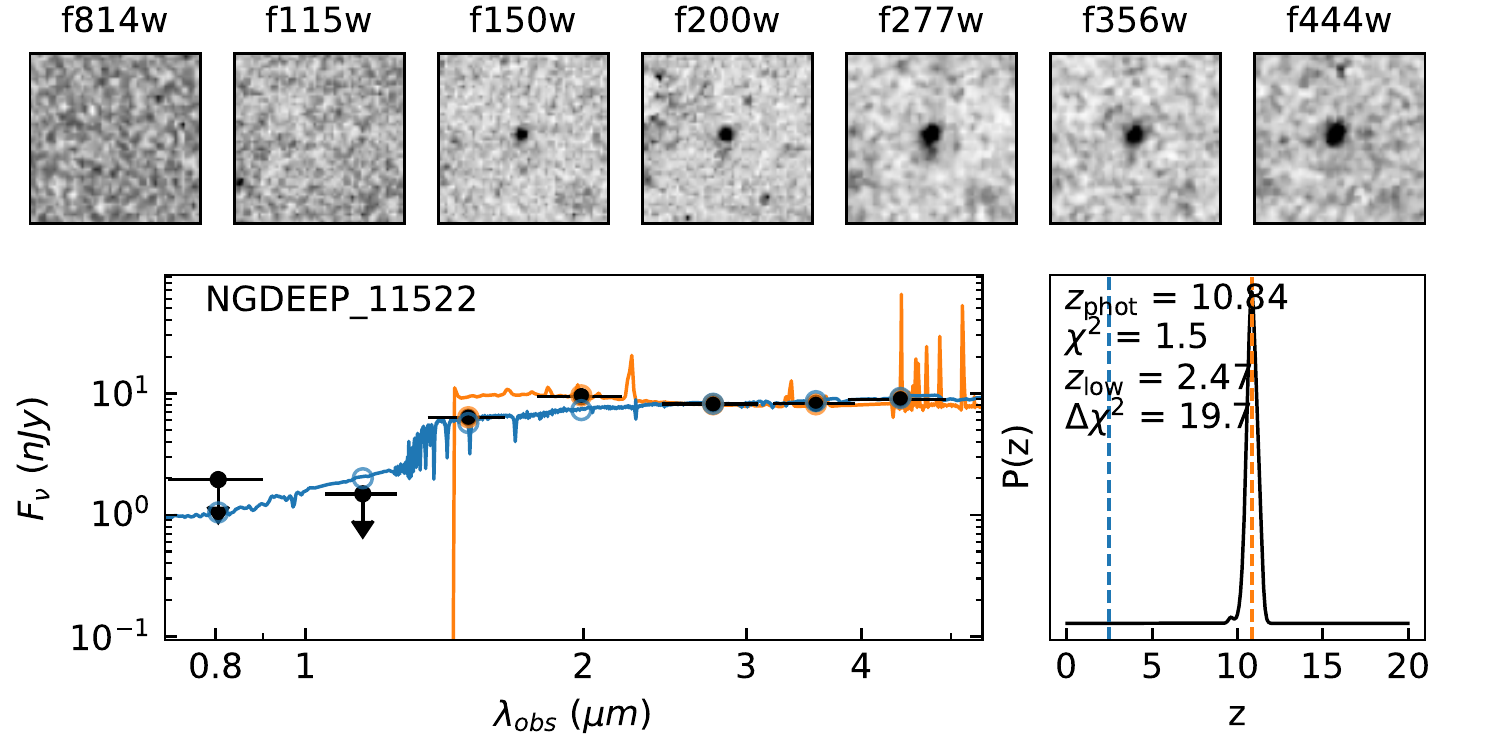}
	\\
	\includegraphics[width=0.33\textwidth]{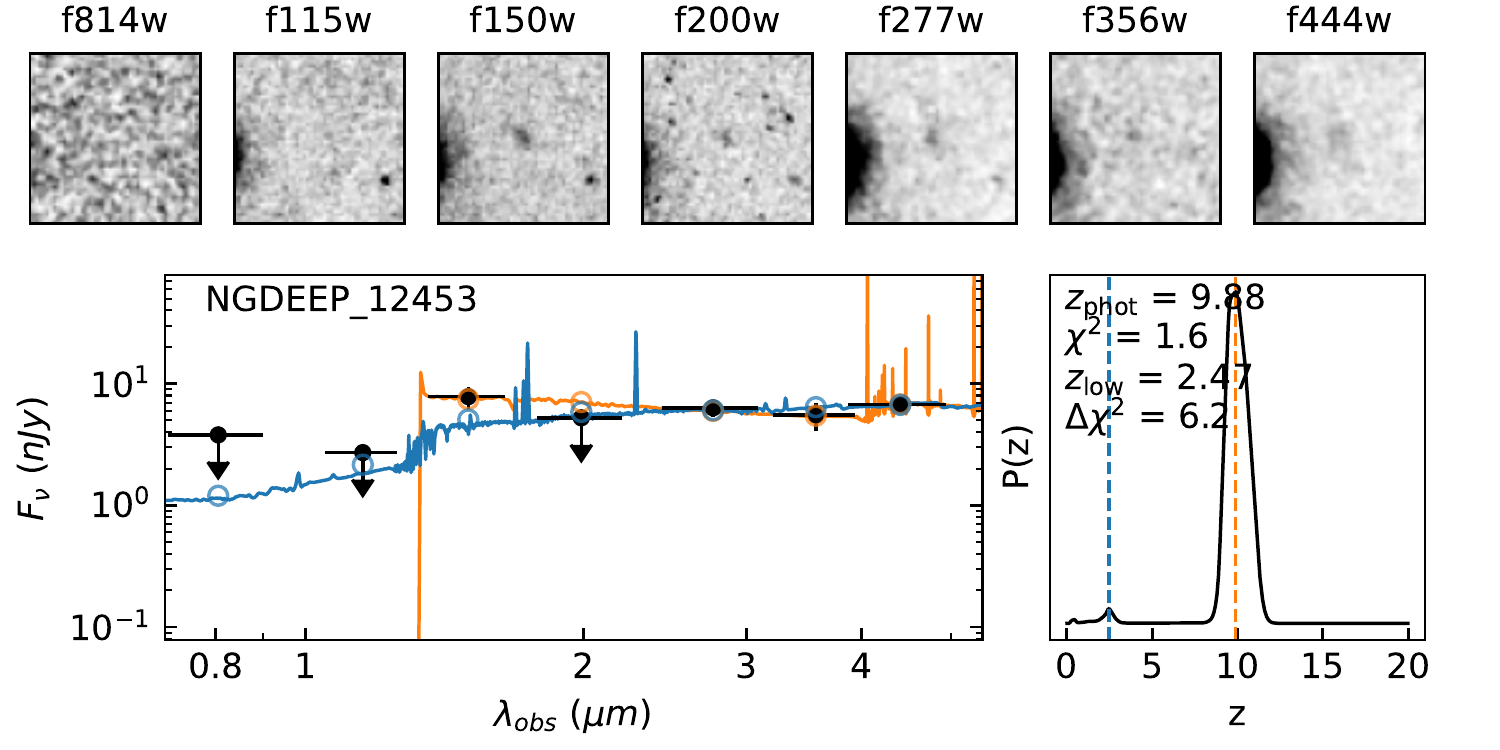}
    \caption{Same as Figure \ref{fig:bio_z15}, but for sources at $z \sim 11$.}\label{fig:bio_z11}
\end{figure*}

\begin{figure*}
	\includegraphics[width=0.33\textwidth]{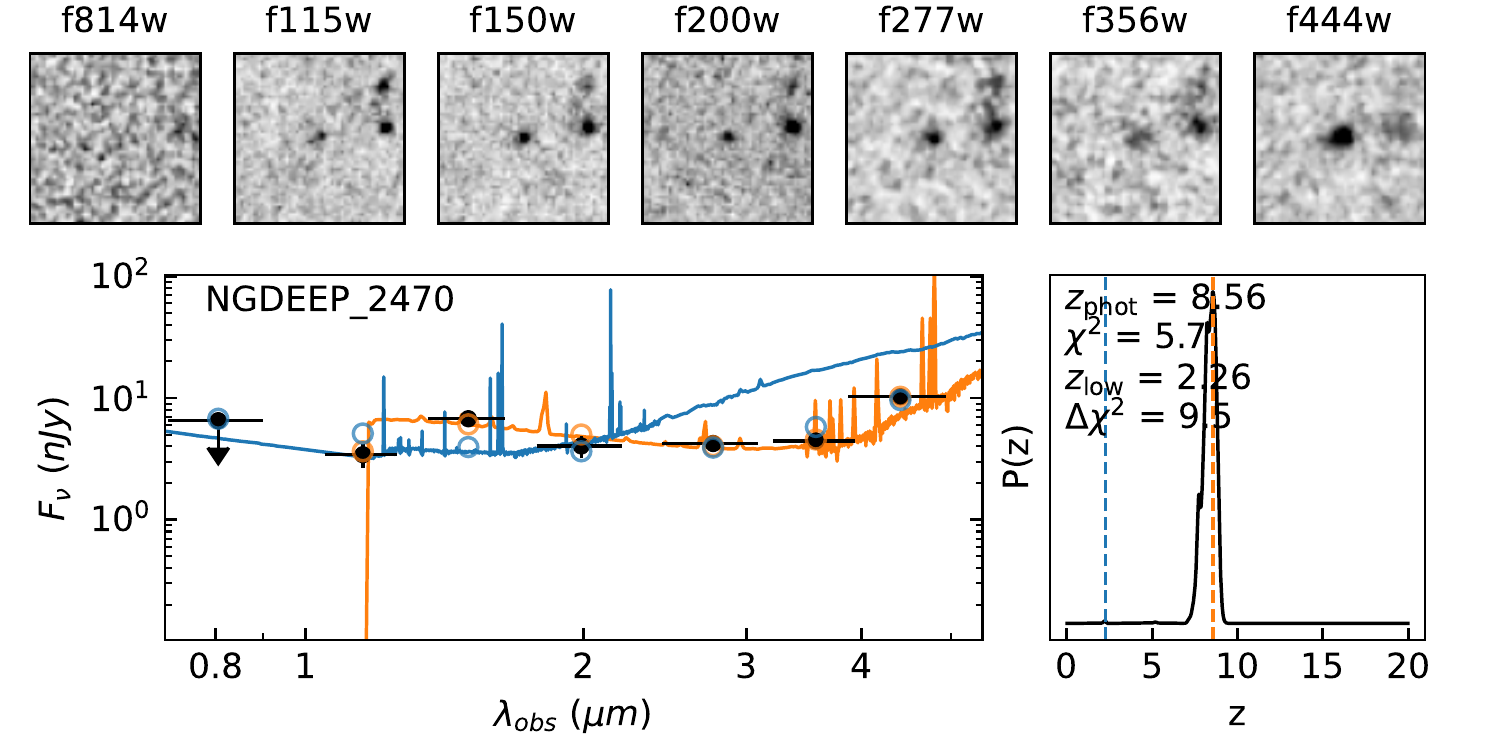}
	\includegraphics[width=0.33\textwidth]{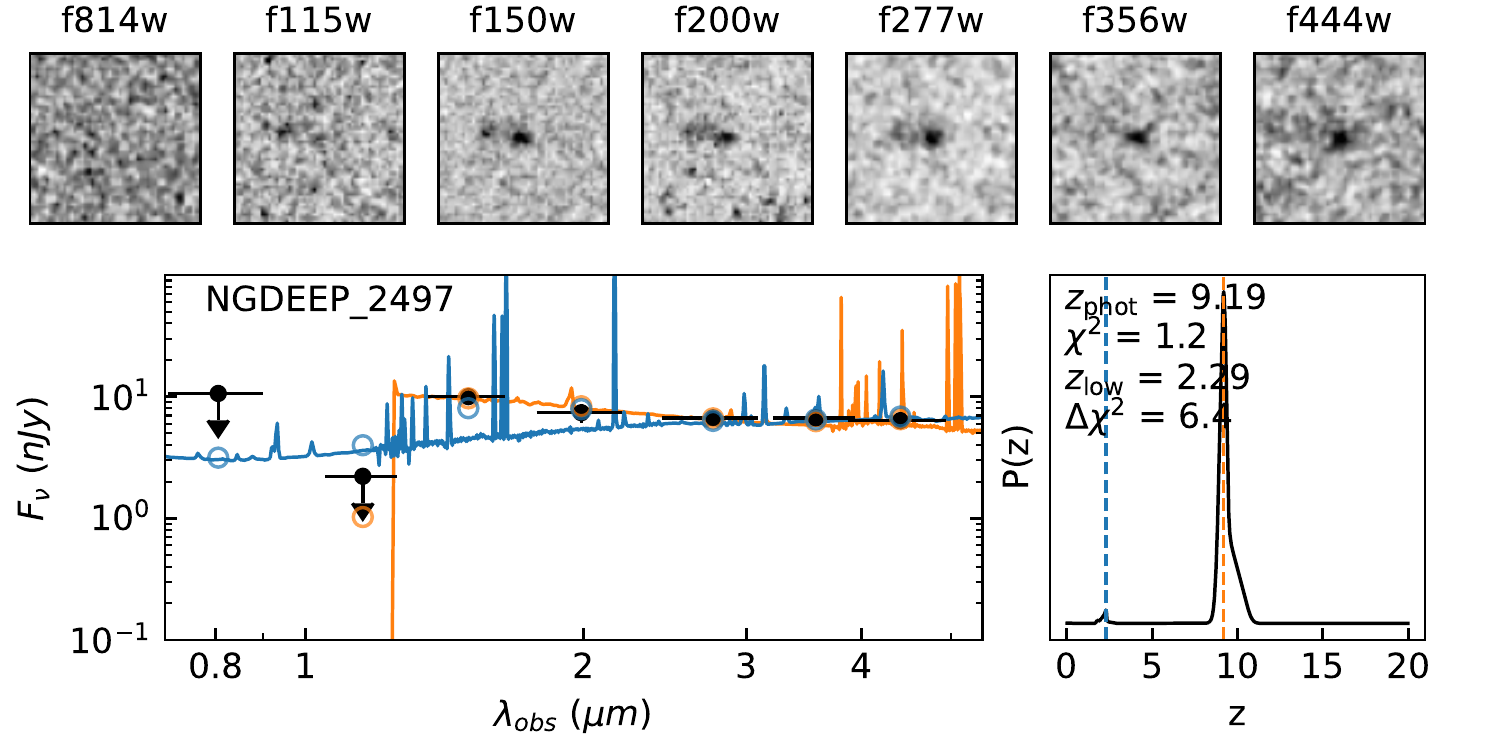}
	\includegraphics[width=0.33\textwidth]{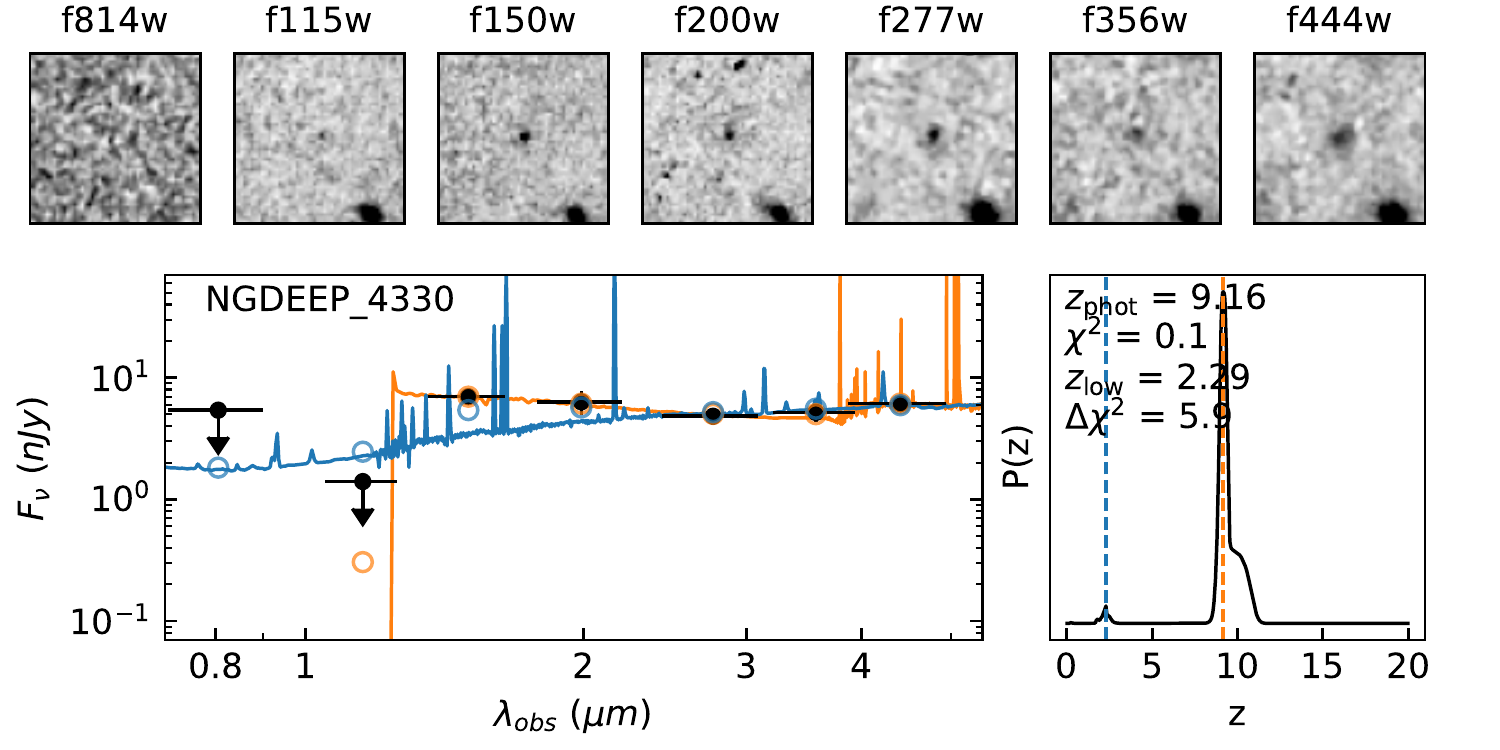}
	\\
	\includegraphics[width=0.33\textwidth]{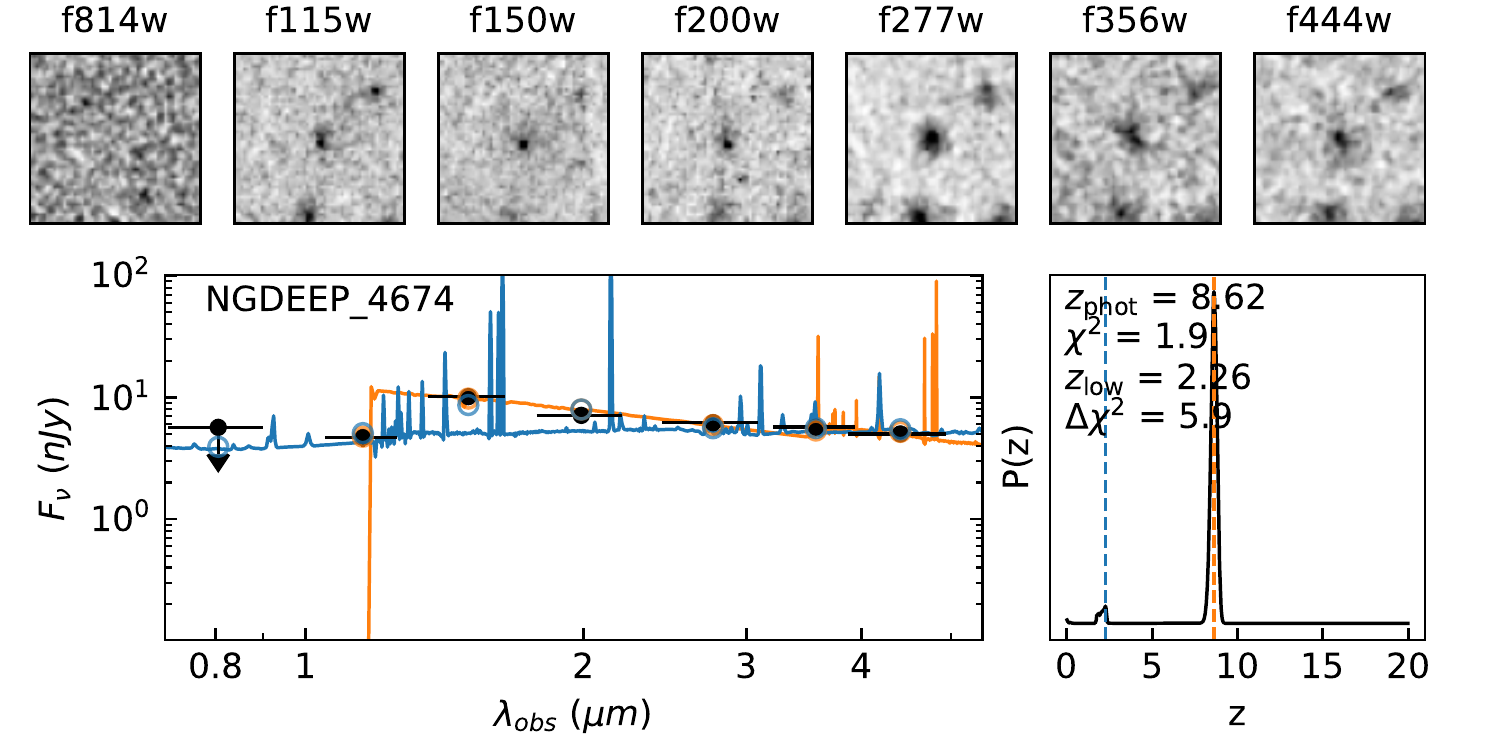}
	\includegraphics[width=0.33\textwidth]{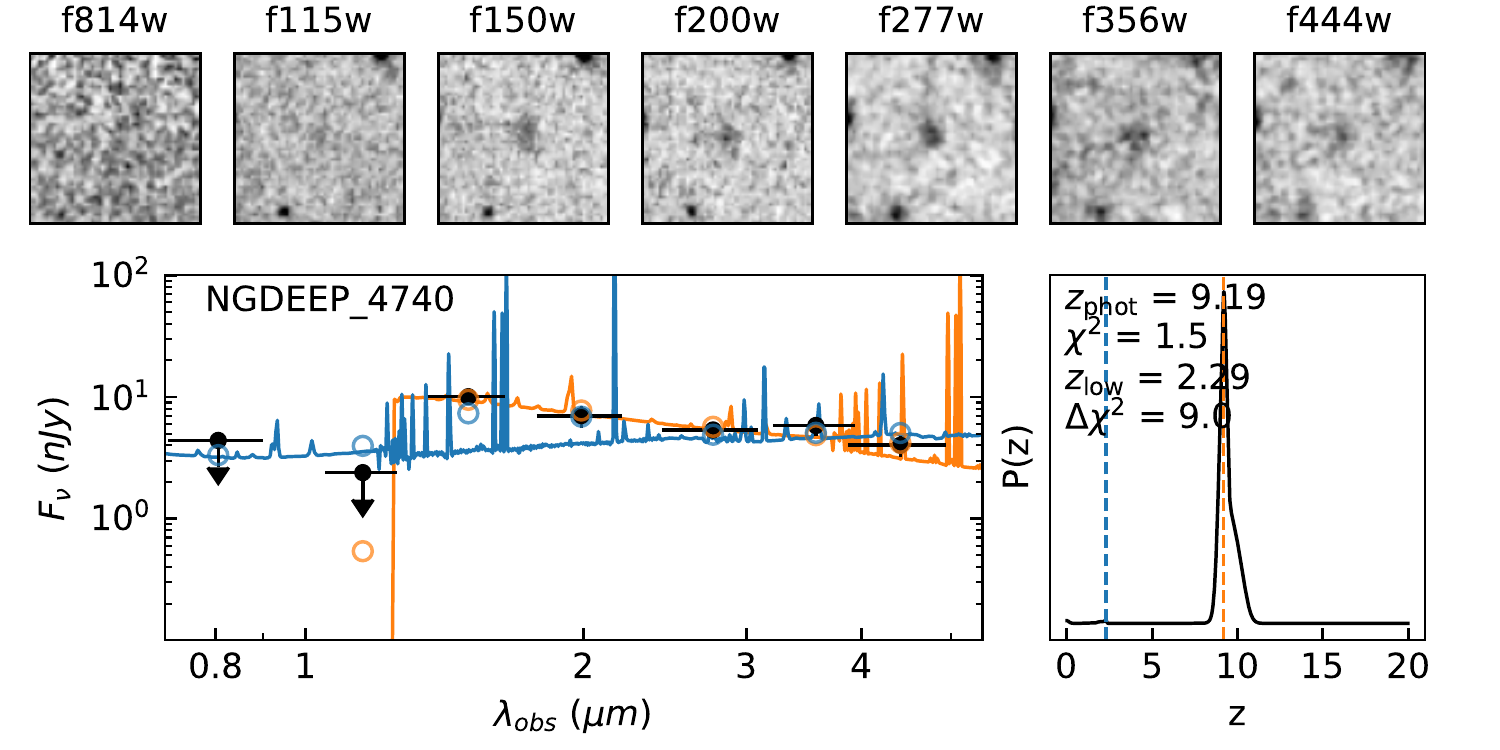}
	\includegraphics[width=0.33\textwidth]{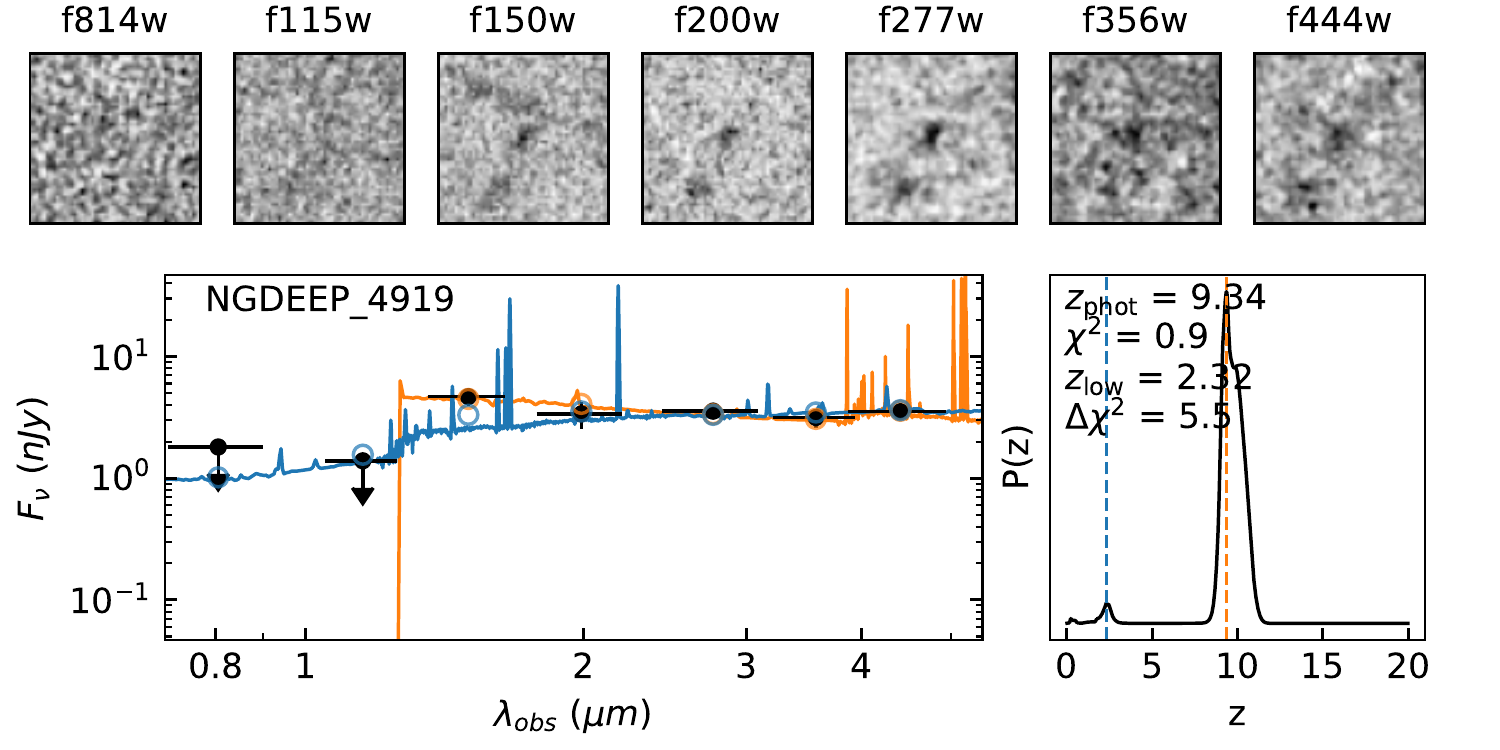}
	\\
	\includegraphics[width=0.33\textwidth]{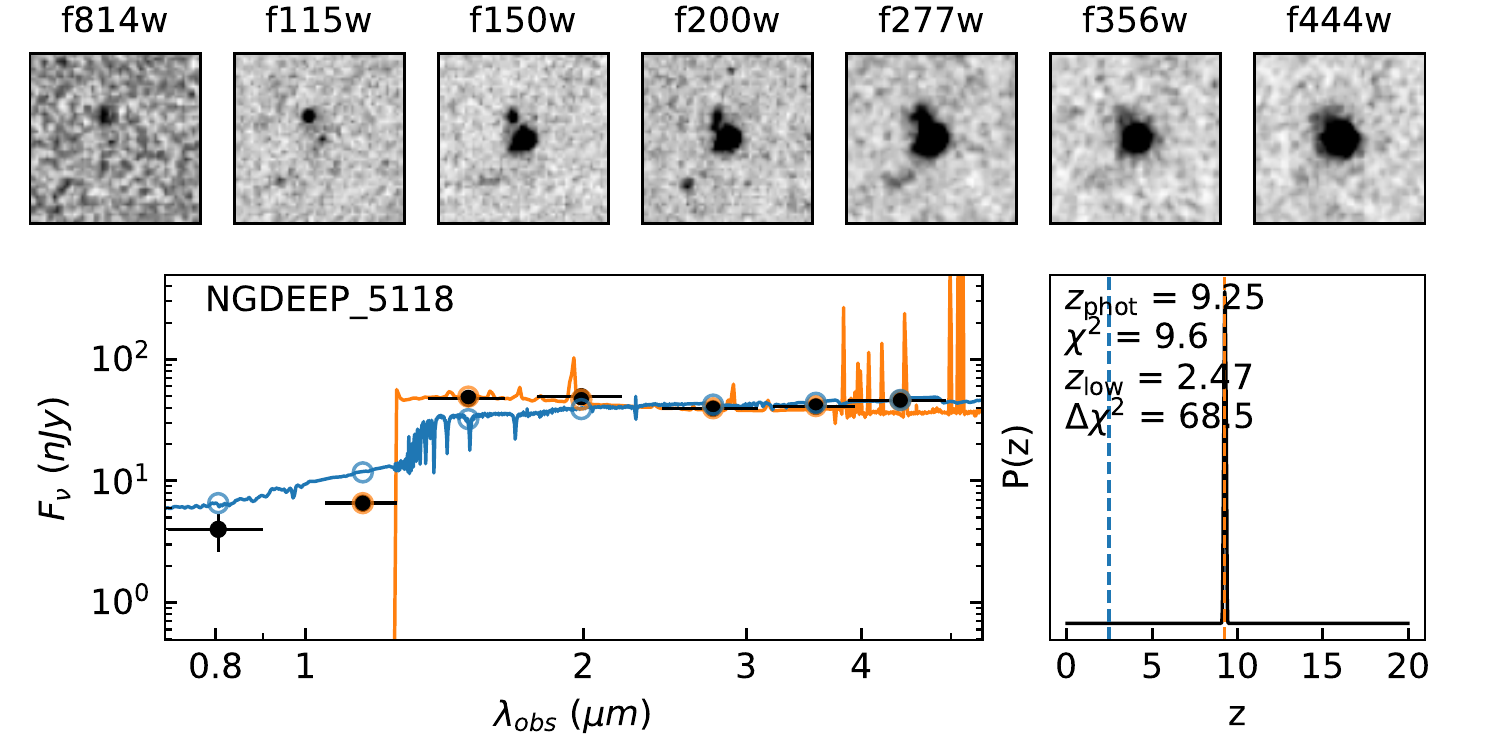}
	\includegraphics[width=0.33\textwidth]{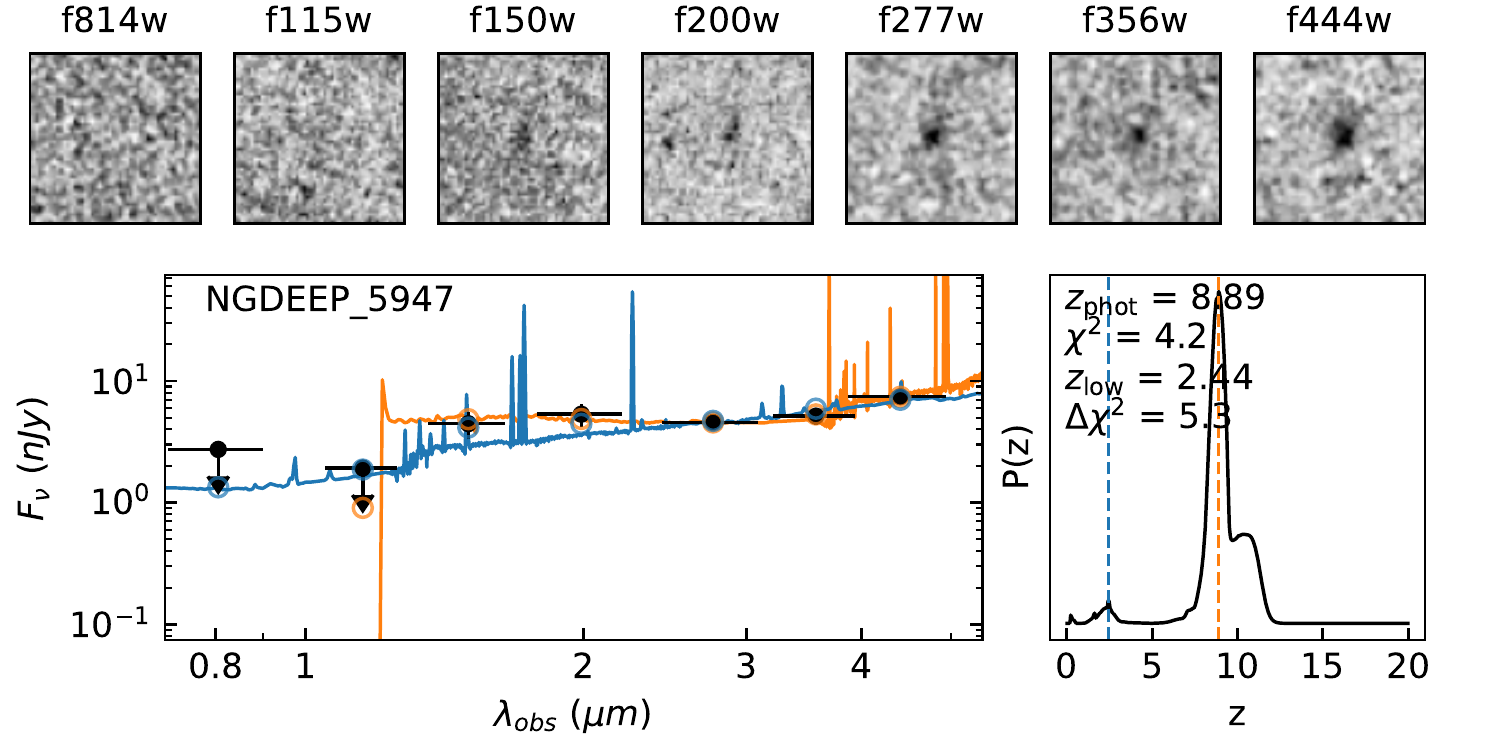}
	\includegraphics[width=0.33\textwidth]{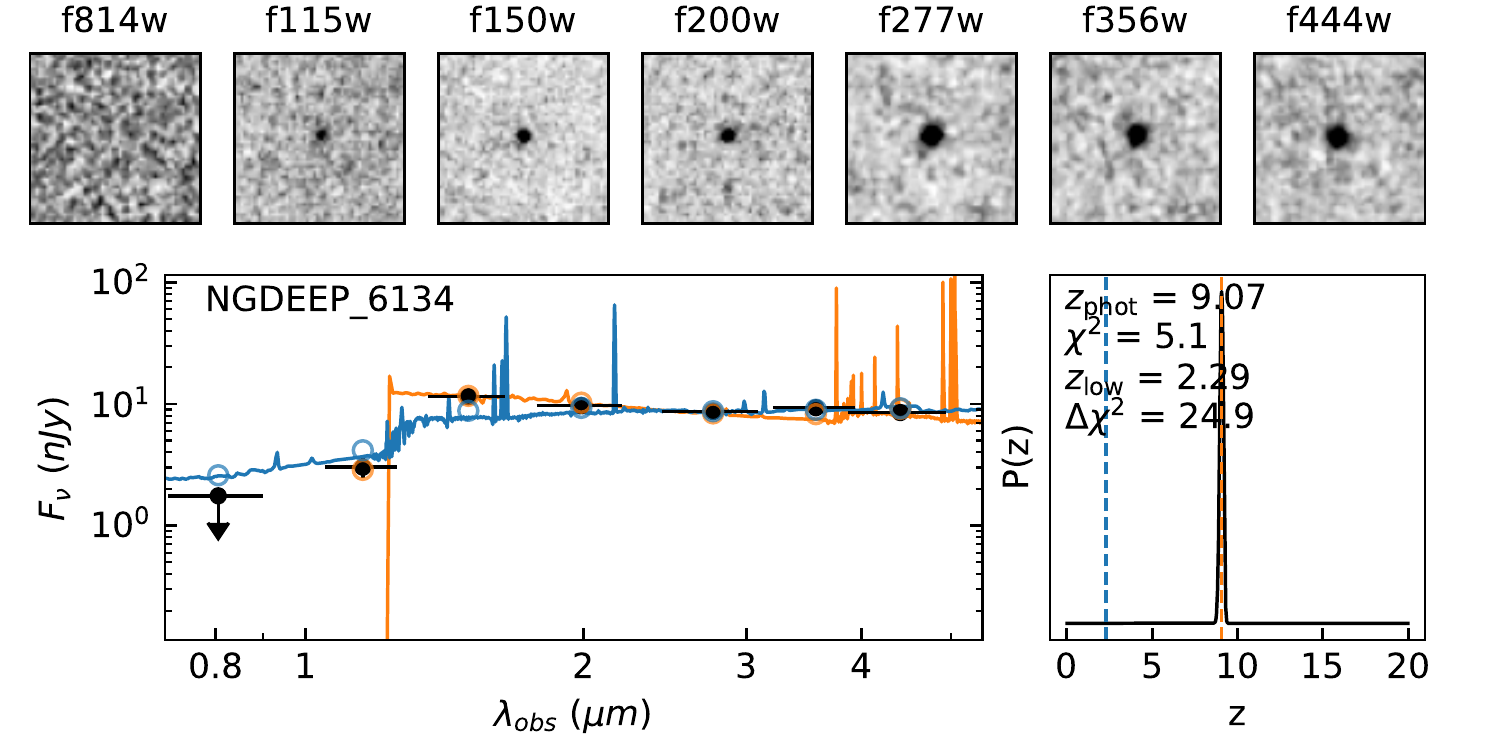}
	\\
	\includegraphics[width=0.33\textwidth]{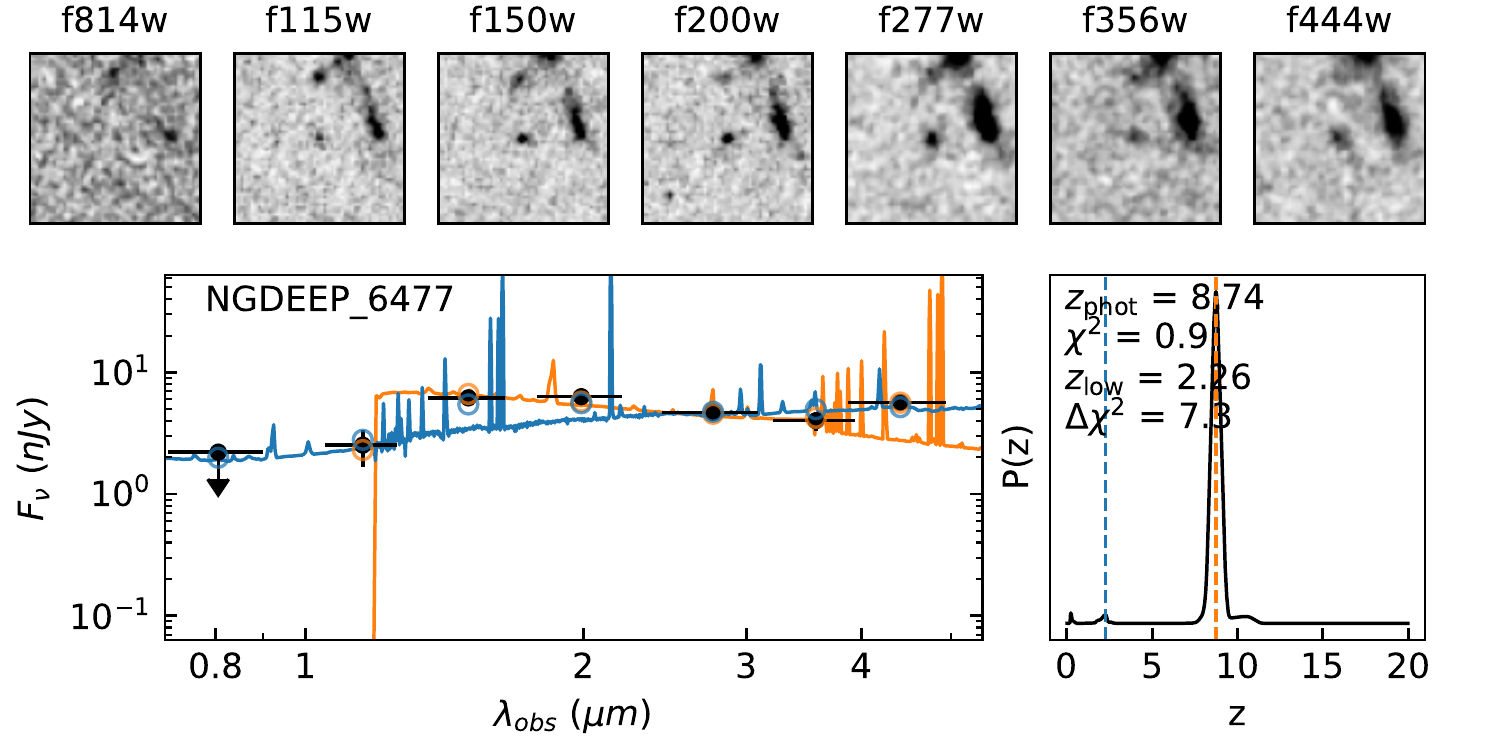}
	\includegraphics[width=0.33\textwidth]{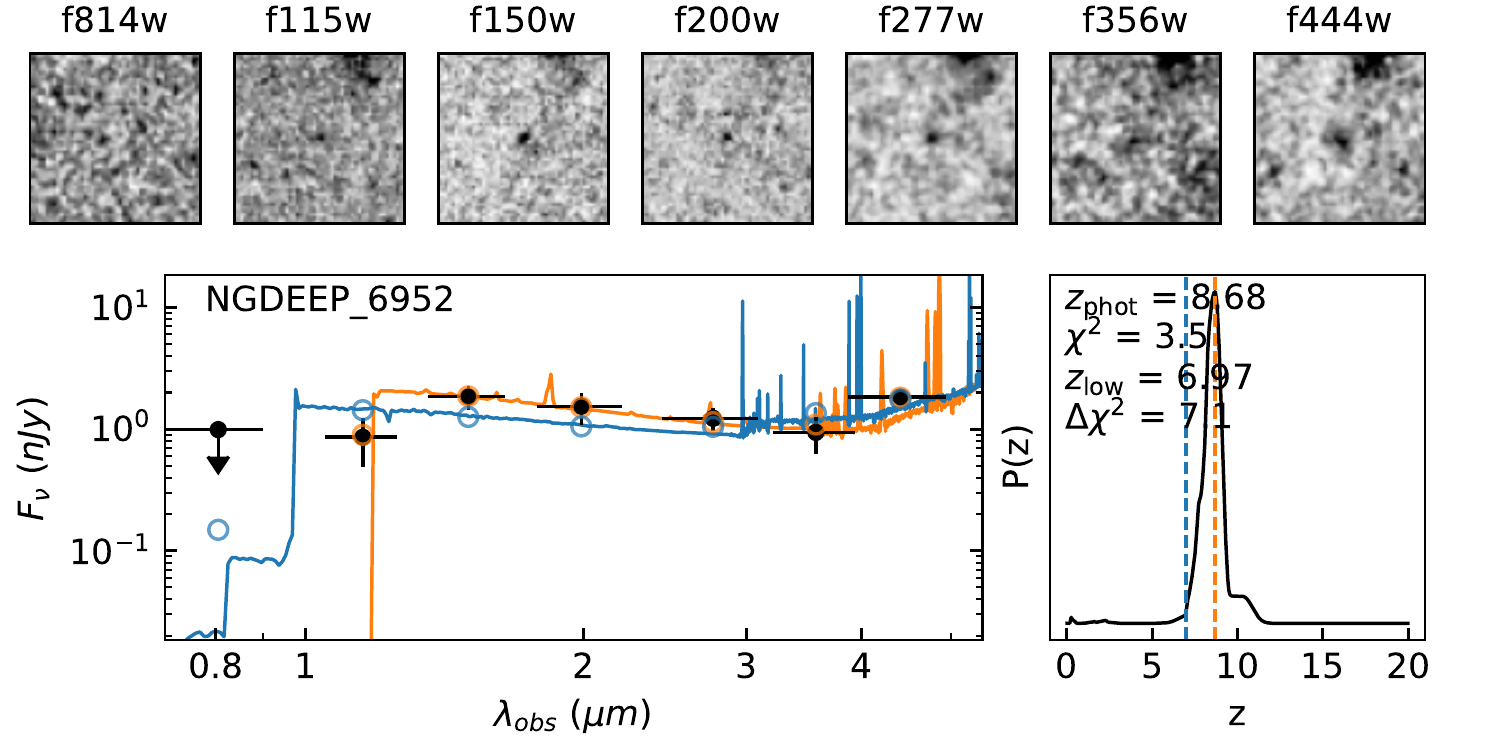}
	\includegraphics[width=0.33\textwidth]{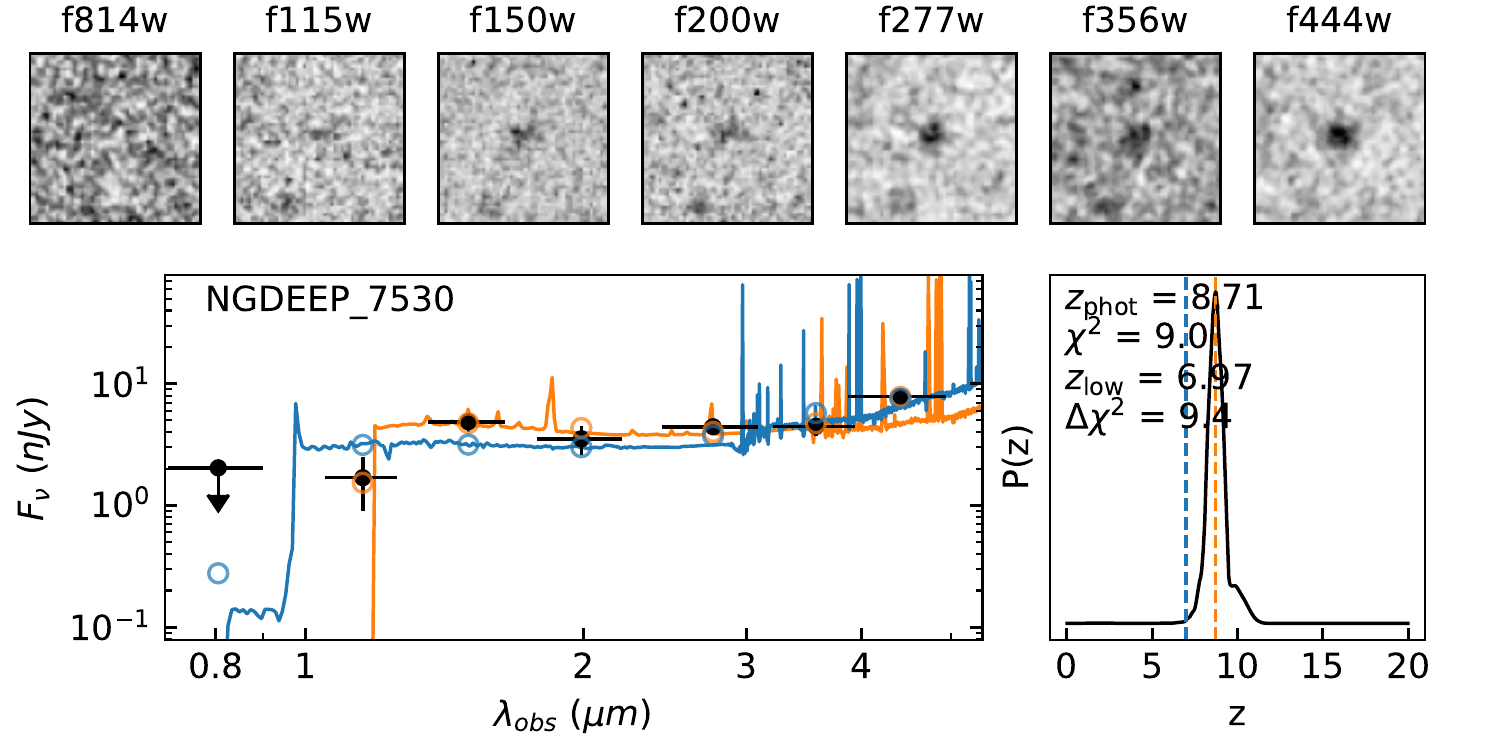}
	\\
	\includegraphics[width=0.33\textwidth]{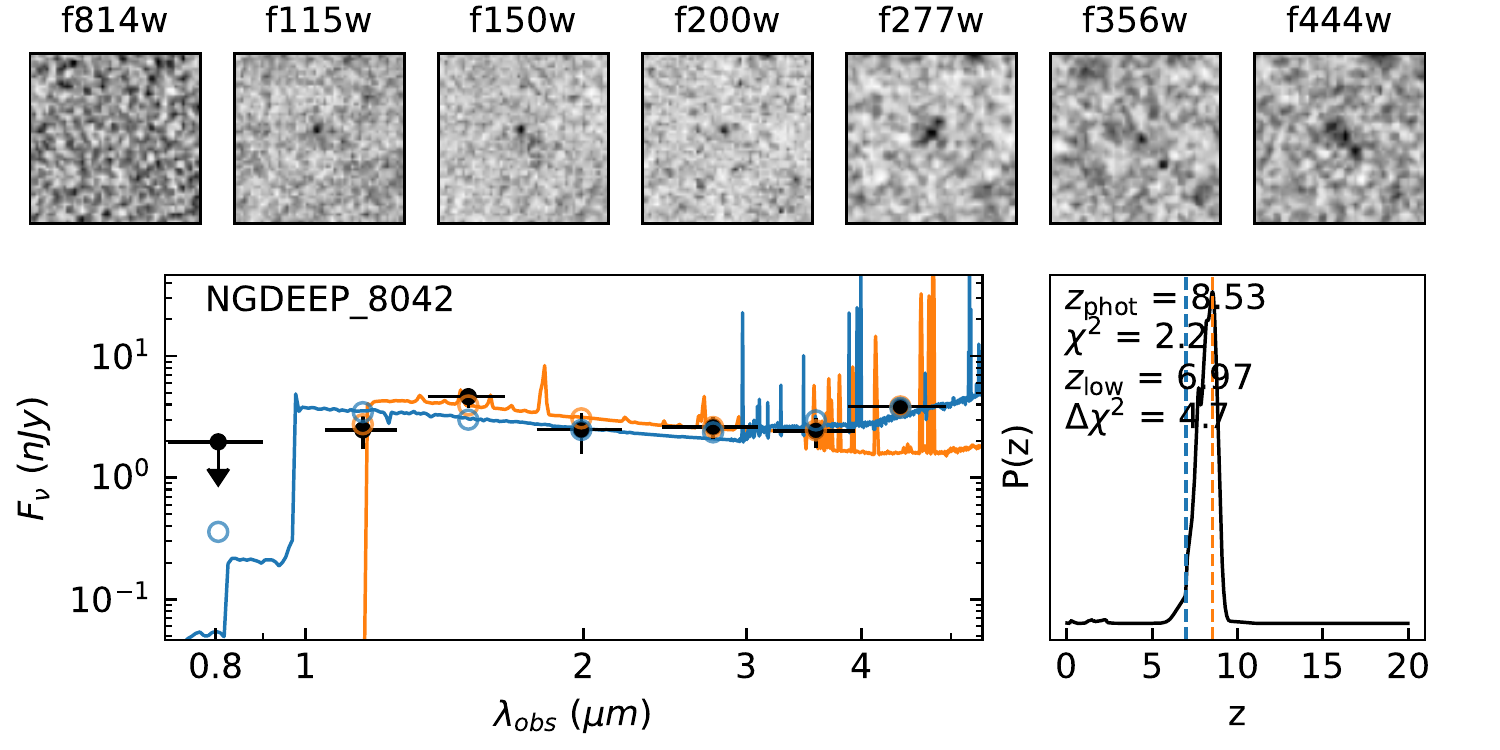}
	\includegraphics[width=0.33\textwidth]{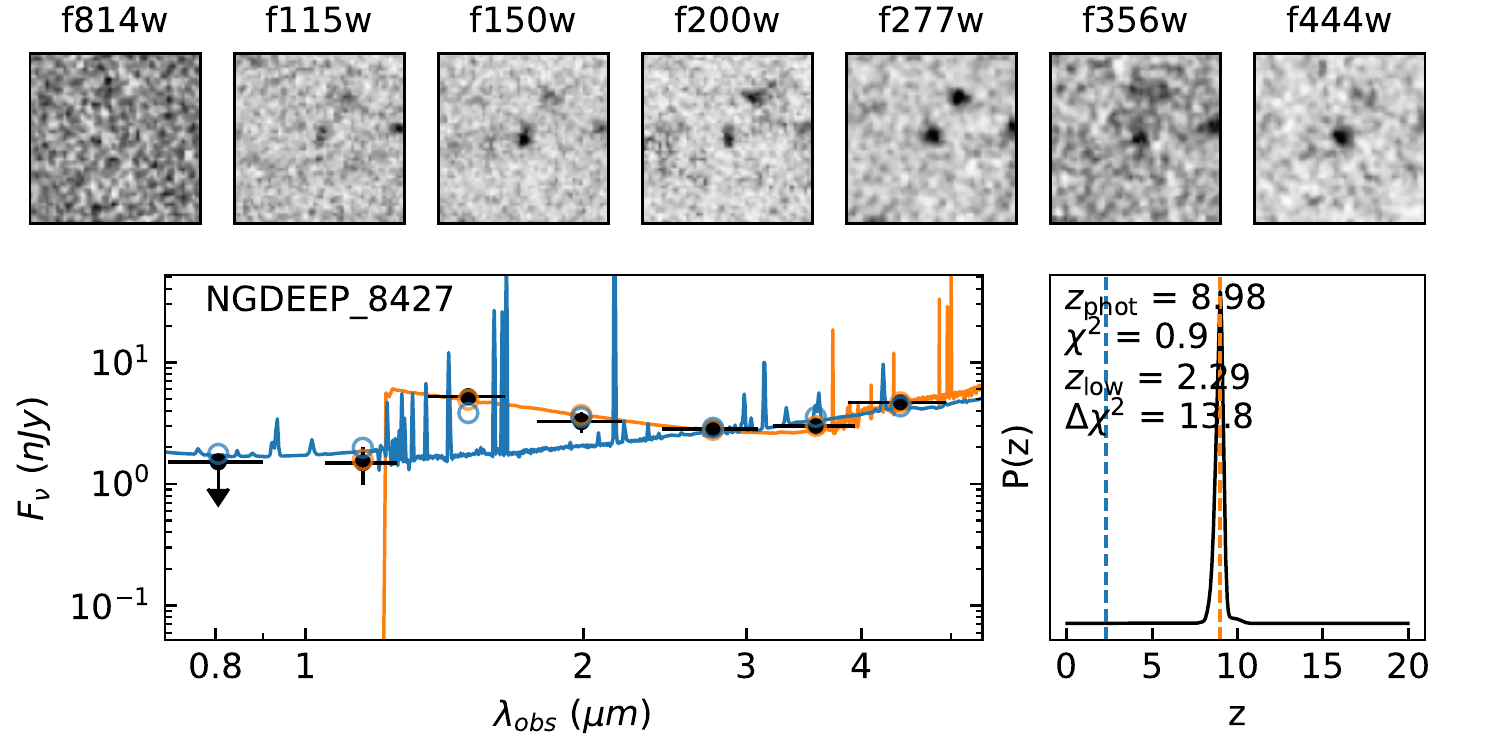}
	\includegraphics[width=0.33\textwidth]{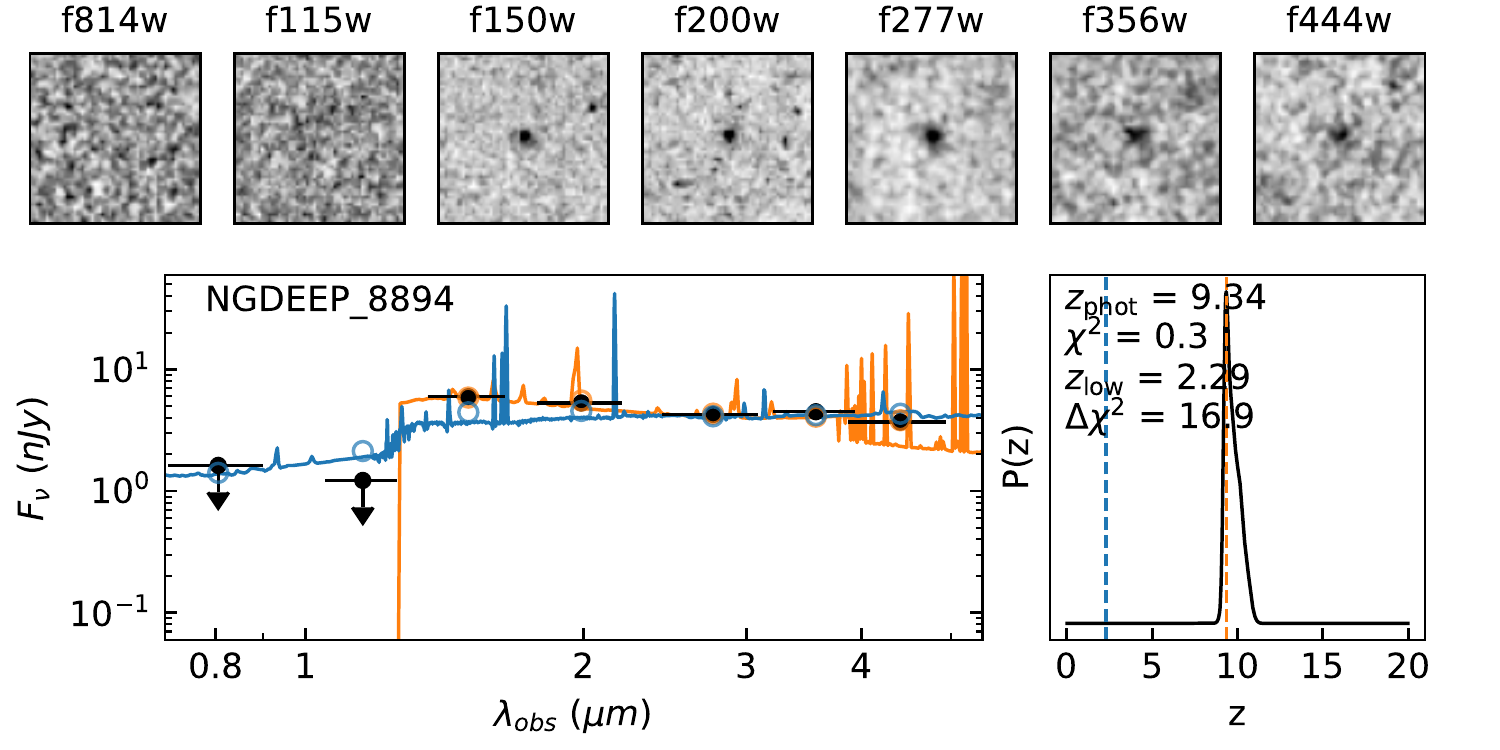}
	\\
	\includegraphics[width=0.33\textwidth]{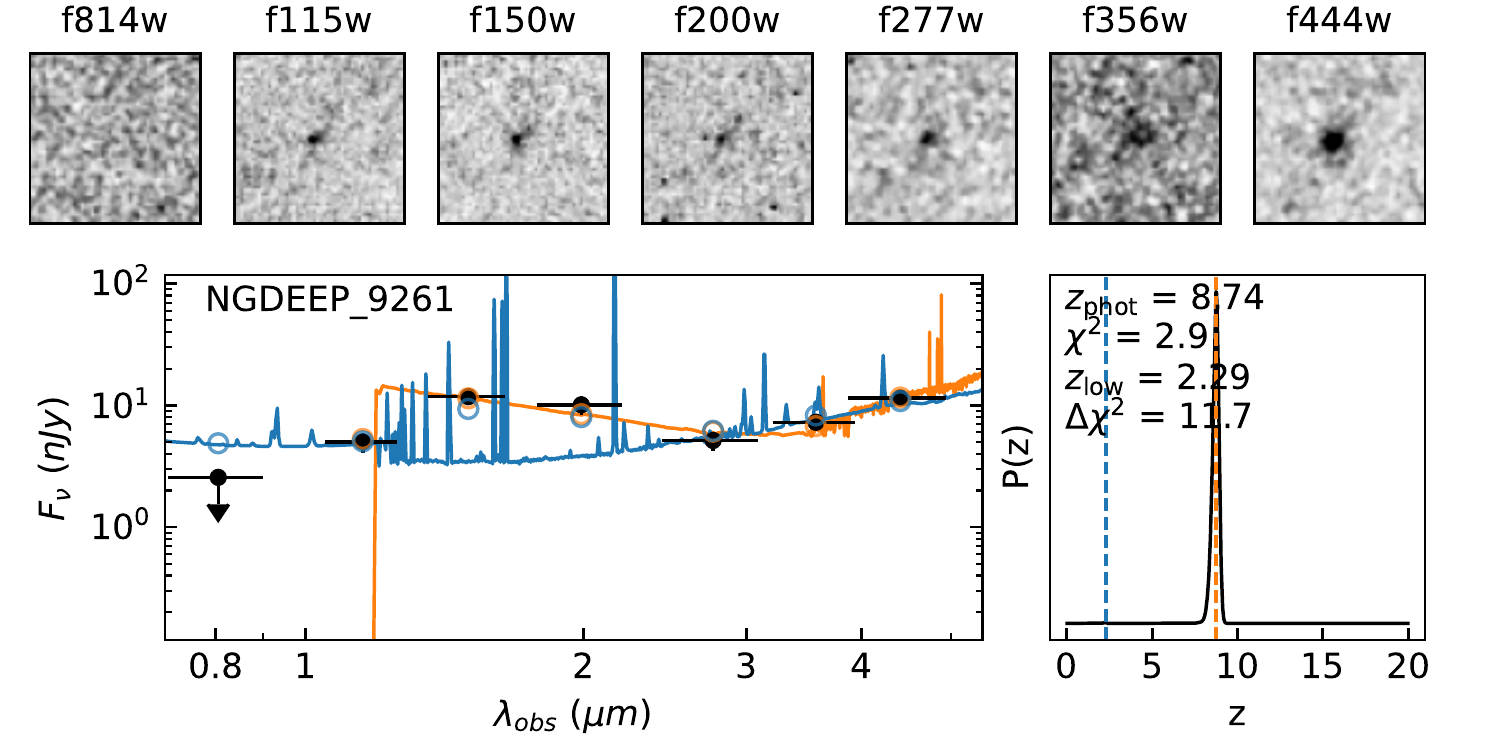}
	\includegraphics[width=0.33\textwidth]{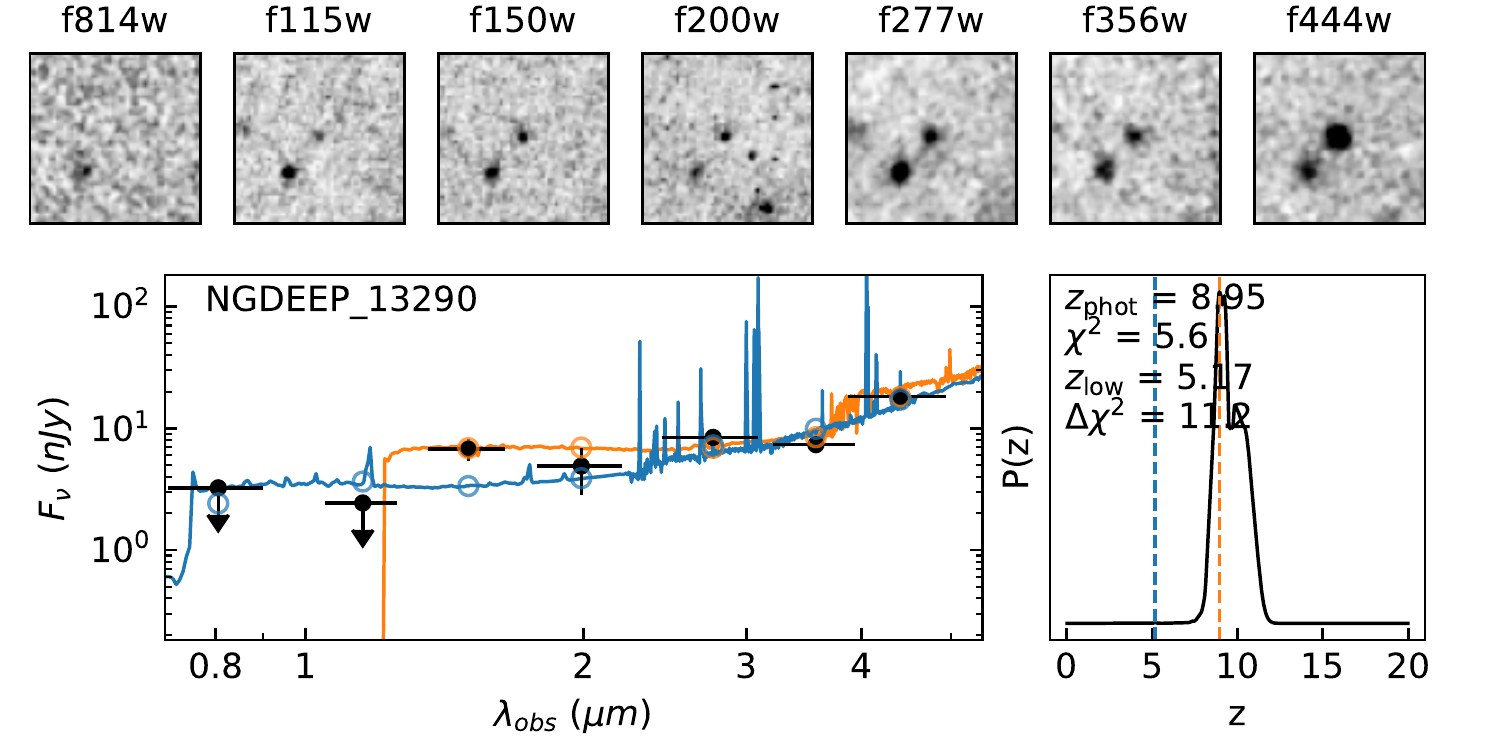}
	\includegraphics[width=0.33\textwidth]{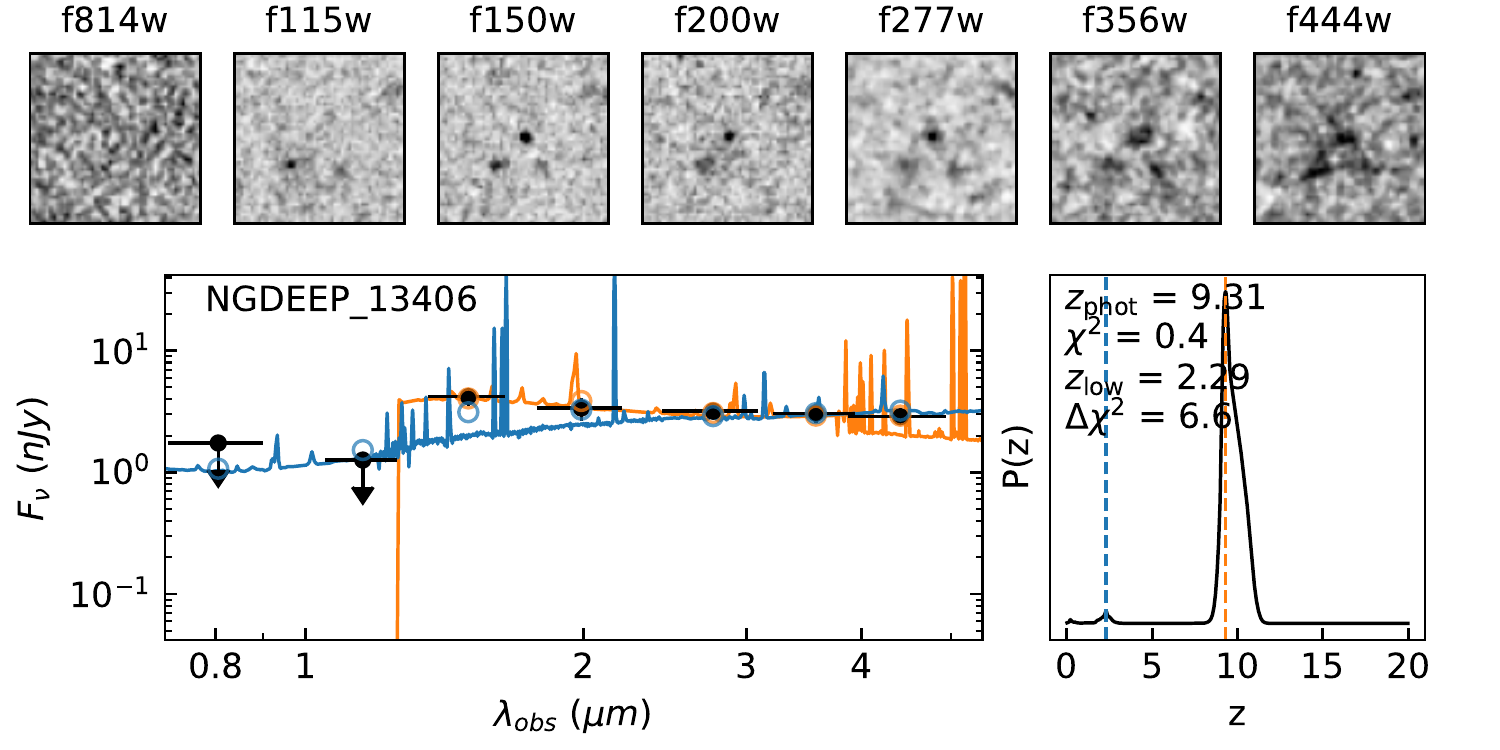}
    \\
    \includegraphics[width=0.33\textwidth]{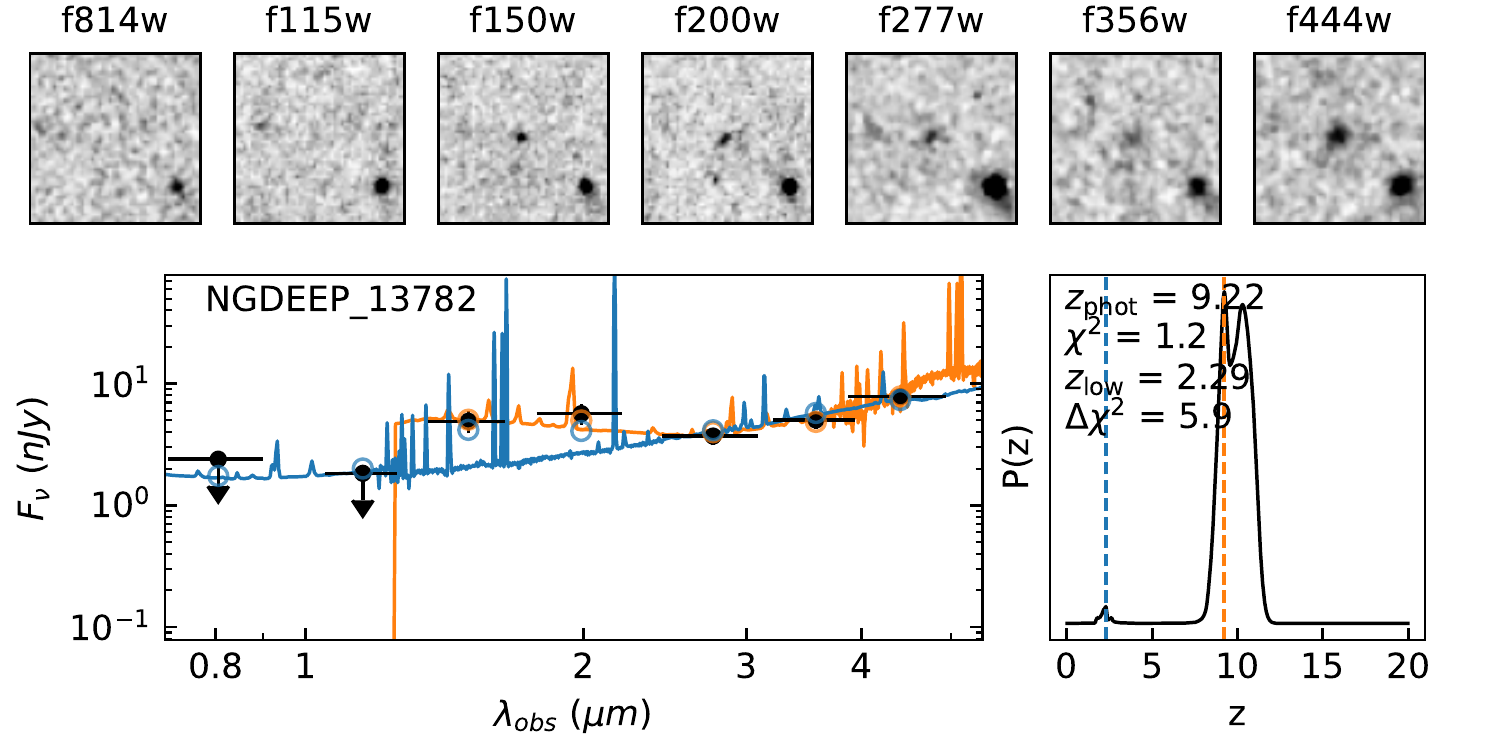}
	\includegraphics[width=0.33\textwidth]{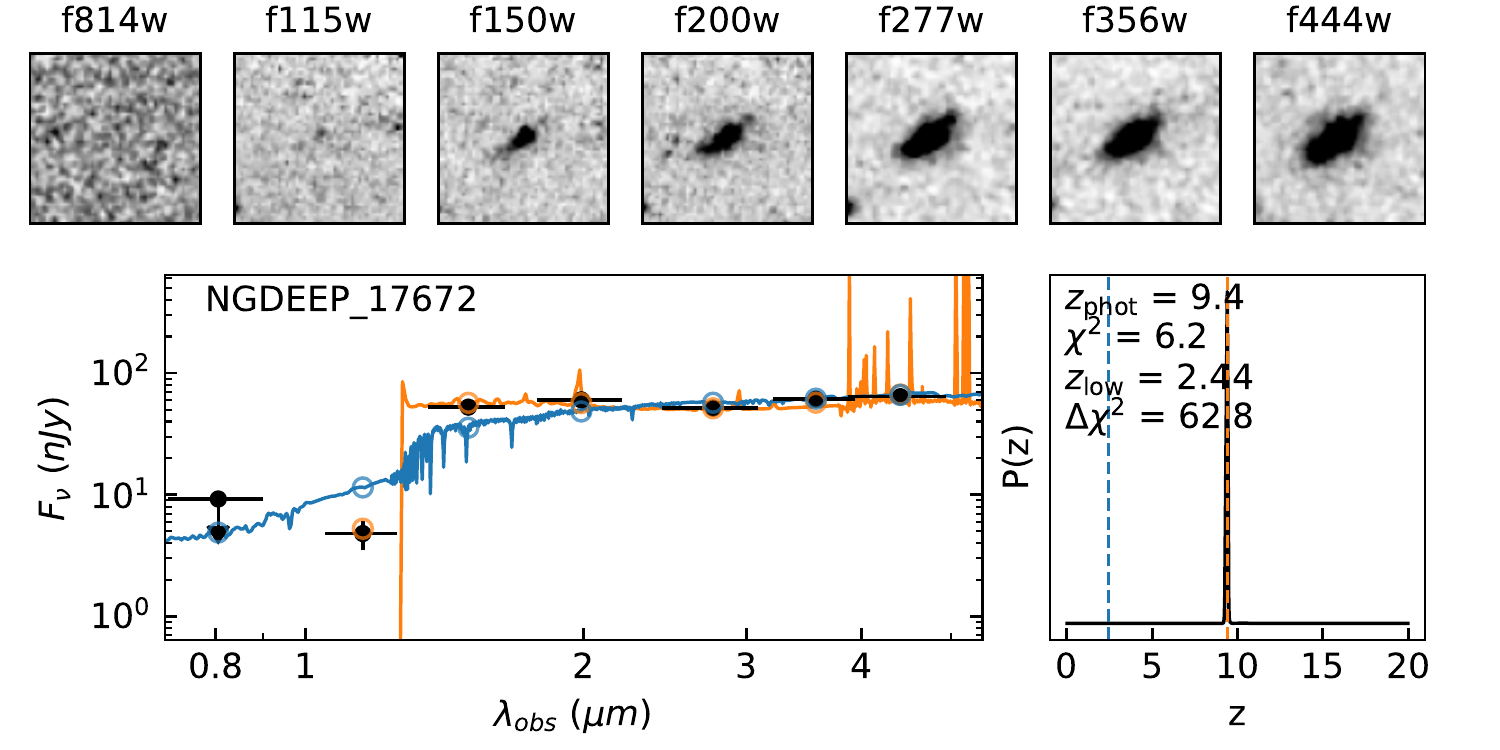}
	\includegraphics[width=0.33\textwidth]{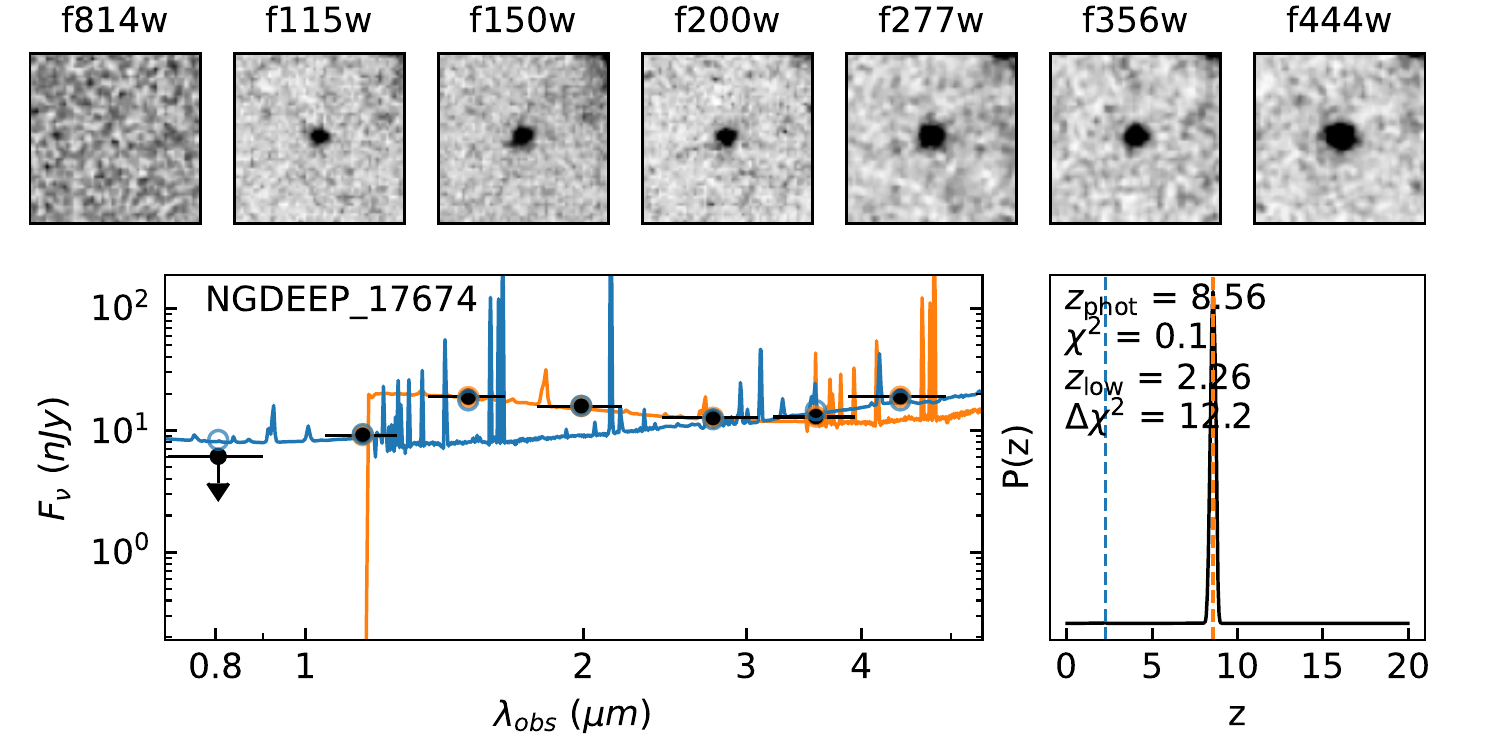}
    \caption{Same as Figure \ref{fig:bio_z15}, but for sources at $z \sim 9$.}\label{fig:bio_z9}
\end{figure*}

\begin{deluxetable*}{lRRRRRR}
\tablewidth{\textwidth}
\tablecaption{NGDEEP $z \ge 9$ Galaxy Sample}\label{tab:samp}
\tablehead{\colhead{ID} & \colhead{R.A.} & \colhead{Dec.} & \colhead{$\m$} & \colhead{$\Delta \chi^2$} & \colhead{$\int P(z>7)$} & \colhead{Photometric}\\
\colhead{} & \colhead{(J2000)} & \colhead{(J2000)} & \colhead{AB mag} & \colhead{} & \colhead{} & \colhead{Redshift}}
\startdata
NGDEEP 250 & 53.249451 & -27.883313 & 29.6 & 15.6 & 0.9994 & 11.62^{+0.21}_{-0.33}\\
NGDEEP 1191 & 53.266583 & -27.876581 & 29.2 & 8.7 & 0.9938 & 12.31^{+1.23}_{-0.30}\\
NGDEEP 1369 & 53.249467 & -27.875710 & 29.1 & 14.1 & 0.9997 & 15.82^{+0.12}_{-0.81}\\
NGDEEP 1716 & 53.251057 & -27.796992 & 30.7 & 6.8 & 0.9658 & 11.29^{+0.30}_{-1.47}\\
NGDEEP 2067 & 53.239797 & -27.800244 & 29.3 & 7.5 & 0.9863 & 10.75^{+0.42}_{-0.66}\\
NGDEEP 2470 & 53.248546 & -27.802743 & 29.8 & 9.5 & 0.9949 & 8.56^{+0.15}_{-0.63}\\
NGDEEP 2497 & 53.248775 & -27.803090 & 29.3 & 6.3 & 0.9808 & 9.19^{+0.60}_{-0.21}\\
NGDEEP 3514 & 53.256875 & -27.807957 & 29.0 & 7.5 & 0.9853 & 11.23^{+0.27}_{-0.90}\\
NGDEEP 4134 & 53.245546 & -27.814372 & 28.4 & 51.7 & 1.0000 & 10.69^{+0.18}_{-0.30}\\
NGDEEP 4330 & 53.264834 & -27.816024 & 29.7 & 5.8 & 0.9736 & 9.16^{+0.96}_{-0.18}\\
NGDEEP 4674 & 53.245601 & -27.817588 & 29.4 & 5.9 & 0.9522 & 8.62^{+0.18}_{-0.21}\\
NGDEEP 4740 & 53.257473 & -27.817828 & 29.6 & 8.9 & 0.9936 & 9.19^{+0.72}_{-0.21}\\
NGDEEP 4919 & 53.262137 & -27.818774 & 30.0 & 5.5 & 0.9684 & 9.34^{+1.02}_{-0.24}\\
NGDEEP 5118 & 53.261009 & -27.819895 & 27.4 & 68.4 & 1.0000 & 9.25^{+0.09}_{-0.06}\\
NGDEEP 5947 & 53.248954 & -27.822988 & 29.8 & 5.3 & 0.9552 & 8.89^{+1.50}_{-0.57}\\
NGDEEP 6134 & 53.248902 & -27.823695 & 29.1 & 24.6 & 1.0000 & 9.07^{+0.09}_{-0.12}\\
NGDEEP 6477 & 53.250469 & -27.825099 & 29.7 & 7.3 & 0.9782 & 8.74^{+0.33}_{-0.30}\\
NGDEEP 6952 & 53.239303 & -27.827168 & 31.2 & 7.0 & 0.9818 & 8.68^{+0.45}_{-0.75}\\
NGDEEP 6980 & 53.256828 & -27.827244 & 30.3 & 10.5 & 0.9985 & 10.06^{+0.45}_{-0.60}\\
NGDEEP 7530 & 53.237017 & -27.829892 & 29.8 & 9.2 & 0.9969 & 8.71^{+0.57}_{-0.45}\\
NGDEEP 7722 & 53.242580 & -27.830637 & 29.5 & 14.9 & 0.9994 & 10.93^{+0.30}_{-0.39}\\
NGDEEP 8024 & 53.235202 & -27.832248 & 30.0 & 18.9 & 1.0000 & 9.79^{+3.48}_{-0.18}\\
NGDEEP 8042 & 53.237915 & -27.832330 & 30.4 & 4.4 & 0.9549 & 8.53^{+0.18}_{-0.99}\\
NGDEEP 8165 & 53.234410 & -27.833172 & 30.0 & 10.2 & 0.9949 & 10.48^{+0.36}_{-0.96}\\
NGDEEP 8427 & 53.235623 & -27.834282 & 30.3 & 13.8 & 0.9996 & 8.98^{+0.18}_{-0.30}\\
NGDEEP 8461 & 53.235797 & -27.834499 & 30.4 & 6.7 & 0.9619 & 10.36^{+0.27}_{-1.26}\\
NGDEEP 8894 & 53.241008 & -27.828822 & 29.8 & 16.7 & 0.9999 & 9.34^{+0.87}_{-0.06}\\
NGDEEP 9261 & 53.270902 & -27.841204 & 29.6 & 11.8 & 0.9988 & 8.74^{+0.15}_{-0.24}\\
NGDEEP 9555 & 53.258685 & -27.847324 & 30.5 & 10.0 & 0.9943 & 10.45^{+0.39}_{-0.84}\\
NGDEEP 10296 & 53.276910 & -27.850568 & 28.5 & 55.9 & 1.0000 & 10.48^{+0.18}_{-0.33}\\
NGDEEP 11522 & 53.242062 & -27.855079 & 29.1 & 19.7 & 1.0000 & 10.84^{+0.36}_{-0.30}\\
NGDEEP 12453 & 53.280001 & -27.858265 & 29.4 & 6.2 & 0.9766 & 9.88^{+0.78}_{-0.54}\\
NGDEEP 13290 & 53.258417 & -27.861651 & 29.1 & 11.2 & 0.9991 & 8.95^{+1.59}_{-0.18}\\
NGDEEP 13406 & 53.240619 & -27.862122 & 30.1 & 6.5 & 0.9809 & 9.31^{+1.05}_{-0.15}\\
NGDEEP 13782 & 53.244589 & -27.863587 & 30.0 & 5.9 & 0.9878 & 9.22^{+1.53}_{-0.18}\\
NGDEEP 15166 & 53.267688 & -27.869336 & 27.8 & 111.3 & 1.0000 & 8.95^{+0.09}_{-0.15}\\
NGDEEP 17672 & 53.249058 & -27.815575 & 27.1 & 62.7 & 1.0000 & 9.4^{+0.03}_{-0.09}\\
NGDEEP 17674 & 53.233517 & -27.816677 & 28.6 & 12.1 & 0.9990 & 8.56^{+0.15}_{-0.15}\\
\enddata
\end{deluxetable*}

\begin{figure*}
	\includegraphics[width=0.33\textwidth]{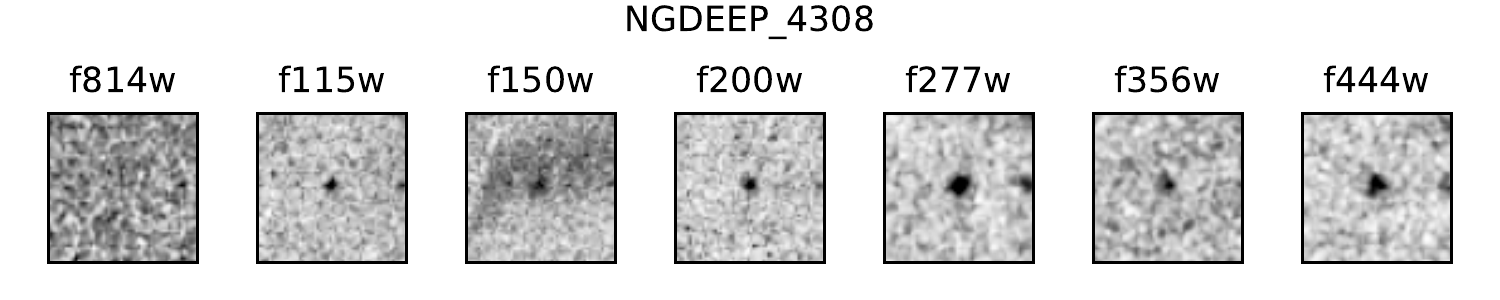}
	\includegraphics[width=0.33\textwidth]{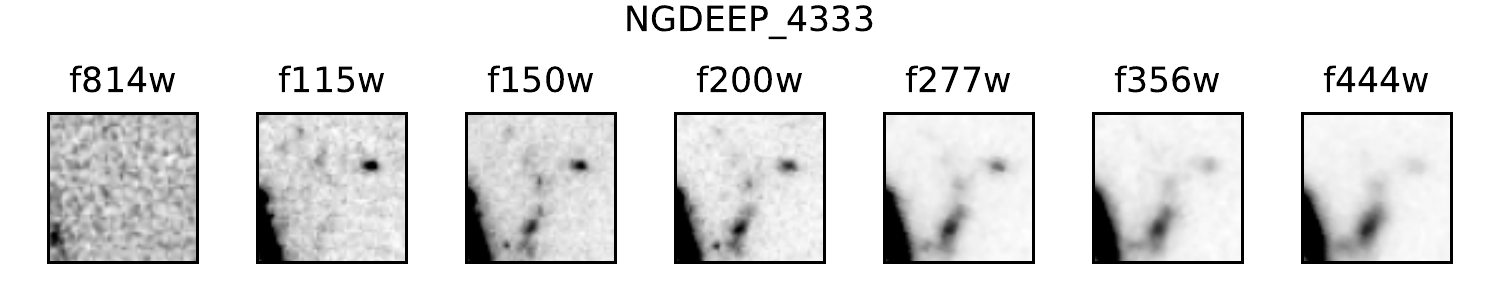}
	\includegraphics[width=0.33\textwidth]{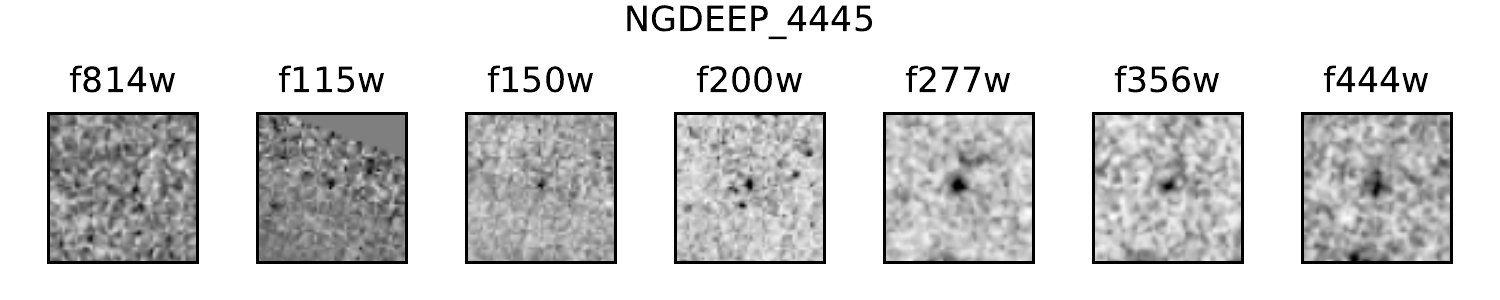}
	\\
	\includegraphics[width=0.33\textwidth]{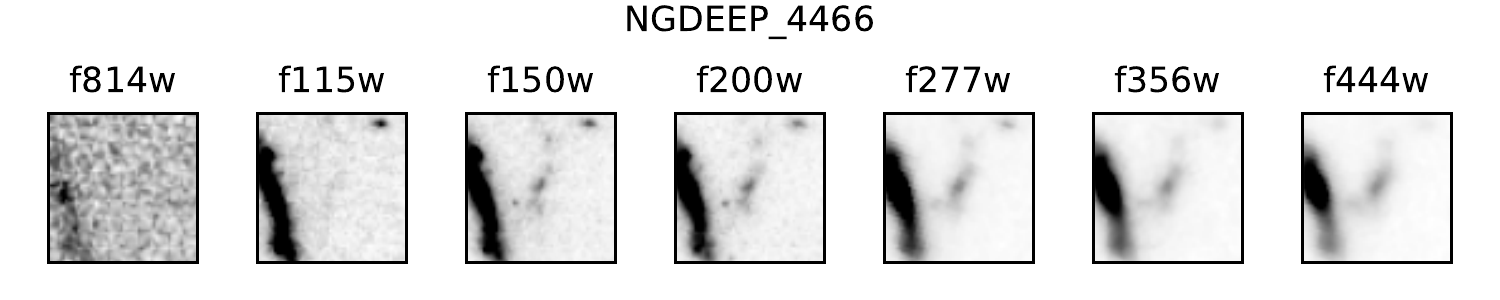}
	\includegraphics[width=0.33\textwidth]{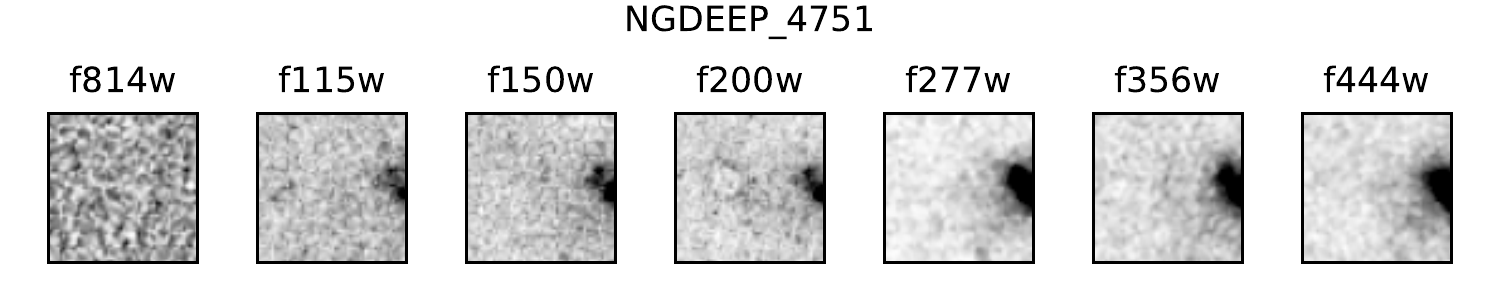}
	\includegraphics[width=0.33\textwidth]{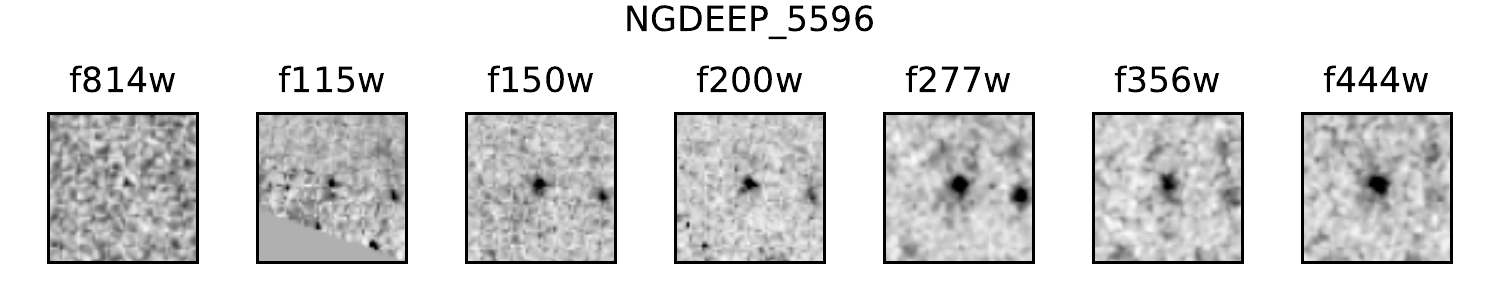}
	\\
	\includegraphics[width=0.33\textwidth]{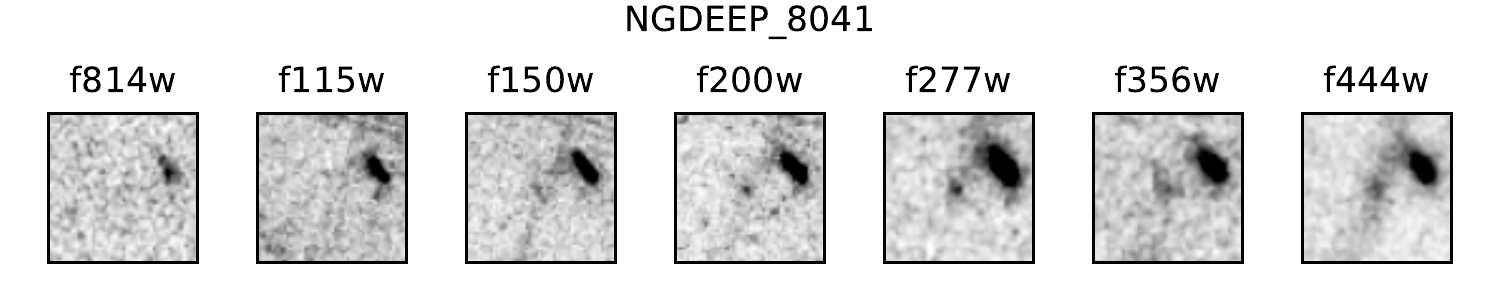}
	\includegraphics[width=0.33\textwidth]{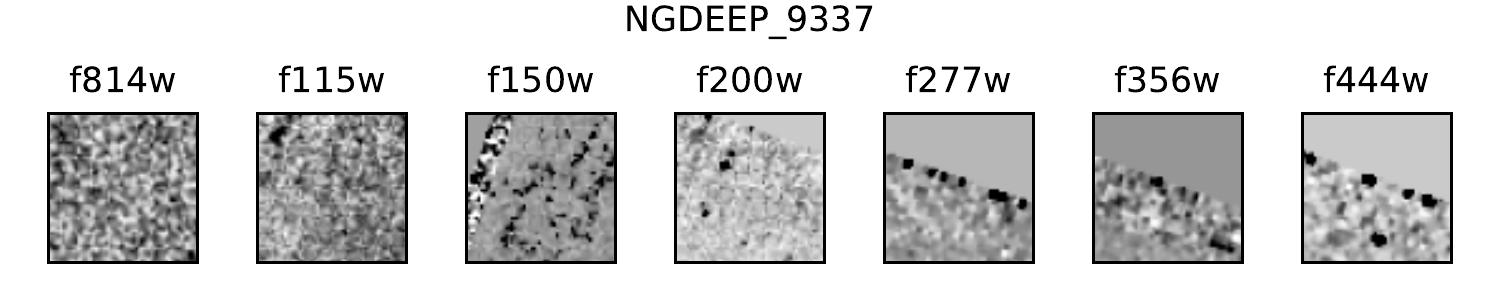}
	\includegraphics[width=0.33\textwidth]{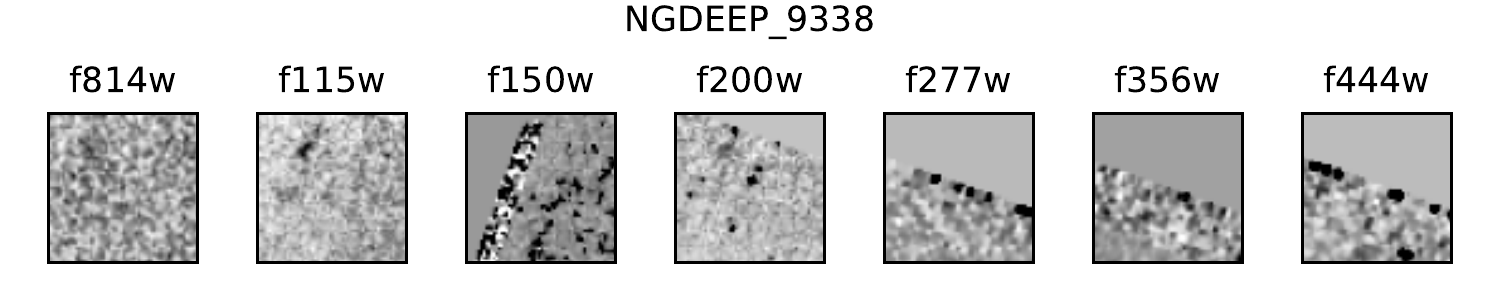}
	\\
	\includegraphics[width=0.33\textwidth]{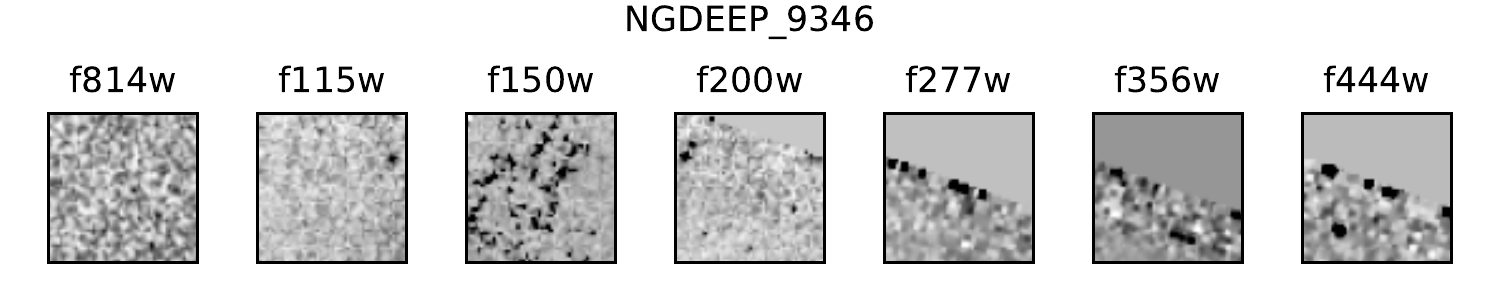}
	\includegraphics[width=0.33\textwidth]{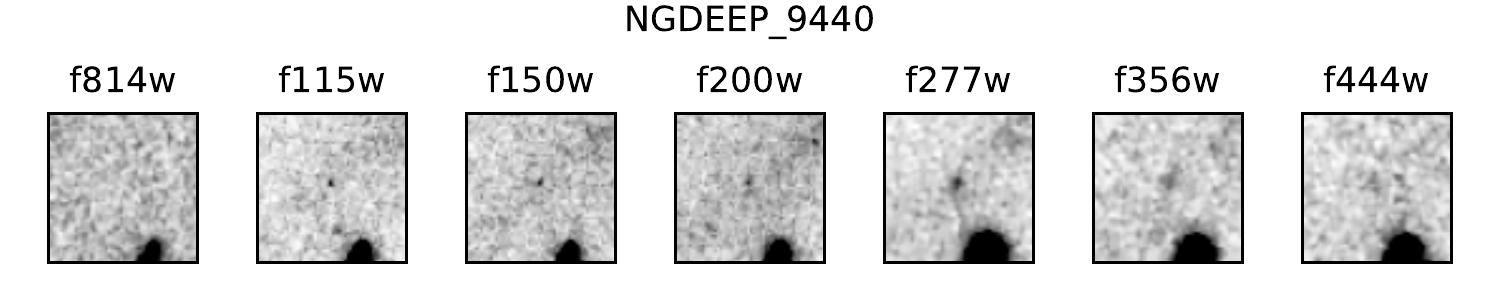}
	\includegraphics[width=0.33\textwidth]{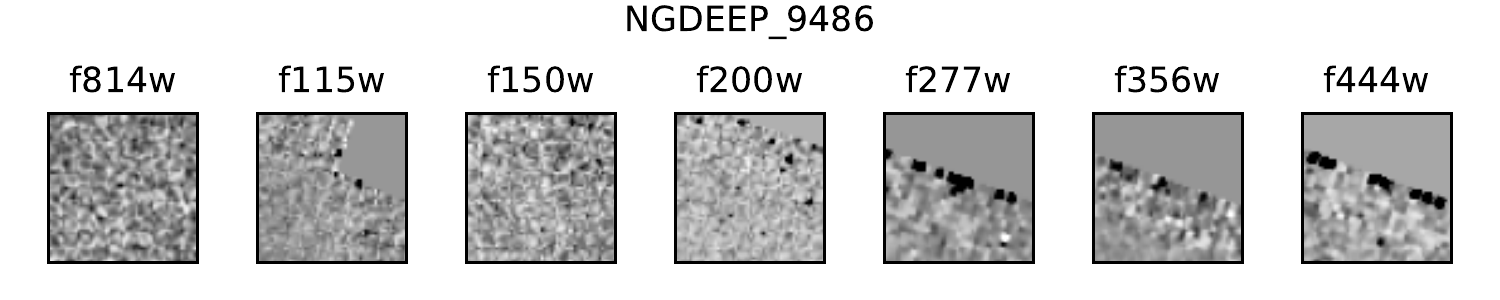}
	\\
	\includegraphics[width=0.33\textwidth]{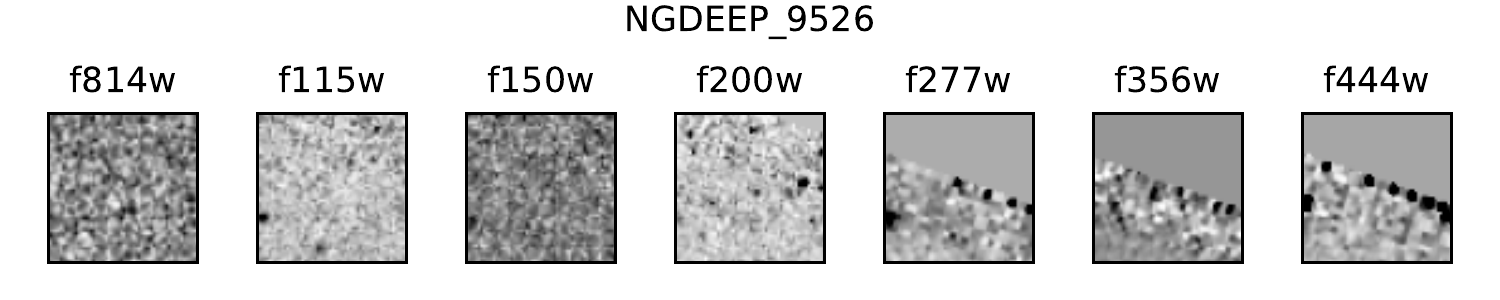}
	\includegraphics[width=0.33\textwidth]{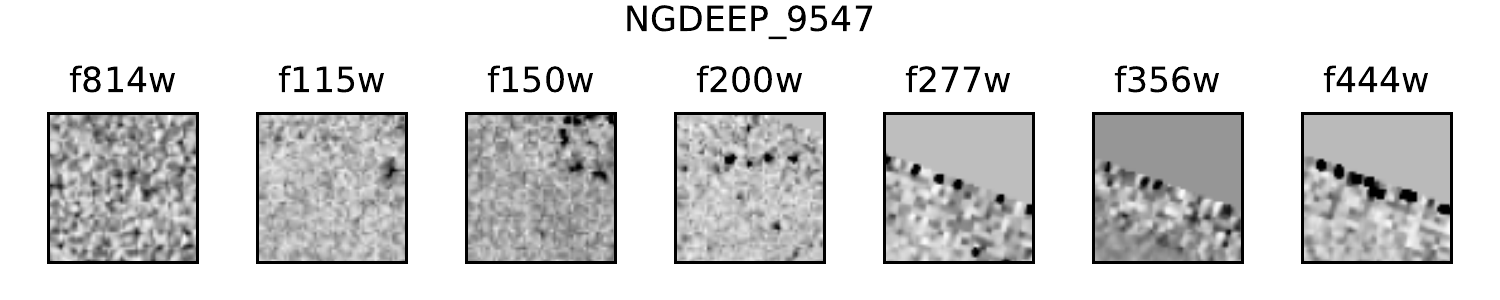}
	\includegraphics[width=0.33\textwidth]{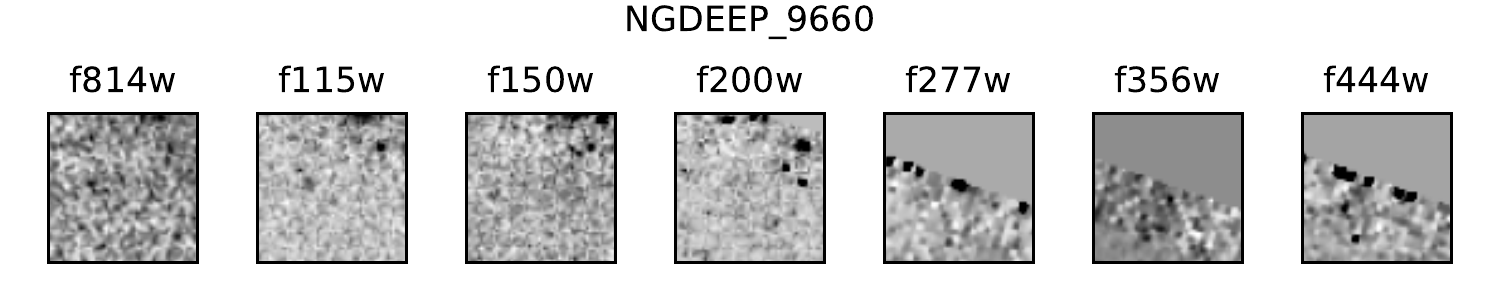}
	\\
	\includegraphics[width=0.33\textwidth]{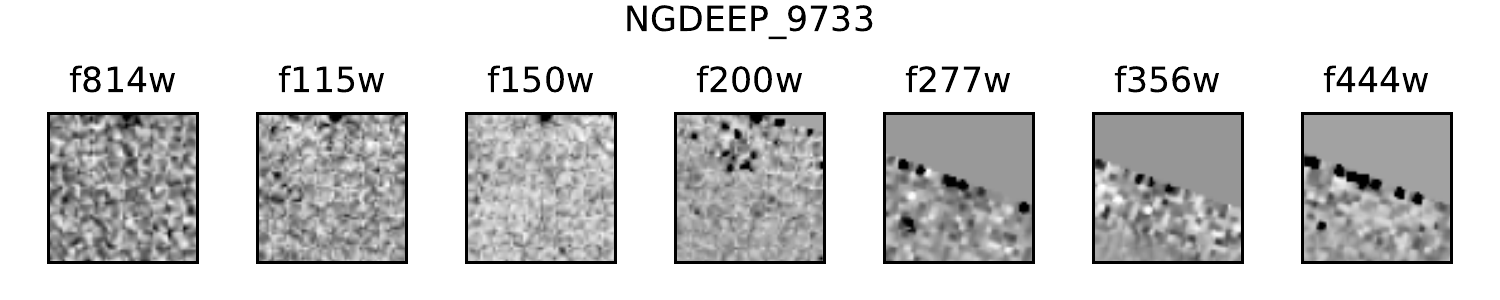}
	\includegraphics[width=0.33\textwidth]{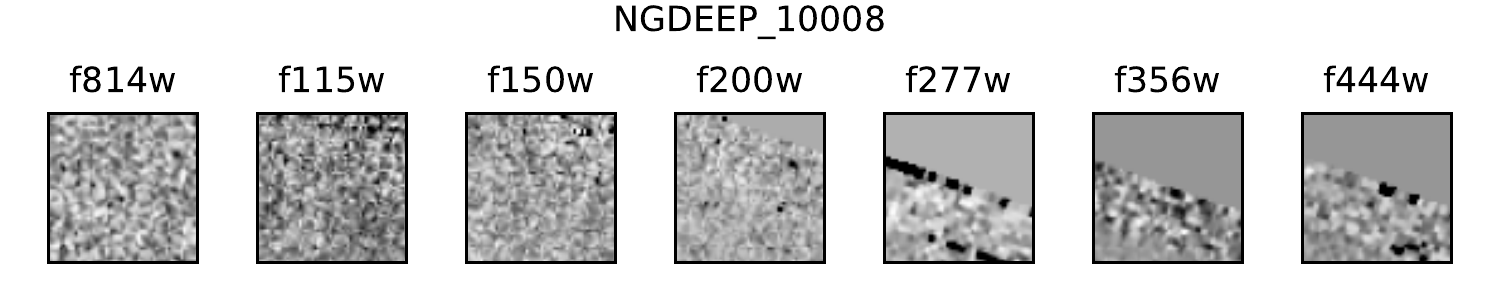}
	\includegraphics[width=0.33\textwidth]{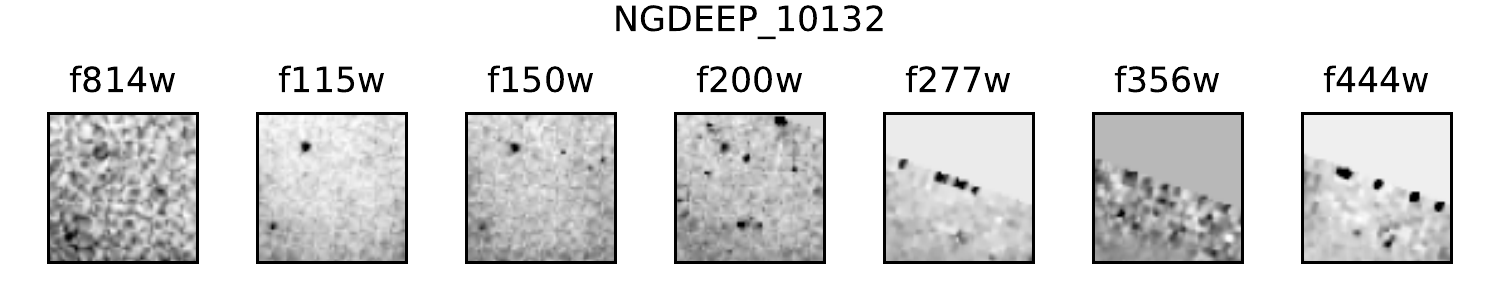}
	\\
	\includegraphics[width=0.33\textwidth]{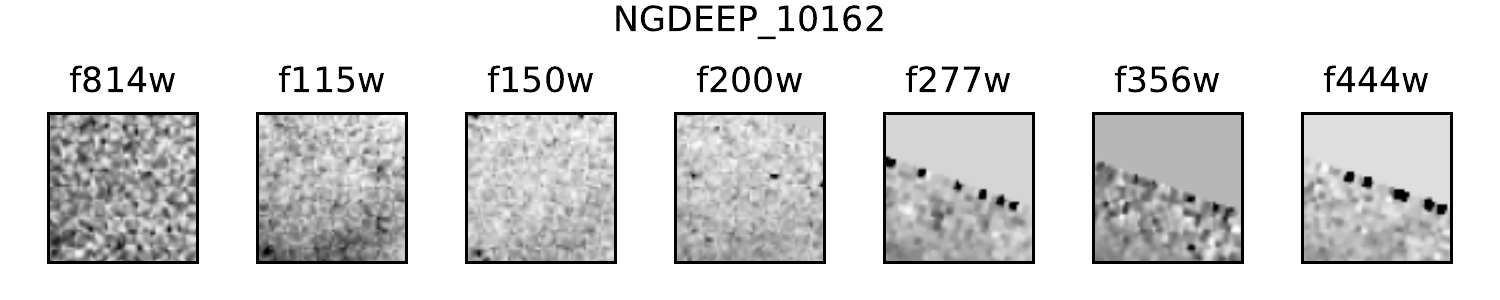}
	\includegraphics[width=0.33\textwidth]{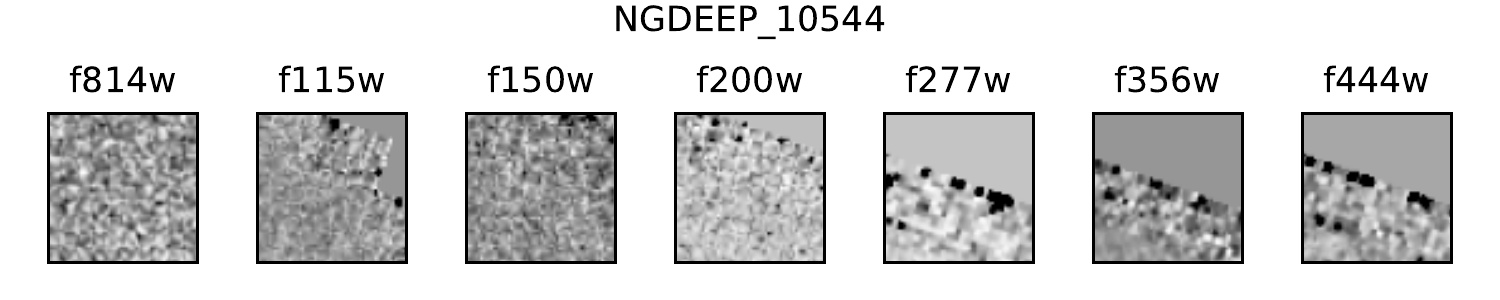}
	\includegraphics[width=0.33\textwidth]{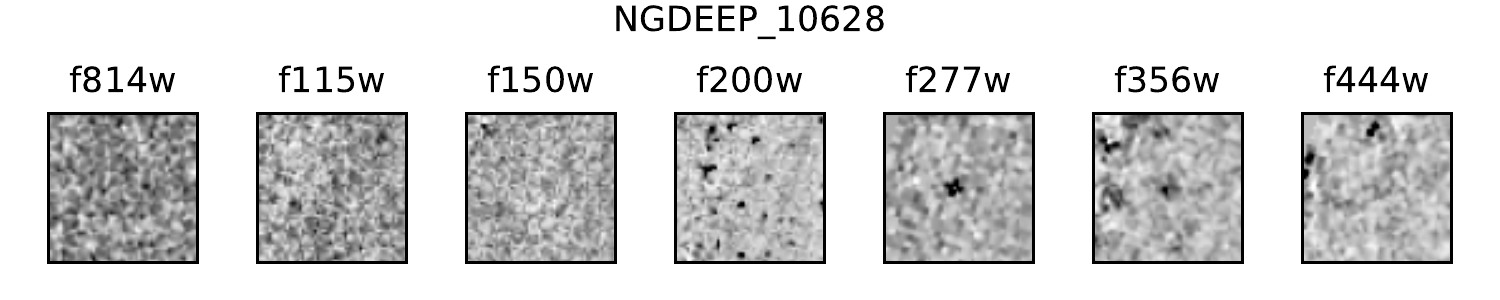}
	\\
	\includegraphics[width=0.33\textwidth]{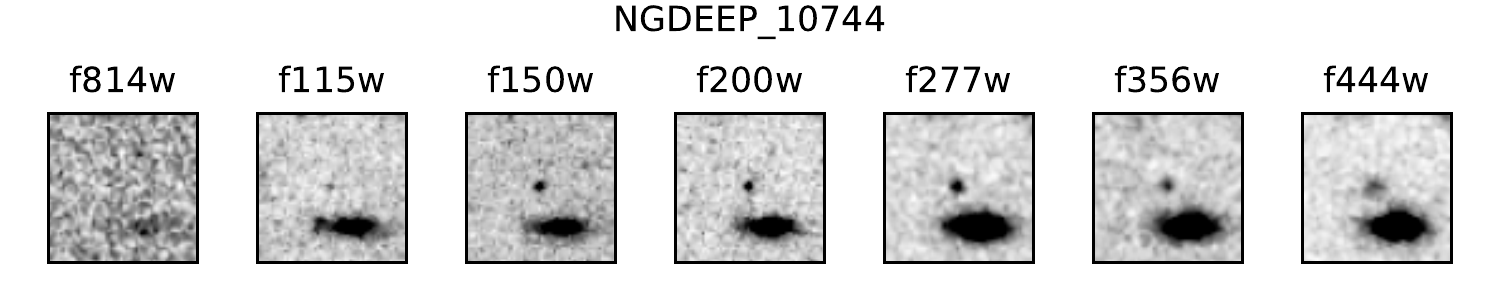}
	\includegraphics[width=0.33\textwidth]{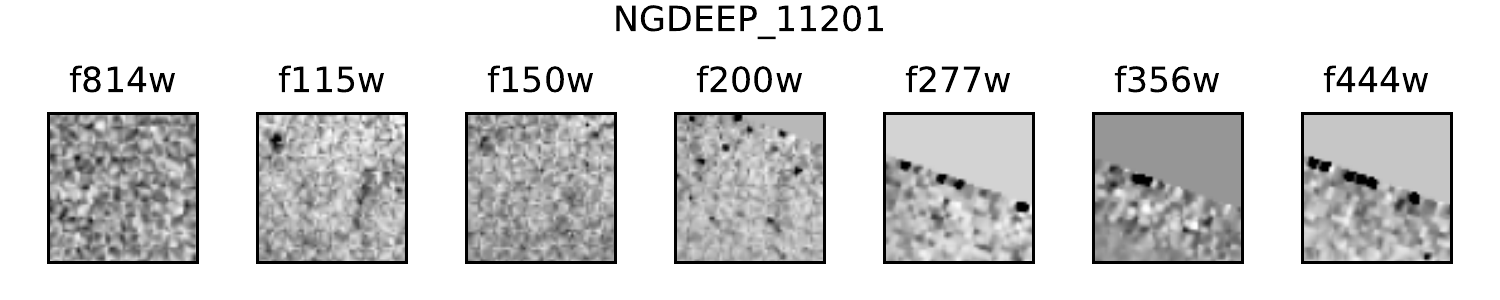}
	\includegraphics[width=0.33\textwidth]{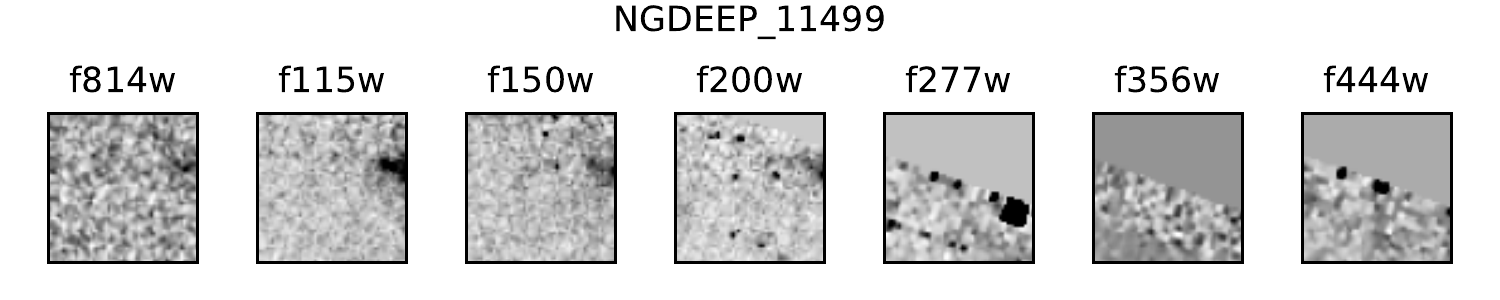}
	\\
	\includegraphics[width=0.33\textwidth]{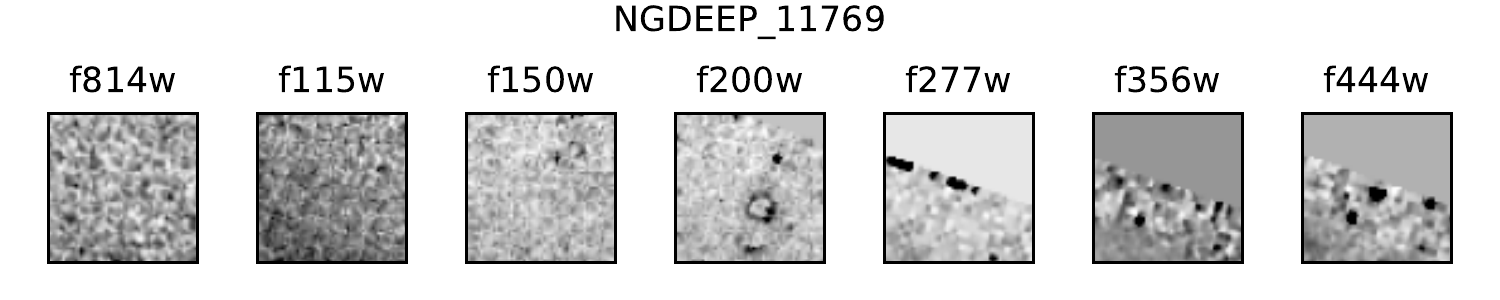}
	\includegraphics[width=0.33\textwidth]{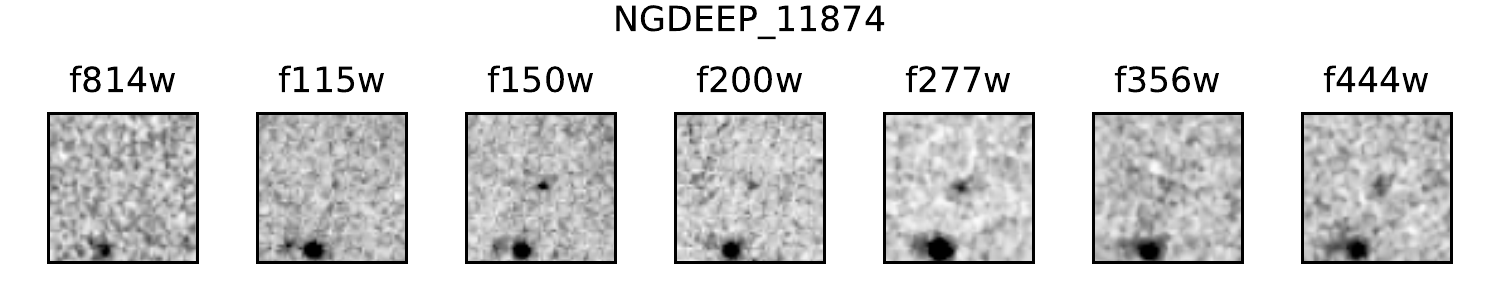}
	\includegraphics[width=0.33\textwidth]{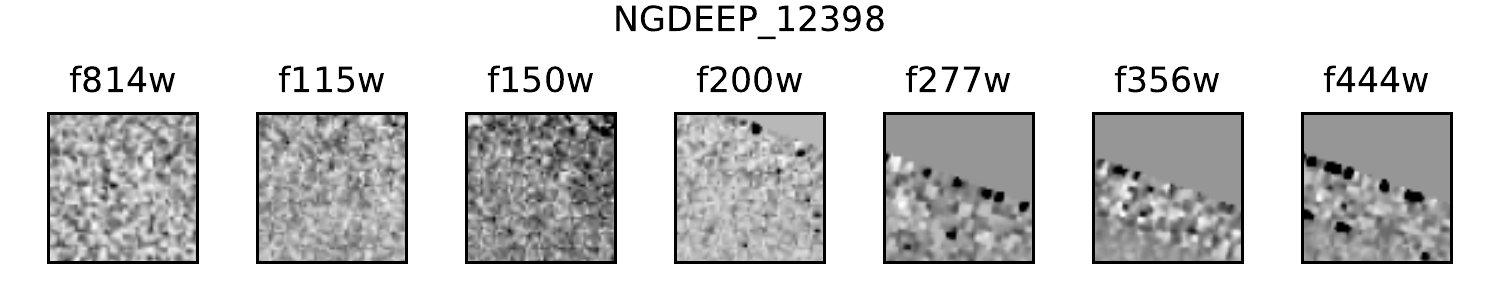}
	\\
	\includegraphics[width=0.33\textwidth]{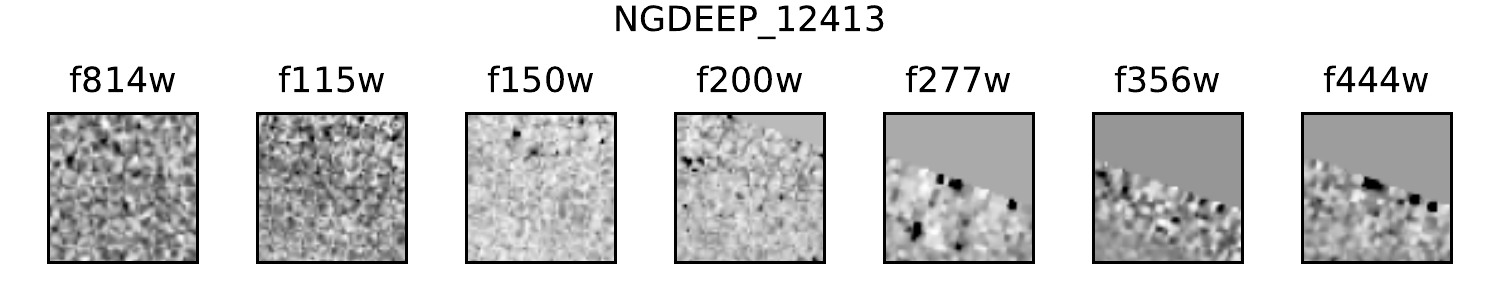}
	\includegraphics[width=0.33\textwidth]{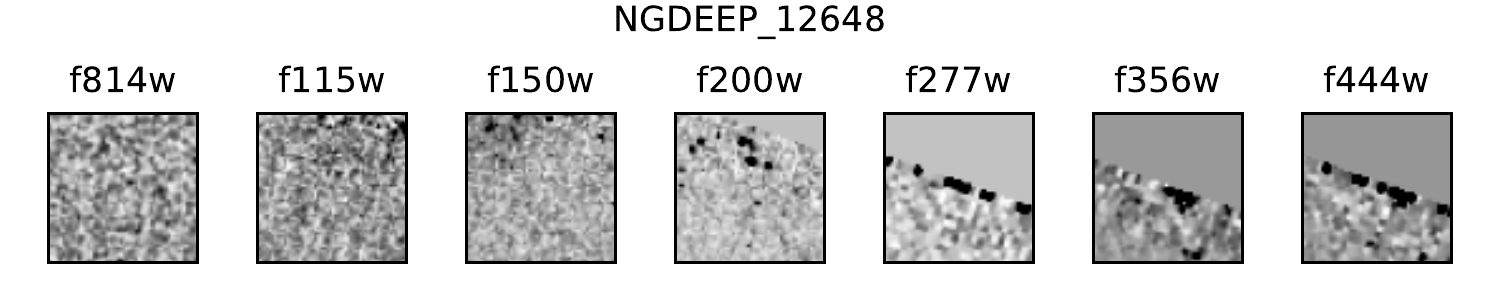}
	\includegraphics[width=0.33\textwidth]{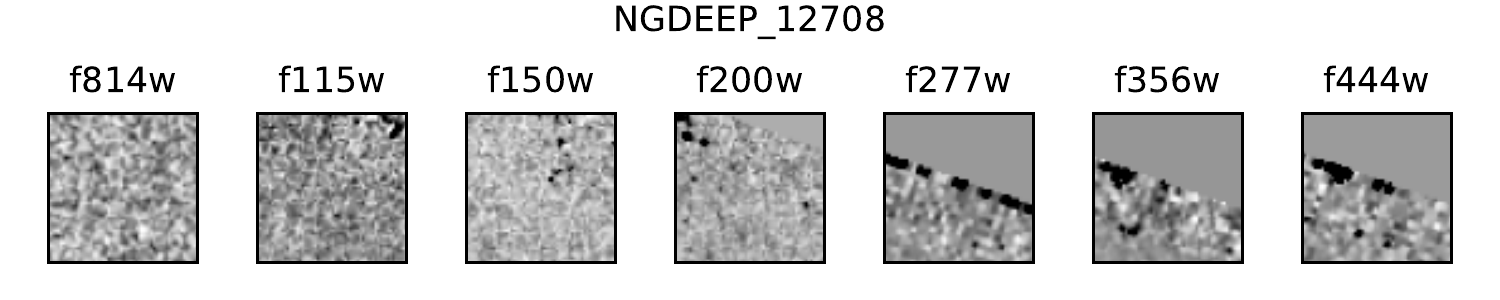}
	\\
	\includegraphics[width=0.33\textwidth]{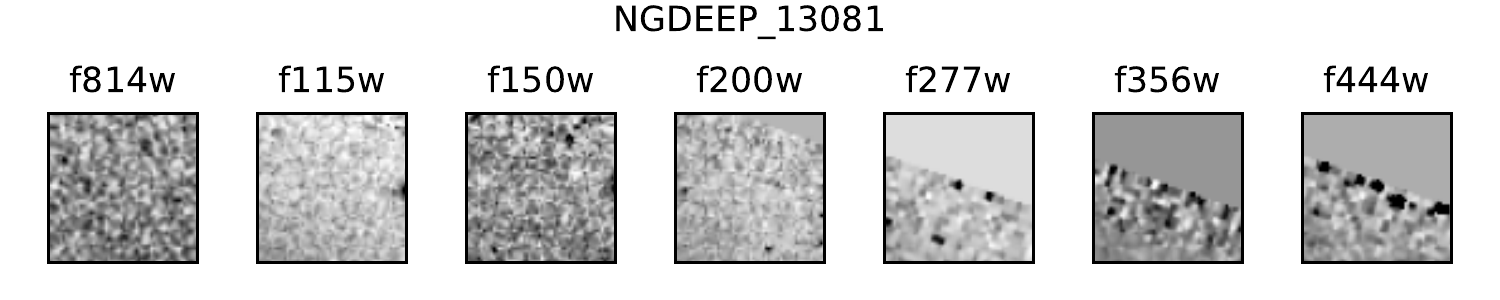}
    \caption{Image cutouts for the 31 sources rejected after visual inspection.}\label{fig:reject}
\end{figure*}

\end{CJK*}
\end{document}